\documentclass[%
reprint,
amsmath,amssymb,
aps,
]{revtex4-1}
\usepackage{graphicx}
\usepackage{amsthm,amsmath}
\usepackage[top=25.4mm, bottom=20.9mm, left=19.1mm, right=19.1mm]{geometry}
\theoremstyle{definition} % age in karo nakonim az jayi ke theorem(inja example) estefade konim, ta akare file fonte hame kharab mishe

\newcommand{\DD}{\, \displaystyle}
\newcommand{\deff}{\, \stackrel{\text{def}}{=}}

\newcommand{\eq}[1]{\,\begin{equation}
#1 
\end{equation}
}

\newcommand{\rond}[2]{\, \frac{\partial #1}{\partial #2}}
\newcommand{\fracc}[2]{\, \displaystyle \frac{ #1}{ #2}}

\newcommand{\morabba}[1]{\,\begin{flushright}
\Rectsteel \\
\end{flushright}}

\newcommand{\eqq}[2]{\,\begin{equation} \label{#2}
#1 
\end{equation}
}

\newcommand{\CC}[2]{\, \binom{#1} {#2} 
}

\newcommand{\all}[2]{\,\begin{align}
#1 
\label{#2}
\end{align}
}

\makeatletter
\newcommand{\vast}{\bBigg@{4}}
\newcommand{\Vast}{\bBigg@{5}}

\makeatother

\begin{document}
\preprint{APS/123-QED}
%\date{}
%opening
\title{  Growing Multiplex Networks with Arbitrary  Number of Layers }

\author{Babak Fotouhi$^{1 }$ and Naghmeh Momeni  $^{2}$   \\
$^{1}$\emph{Department of Sociology, 
McGill University, Montr\'eal, Qu\'ebec, Canada} \\
%$^{\dagger}$Email:\texttt{ babak.fotouhi@mail.mcgill.ca }\\
$^2$\emph{Department of Electrical and Computer Engineering, McGill University, Montr\'eal, Qu\'ebec, Canada } 
}

%\newgeometry{top=19.1mm, bottom=19.1mm, left=19.1mm, right=19.1mm}

\begin{abstract}
This paper focuses on the problem of growing multiplex networks. Currently, the   results on the joint degree distribution of growing multiplex networks  present  in the literature pertain  to the case of two layers,  and  are  confined to  the special case of homogeneous growth, and are limited to the state state (that is,   the limit of infinite size).  In the present paper, we obtain closed-form solutions for the joint degree distribution of heterogeneously growing multiplex networks with arbitrary number of layers in the steady state. Heterogeneous growth means that each incoming node establishes different numbers of links in different layers.   We consider both uniform and preferential growth. We then extend the analysis  of  the uniform growth mechanism to arbitrary times. We obtain a closed-form solution for the time-dependent joint degree distribution of a growing multiplex network with  arbitrary initial conditions. Throughout, theoretical findings are corroborated with Monte Carlo simulations.  The results shed light on   the effects of the initial network on the transient dynamics of growing multiplex networks, and takes a step towards characterizing the temporal variations of the connectivity of growing multiplex networks, as well as predicting their future structural properties. 
\end{abstract}

%\author{Babak Fotouhi, Michael Rabbat}
\maketitle

%\newgeometry{top=19.1mm, bottom=19.1mm, left=19.1mm, right=19.1mm}

\section{Introduction} \label{sec:intro}

The network framework is widely used for   studying complex systems and their properties, through mapping units onto nodes and interactions onto links. The conventional conceptualization consists of a single graph, while many real-world systems exhibit different types of interactions between constituents. Such networked systems in which units have heterogeneous types of interactions can be mathematically modeled under the framework of multiplex networks. In these settings, units (nodes) are members of distinct networks simultaneously. For example, it would be an oversimplified depiction to talk about  \emph{ the }  social network  of a given group of people, because the same pair of individuals can have distinct types of relations at the same time: they can be kins, friends, coworkers, they can be connected on the social media, etc.   In other words, there can be  multiple networks between the same set of people. This  is the basic rationale    behind  the  multiplex  representation of networked systems. 

The multiplex framework envisages different layers housing different types of links between the same set of nodes (so there is one set of nodes, multiples sets of links). For example, we can take a sample of individuals and constitute a social media layer, in which links represent interactions on social media, a kinship layer, a geographical proximity layer, and so on. Examples of studies that have    conceptualized real networked systems  under  the multiplex framework include citation networks~\cite{citation_aps,aps_imdb}, online social media~\cite{facebook,online}, interbank networks~\cite{interbank}, airline networks~\cite{air}, scientific collaboration networks~\cite{citation_aps,battiston}, and web of connections and interactions in online games~\cite{game}. 
%been conceptualized so far using the multiplex framework include citation networks, online social media, airline networks, scientific collaboration networks, and online games~\cite{survey}.

Theoretical tools for characterizing and quantifying the properties of multiplex networks are generalizations of the single-layer scenario to multiple layers~\cite{pion1,pion2,pion3}. For example, the adjacency matrix is generalized to the adjacency tensor, whose $ijk$ element is the weight of the link from node $i$ to node $j$ in layer $k$. Similarly, all nodal attributes which were scalars in the single-layer picture (such as various types of centralities, degree, and  clustering) are generalized to vectors in the multiplex scenario. These new theoretical measures enable  studying various phenomena on top of multiplex networks analytically. Examples include epidemics~\cite{epid1,epid2}, pathogen-awareness interplay~\cite{interplay,interplay2},   percolation processes~\cite{percolation1,percolation2,epid1}, random walks~\cite{RW,RW2}, evolution of cooperation~\cite{cooperation3,cooperation1,cooperation2,matjaz}, diffusion processes~\cite{diffusion} and social contagion~\cite{information}.
 For thorough reviews, see~\cite{survey,matjaz,matjaz2}.

%In the present paper we focus on the problem of growing multiplex networks.

%In~\cite{Bian1}, the number of links that each new incoming node establishes in both layers is equal to $m$. 

Since many real networks are growing in size,  growing multiplex networks have also attracted attention in the literature. 
The mean-field approach is a potent method for investigating the temporal evolution of the degrees of individual nodes, extracting its asymptotic behavior  in order to find the asymptotic (tail behavior) degree distribution of each of the individual layers. This approach is undertaken in~\cite{Bian2,Bian1,coevolution,SUSY}. An alternative approach of tackling network growth problems is the rate equation approach, undertaken in~\cite{Bian2,Bian1,nameni}. The rate equation enables solving for the joint degree distribution of the system, so that we can obtain the fraction of nodes that have a given degree  vector  across layers. 

Previous results on the joint degree distribution of growing multiplex networks---attainable through the rate equation approach---are confined to the case of  homogeneous  two-layer growth, where the number of links established by each new incoming node is the same across layers~\cite{Bian1,Bian2}. Note that the possibility of heterogeneous growth is envisaged in~\cite{Bian2} (in the Supplemental Material therein), and their implications  for  the mean-field scenario are correctly alluded to. In the present paper, we consider different rates of link growth across layers explicitly, and obtain the joint degree distributions. Moreover, we extend the problem to general $M$ layers. To our knowledge, no solution for the  joint degree distribution of growing multiplex networks with  arbitrary number of layers  exists in the literature. Furthermore, the existing results on the joint degree distribution are confined to the steady-state, that is, the limit as ${t \rightarrow \infty}$ where the network has infinite size. The evolution of the joint degree distribution at arbitrary times is hitherto unknown. In this paper, we take a step towards alleviating these two gaps.

%%Previous results on growing multiplex networks are confined to homogeneously-growing layers~\cite{survey,Bian2,Bian1,coevolution}.
% In~\cite{Bian1}, the case where two layers are homogeneously growing (that is, the number of links that each newly-born node establishes is the same for both layers)
%according to preferential attachment is considered, and it is shown that $\overline{\ell}(k)$ (which is the average layer-2 degree of nodes whose layer-1 degree is $k$) is a function of $k$. The joint coupling of degrees is a consequence of arrival times: a node whose degree is high in one layer is likely to have existed in the network for a long time, thus its expected degree in the other layer is also high.

The assumption of link growth heterogeneity is motivated by the empirical studies on the structure of multi-layered interacting systems, which report that different layers generally exhibit different connectivity patterns (consequently, different average degrees, and other structural properties). For example, in~\cite{game}, the connectivity structure of the players of a massive online game is mapped onto a multiplex network of six layers, and the average degrees of the layers are different (ranging from 3 in the most sparse layer to 61  in  the  densest layer). In~\cite{facebook}, the friendship ties of a group of students is mapped onto Facebook friends, picture friends and cohabitation, and these layers are shown to have different connectivity distributions, hence different  average degrees. In~\cite{india}, the Indian airline and railway transportation networks are mapped onto two layers, representing two distinct modes of transportation between geographic locations. The degree distributions of these layers are then depicted, and it is observed that they have different degree distributions (as well as different nearest-neighbor degree distributions). In~\cite{trade}, the international trade network is mapped onto 97 layers, each layer pertaining to one distinct commodity. The connectivity patterns are different across layers. Although in this example the layers represent weighted networks, the assertion that the connectivity patterns are heterogeneous still holds.

In the present paper, we consider heterogeneously-growing layers.  First, we consider a simple two-layer system for expository purposes. We obtain $n(k,\ell)$, the fraction of nodes with degree $k$ in the first layer and degree $\ell$ in the second layer. We also use this result to find $\overline{\ell}(k)$, the expected  layer-2 degree of a node whose degree in layer 1 is $k$. We solve the problem for  the cases of  preferential and uniform growth, separately. We demonstrate that the expression for $\overline{\ell}(k)$ is identical under these two settings. For the special case of homogeneous growth, this result agrees with those of ~\cite{Bian1}. 

We then generalize both the preferential and the uniform setups to $M>2$ layers. Each incoming node establishes $\beta_1,\beta_2,\ldots,\beta_M$ links in layers $1,2,\ldots,M$, respectively. For both uniform and preferential growth we obtain $n(\mathbf{k})$, which is the fraction of nodes with vector degree $\mathbf{k}$. That is, the fraction of nodes with degree $k_1$ in layer 1, degree $k_2$ in layer 2, and so on. 
 In all cases, we corroborate  our  theoretical findings with  Monte Carlo simulations. 
%Throughout the paper, we verify the theoretical findings with Monte Carlo simulations. 

In addition to heterogeneity of connectivity across layers, this paper contributes to the literature by taking a step towards extending the analysis of the network growth  process beyond the steady state, by considering arbitrary times. To that end, in Section~\ref{temporal}  we focus on the uniform attachment model and analyze it  in  more detail. In particular, we study the temporal evolution of the joint degree distribution of a given arbitrary multiplex network, with arbitrary number of layers, whose joint degree distribution is known. We find $n_t(\mathbf{k})$, which is the fraction of nodes with degree vector $\mathbf{k}$ at  arbitrary  time $t$. Through several case studies, we verify the accuracy of our theoretical predictions through Monte Carlo simulations on  multiple example topologies. 
 We consider diverse topologies  in order to examine how the initial conditions influence the evolution of degrees and to ascertain the accuracy of the predictions for different structures. 
 Simulation results are consistently in good agreement with theoretical findings.

\section{Model 1: Preferential Attachment in two Layers}

The network is constructed by a set of nodes and two sets of links. There are two layers, each housing one set of links. This means that node $x$ can have a set of neighbors in layer 1 and a different set of neighbors in layer 2. Similarly, the degree of node $x$ in layer 1 can differ from its degree in layer 2. We denote the degree of node $x$ in layer 1 by $k_x$ and in layer 2 by $\ell_x$. We denote by $N_t(k,\ell)$ the number of nodes that have degree $k$ in the layer 1 and degree $\ell$ in layer 2 at time $t$. The fraction of such nodes is denoted by $n_t(k,\ell)$. 

At the outset, the network comprises $N(0)$ nodes, with $L_1(0)$ links in layer 1 and $L_2(0)$ links in layer 2. The network grows by the successive addition of new nodes---one at each timestep. Each new node, upon birth, establishes $\beta_1$ links to the existing nodes in layer 1 and $\beta_2$ links in layer $2$.

In this model, the probability that node $x$ receives a link in layer 1 from an incoming node is proportional to ${k_x }$. Similarly, in layer 2, the probability that node $x$ receives a link from the newcomer is proportional to ${\ell_x }$. 
At time $t$, when a new node is introduced the values of $N(k, \ell)$ can consequently change. If a node with degree $k-1$ in layer 1 and degree $\ell$ in layer 2 receives a link in the first layer, its degree in layer 1 increments and turns into $k$, and $N(k,\ell)$ increments consequently. If a node with degree $k$ in layer 1 and degree $\ell-1$ in layer 2 receives a link in layer 2, its degree in layer 2 increments and consequently, $N(k, \ell)$ increments. Moreover, if a node that has degree $k$ in layer 1 and degree $\ell$ in layer 2 receives a link in either of the layers, $N(k, \ell)$ decrements consequently. Finally, each incoming node has degree $\beta_1$ in layer 1 and degree $\beta_2$ in layer 2 upon birth, so each new incoming node increments $N(\beta1 ,\beta2)$ by one. The following rate equation captures the evolution of the expected value of $N_t(k ,\ell)$ by addressing the said events with their respective probabilities of occurrence: 

\all{
& N_{t+1}( k, \ell )  
= N_t( k, \ell ) 
\nonumber \\
&
+
\beta_1 \fracc{(k-1 ) N_t(k-1,\ell)- k N_t(k,\ell) }{L_1(0)+ 2\beta_1 t}
\nonumber \\
&
+
\beta_2 \fracc{(\ell-1) N_t(k,\ell-1)- \ell N_t(k,\ell) }{L_2(0)+ 2\beta_2 t}
+
\delta_{k \beta_1} \delta_{\ell \beta_2} 
.}{rate_1}

We can use this to write the rate equation for $n(k,\ell)$. Using the substitution ${n(k,\ell)=(N(0)+t) n(k,\ell)}$, we obtain
\all{
& \big[ N(0)+t \big] \big[ n_{t+1}(k,\ell) - n_t(k,\ell) \big] 
+ n _{t+1}(k,\ell) 
= 
\nonumber \\
&
+
\beta_1 \fracc{(k-1 ) N_t(k-1,\ell)- k N_t(k,\ell) }{L_1(0)+ 2\beta_1 t}
\nonumber \\
&
+
\beta_2 \fracc{(\ell-1) N_t(k,\ell-1)- \ell N_t(k,\ell) }{L_2(0)+ 2\beta_2 t}
+
\delta_{k \beta_1} \delta_{\ell \beta_2} 
.}{rate_2}

Now we focus on the steady state, that is, the limit as ${t \rightarrow \infty}$ (the validity of this assumption is verified through simulations below). In this time regime, the time variations of ${n(k,\ell)}$ vanish, and we also have the following simplifications 
\all{
\begin{cases}
\DD \lim_{t \rightarrow \infty} \beta_1 \frac{N(0)+t}{L_1(0)+ 2\beta_1 t} = \fracc{1}{2 } \\ 
\DD \lim_{t \rightarrow \infty} \beta_2 \frac{N(0)+t}{L_2(0)+ 2\beta_2 t} = \fracc{1}{2} 
\end{cases}
.}{limits_1}

So~\eqref{rate_2} transforms into the following equation in the steady state:
\all{
n(k ,\ell) = &\fracc{(k-1 ) n(k-1,\ell) - k n(k, \ell)}{2 } 
\nonumber \\
&
+ \fracc{(\ell-1 ) n(k ,\ell-1) - \ell n(k ,\ell)}{2 } +
\delta_{k \beta_1} \delta_{\ell \beta_2} 
,}{difference_1}
 where we have dropped the $t$ subscript in the steady state. 
This can be equivalently expressed as follows
\all{
n(k,\ell) =& \fracc{k-1}{k+\ell+2} n(k-1,\ell) 
+
\fracc{\ell-1}{k+\ell+2} n(k, \ell-1) 
\nonumber \\ &
+ \fracc{2
\delta_{k \beta_1} \delta_{\ell \beta_2} }{2+\beta_1+\beta_2}
.}{difference_1}

This difference equation is solved in Appendix~\ref{app:sol_1}. The  solution  is
\all{
\resizebox{\linewidth}{!}
{$ n(k,\ell) = 
\fracc{2\beta_1(\beta_1+1) \beta_2(\beta_2+1)}{(2+\beta_1+\beta_2) k(k+1)\ell(\ell+1)}
\fracc{\CC{\beta_1+\beta_2+2}{\beta_1+1} }{\CC{k+\ell+2}{k+1}}
\CC{k-\beta_1+\ell-\beta_2}{k-\beta_1}
$}
}{nkl_FIN_1}
This agrees with the findings in~\cite{nameni}, and those in~\cite{Bian1,Bian2}---in the special case of $\beta_1=\beta_2$. 

This result can be also simplified and equivalently expressed in the following form: 
\all{
n(k,\ell) = &
\fracc{2(1+\beta_1+\beta2 )!}{(2+k + \ell)!}
\fracc{  (k-1)!(\ell-1)!}{(\beta_1-1)!(\beta_2-1)!} 
\nonumber \\ &  \times
\fracc{(k -\beta_1+\ell-\beta_2)!}{(k -\beta_1)!(\ell-\beta_2)!}
.}{nkl_FIN_1_simple}

\begin{figure}[h]
\centering
\includegraphics[width=.95 \columnwidth]{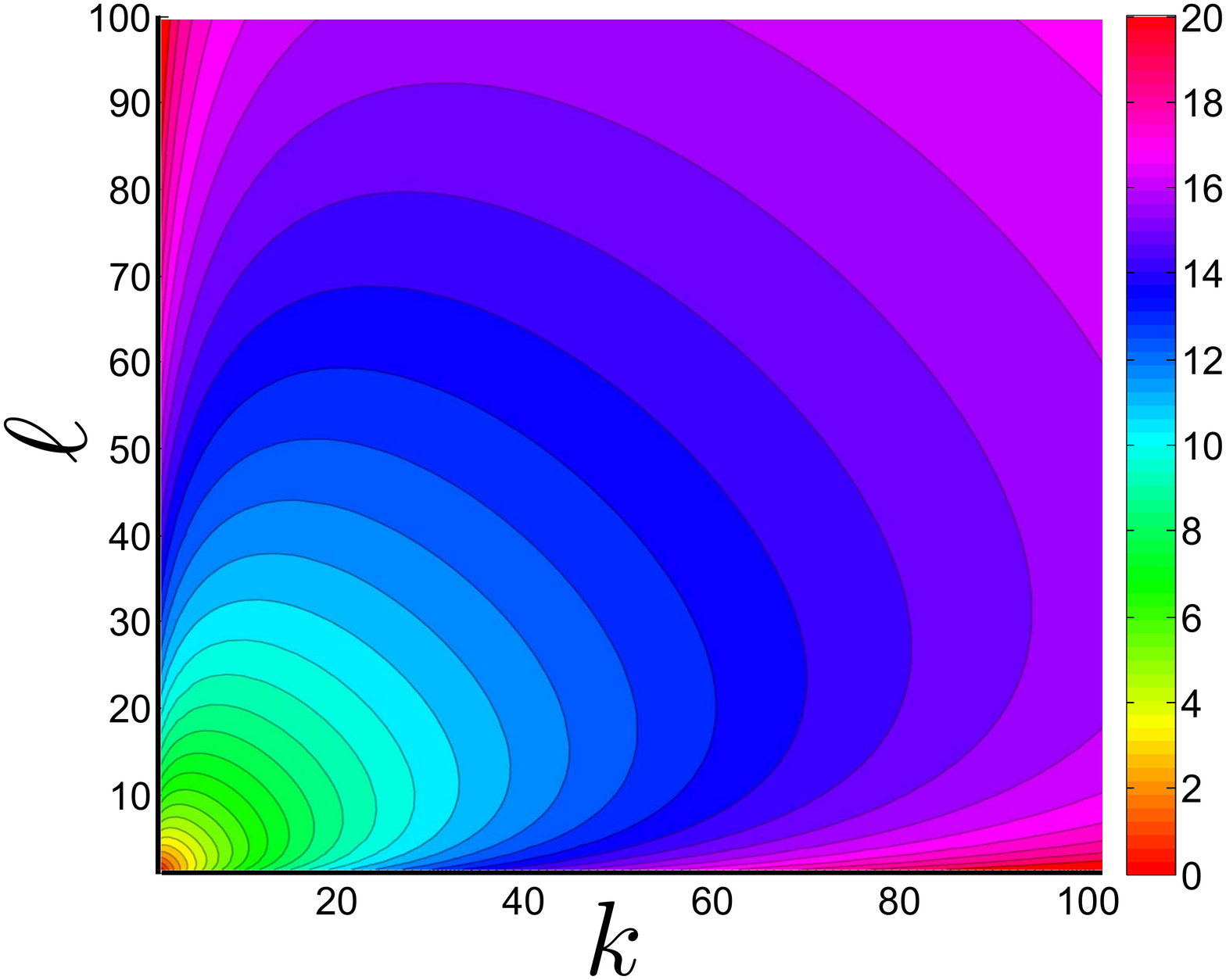}
\caption[Figure ]%
{(Color online) The   joint degree distribution for preferential growth with $\beta_1=3$ and $\beta_2=3$. Theoretical result is presented in~\eqref{nkl_FIN_1}. Since the values decay fast in $k$ and $\ell$, we have depicted the logarithm of the inverse of this function, for a smoother output and better visibility. The joint distribution attains its maximum at $k=\beta_1$ and $\ell=\beta_2$. The contours are symmetric with respect to the bisector because $\beta_1$ and $\beta_2$ are equal. Note that each axis begins at its corresponding value of $\beta$, so the origin is $\beta_1,\beta_2$.
}
\label{fig_3_3}
\end{figure}

\begin{figure}[h]
\centering
\includegraphics[width=.95 \columnwidth]{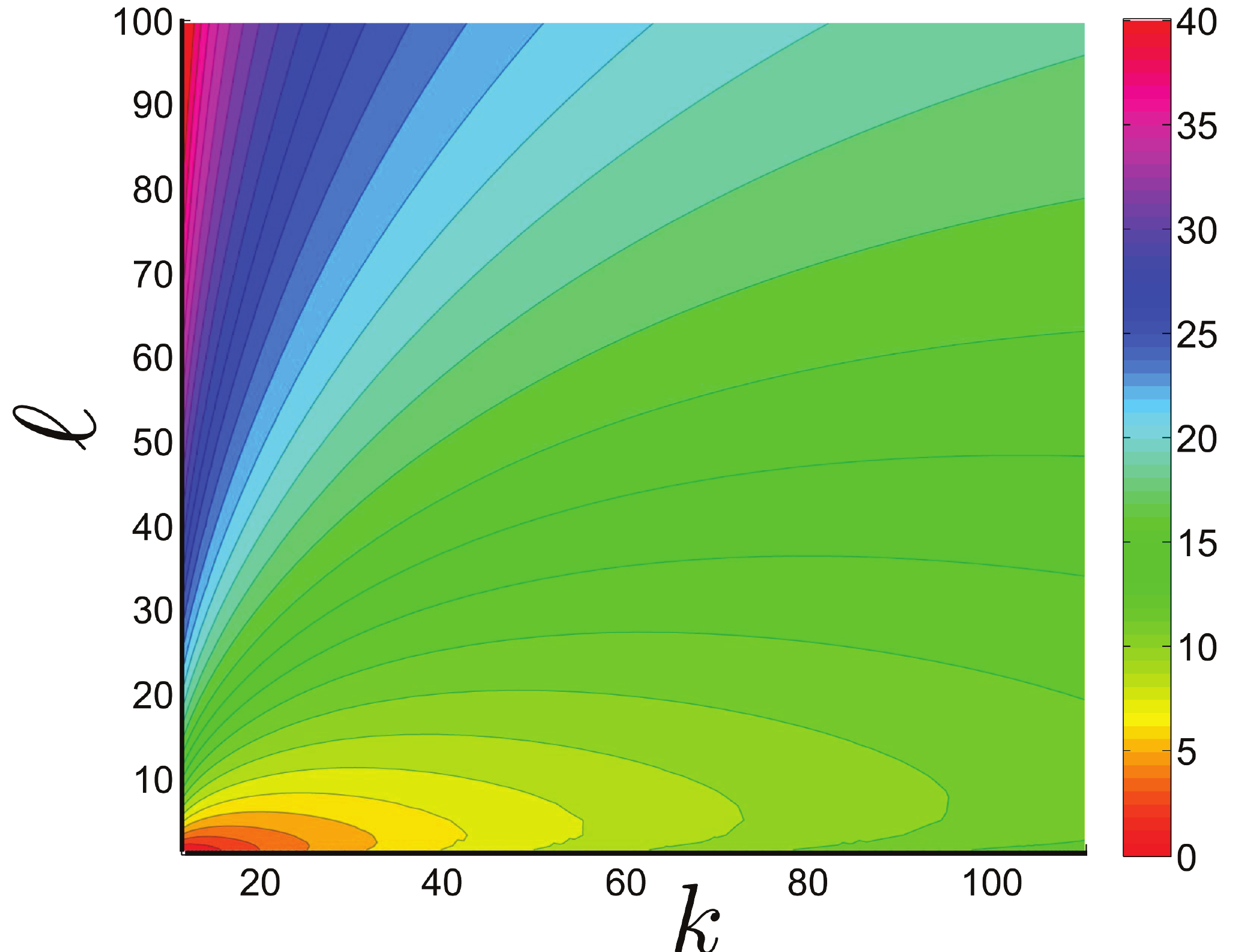}
\caption[Figure ]%
{(Color online) logarithm of the inverse of the  joint degree distribution for preferential growth with $\beta_1=10$ and $\beta_2=2$. Theoretical solution is presented in~\eqref{nkl_FIN_1}. The joint distribution attains its maximum at $k=\beta_1$ and $\ell=\beta_2$. The contours are skewed towards the $k$ axis because $\beta_1>\beta_2$. Note that each axis begins at its corresponding value of $\beta$, so the origin is $\beta_1,\beta_2$.
}
\label{fig_10_2}
\end{figure}

Figure~\ref{fig_3_3} is a depiction of the joint degree distribution for the symmetric case of $\beta_1=\beta_2=3$. Figure~\ref{fig_10_2} pertains to the asymmetric case of $\beta_1=10$ and $\beta_2=10$. Note that in both cases, the origin is at ${(\beta_1,\beta_2)}$, because  the joint degree distribution is zero if for all $k <\beta_1$  or  $\ell<\beta_2$.

%The case of directed networks is not dramatically different. Suppose that the link reception probability of each node is proportional to its in-degree. In this case, every new incoming node has in-degree zero upon birth. The normalization constant (which the degrees of individual nodes must be divided by in order to give their link-reception probabilities) is the sum of in-degrees, which is equal to the number of links in the steady-state (unlike the directed case, for which this was twice the number of links). Taking these considerations into account, the directed analog of~\eqref{difference_1} becomes
%\all{
%\resizebox{\linewidth}{!}{$
%n(k,\ell) = (k-1 ) n(k-1,\ell) - k n(k,\ell) 
%+ (\ell-1 ) n_{k ,\ell-1} - \ell n(k,\ell) +
%\delta_{k 0} \delta_{\ell 0} 
%.
%$}
%}{difference_1_dir}
%This can be rearranged as follows
%
%\all{
%\resizebox{\linewidth}{!}{$
%n(k,\ell) = \fracc{ (k-1 ) n(k-1,\ell)}{k+\ell+1}
%+ (\ell-1 ) n_{k ,\ell-1} - \ell n(k,\ell) +
%\delta_{k 0} \delta_{\ell 0} 
%.
%$}
%}{difference_2_dir}
%

\begin{figure}[h]
\centering
\includegraphics[width=.95 \columnwidth]{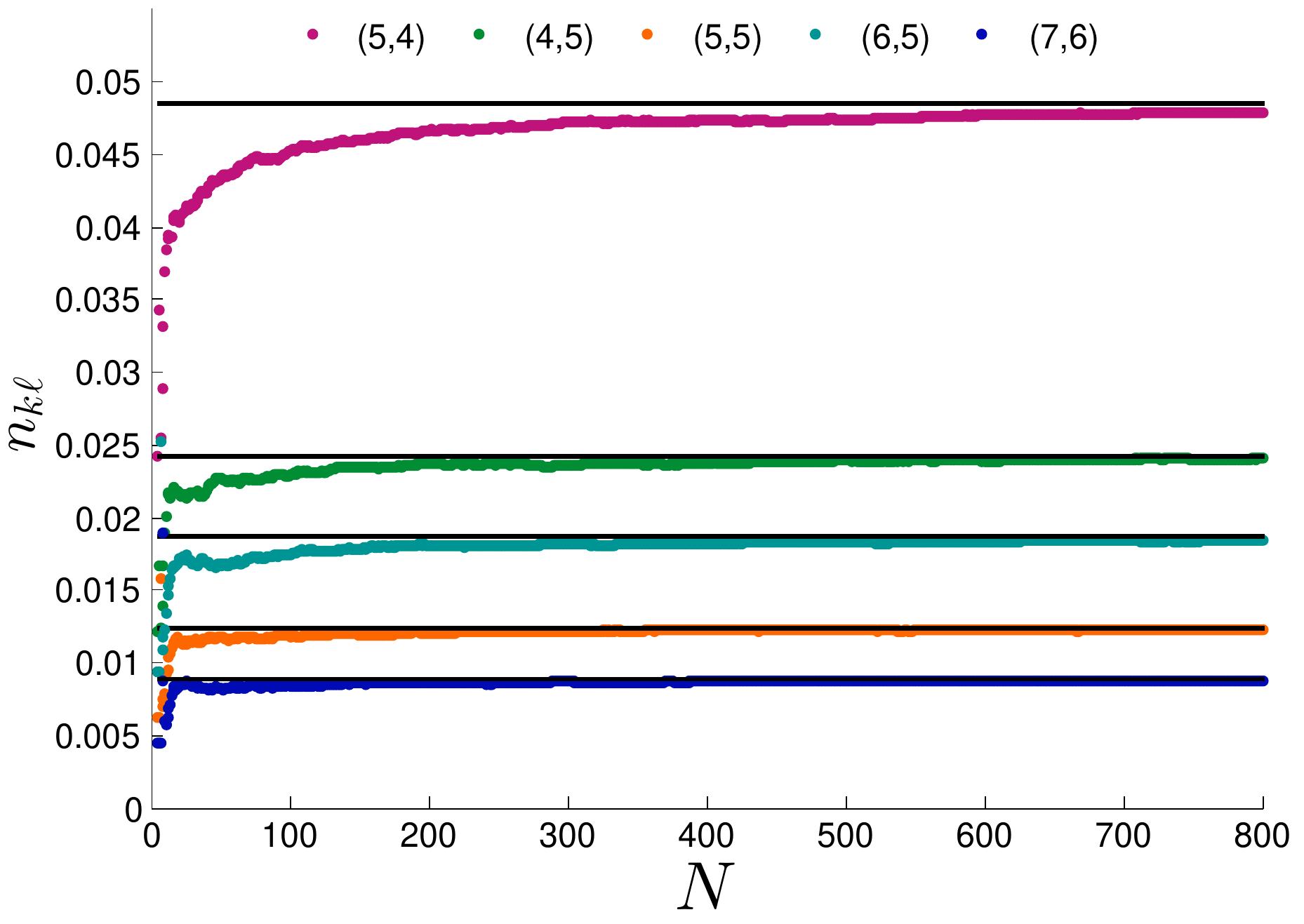}
\caption[Figure ]%
{(Color online) Comparison between the steady-state joint degree distribution and the theoretical prediction of~\eqref{nkl_FIN_1} for growth under preferential attachment with   growth parameters   $\beta_1=4$ and $\beta_2=3$.  The horizontal lines represent the steady-state asymptotes which accord with~\eqref{nkl_FIN_1}, and markers represent simulation results. The initial seed networks are  star  graphs with  6  nodes in both layers.  Several example value of $(k,\ell)$ are chosen for illustrative purposes.  The results are averaged over 100 Monte Carlo trials. 
}
\label{ss_pref_b1_4_b2_3_star_N0_5}
\end{figure}

\begin{figure}[h]
\centering
\includegraphics[width=.95 \columnwidth]{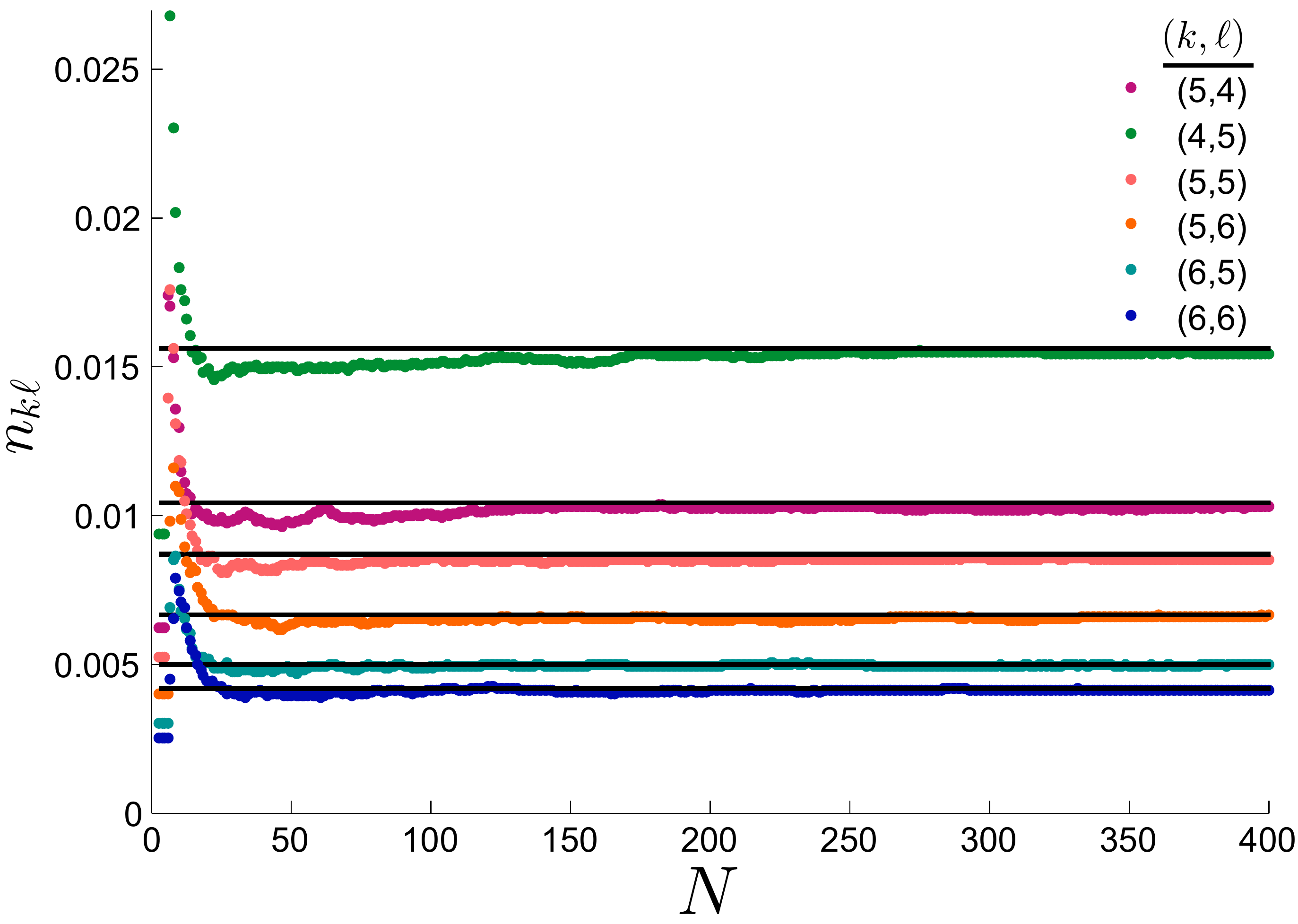}
\caption[Figure ]%
{(Color online) Comparison between the steady-state joint degree distribution and the theoretical prediction of~\eqref{nkl_FIN_1} for growth under preferential  attachment with growth parameters   $\beta_1=2$ and $\beta_2=3$. The horizontal lines represent the steady-state asymptotes which accord with~\eqref{nkl_FIN_1}, and markers represent simulation results. The initial seed networks are  complete graphs with 4 nodes in both layers.  Several example value of $(k,\ell)$ are chosen for illustrative purposes.  The results are averaged over 100 Monte Carlo trials. 
}
\label{ss_pref_complete_b1_2_b2_3_N0_3_GOOD}
\end{figure}

Let us verify~\eqref{nkl_FIN_1} via simulations. For the first setup, we consider $\beta_1=4$ and $\beta_2=3$, and  plot $n(k,\ell)$ for multiple example instances  of $k$ and $\ell$. The initial seed networks are identical star graphs with 6 nodes in both layers. The results are depicted in Figure~\ref{ss_pref_b1_4_b2_3_star_N0_5}. We observe that the simulation results reach the horizontal asymptotes predicted by~\eqref{nkl_FIN_1} relatively early on in the process. When the size of the network is $N=300$, the effects of the initial seed graph  are already negligible. 
 For the second example, we consider   $\beta_1=2$ and $\beta_2=3$. For the initial seed  network, we consider  complete graphs with 4 nodes in both layers. The results are depicted in Figure~\ref{ss_pref_complete_b1_2_b2_3_N0_3_GOOD}. We again observe convergence to the theoretical predictions. Moreover, we observe that equilibrium emerges when the size of the network is 20 times the size of the initial seed network.  Note that the asymptotes  in Figure~\ref{ss_pref_b1_4_b2_3_star_N0_5} and those of Figure~\ref{ss_pref_complete_b1_2_b2_3_N0_3_GOOD} differ, because they pertain to different growth parameters.

We can see the entanglement between the two layers through the conditional average degree, which can be derived from~\eqref{nkl_FIN_1}. For the nodes who have degree $k$ in layer 1, we can find their average degree in layer 2. Let us denote this quantity by $\bar{\ell}(k)$. In the calculations, the marginal degree distribution of layer 1   is invoked. This is obtained in~\cite{dorog_nk,redner2}, and we also obtain it   in Equation~\eqref{1D} in Section~\ref{prefM}---as a byproduct of our calculations. We use this expression in the calculation of the conditional degree. To find the conditional degree,  we need to perform the following summation:
\all{
\bar{\ell}(k) &= \DD \sum_{\ell} \ell n(\ell|k)= \DD \sum_{\ell} \ell \fracc{n(k,\ell)}{n_{k}}
\nonumber \\
&=
\DD \sum_{\ell} \ell \fracc{\frac{2\beta_1(\beta_1+1) \beta_2(\beta_2+1)}{(2+\beta_1+\beta_2) k(k+1)\ell(\ell+1)}
\frac{\CC{\beta_1+\beta_2+2}{\beta_1+1} }{\CC{k+\ell+2}{k+1}}
\CC{k-\beta_1+\ell-\beta_2}{k-\beta_1}}{\frac{2\beta_1(\beta_1+1)}{k(k+1)(k+2)}}
\nonumber \\ &
= 
\DD \sum_{\ell} 
\frac{ \beta_2(\beta_2+1)(k+2) }{(2+\beta_1+\beta_2) (\ell+1) }
\frac{\CC{\beta_1+\beta_2+2}{\beta_1+1}\CC{k-\beta_1+\ell-\beta_2}{k-\beta_1 } }{\CC{k+\ell+2}{k+1}}
\nonumber \\ &
= 
\DD \sum_{\ell} 
\frac{ \beta_2(\beta_2+1) }{(2+\beta_1+\beta_2) }
\frac{\CC{\beta_1+\beta_2+2}{\beta_1+1}\CC{k-\beta_1+\ell-\beta_2}{k-\beta_1 } }{\CC{k+\ell+2}{\ell}}
}{lbar_k_1}

In Appendix~\ref{app:nk_1}, we perform this summation. The answer is
\all{
\bar{\ell}(k)= \fracc{\beta_2}{\beta_1+1} (k+2)
.}{lbar_1}
%we prove that the following holds
%\all{
%\sum_{\ell=0}^{\infty} n(k,\ell) = \fracc{2 \beta_1 (\beta_1+1)}{k (k+1)(k+2)} u(k-\beta_1)
%,}
%which is the degree distribution of layer 1, and is consistent with the previous results in the literature~\cite{}. 

In the special case of $\beta_1=\beta_2=m$, this reduces to $\frac{m(k+2)}{1+m}$, which is consistent with the previous result in the literature~\cite{Bian1}.

We now focus on the distribution of total degree (i.e., the sum of degrees in the two layers).  Let us denote $k+\ell$ by  $q$. The joint distribution of $q,k$ is simply ${n(k,q-k)}$. If we sum over all possible values of $k$, we get the distribution of $q$. Note that $k$ is at least $\beta_1$, because  every incoming node has an initial degree of $\beta_1$  in the first layer upon birth. Similarly, note that $q-k_1$ is at least $\beta_2$. Taking these two into account for the summation bounds, we have: 
\all{
n(q)=&\fracc{2\beta_1(\beta_1+1) \beta_2(\beta_2+1)\CC{\beta_1+\beta_2+2}{\beta_1+1}}{(2+\beta_1+\beta_2)}
\nonumber \\ &
\times \DD \sum_{k=\beta_1}^{q-\beta_2}
\fracc{ \CC{q-\beta_1 -\beta_2}{k-\beta_1}}{ k(k+1)(q-k)(q-k+1)\CC{q+2}{k+1}}
\nonumber \\ 
=&
\resizebox{.85\linewidth}{!}{$
\fracc{2\beta_1(\beta_1+1) \beta_2(\beta_2+1)\CC{\beta_1+\beta_2+2}{\beta_1+1}}{(2+\beta_1+\beta_2)(q+2)(q+1)q(q-1)}
  \DD \sum_{k=\beta_1}^{q-\beta_2}
\fracc{  \CC{q-\beta_1 -\beta_2}{k-\beta_1}}{\CC{q-2}{k-1}}
.
$}
}{sum_nq1}

We use the following identity:
\all{
 \DD \sum_{k=\beta_1}^{q-\beta_2}
\fracc{  \CC{q-\beta_1 -\beta_2}{k-\beta_1}}{\CC{q-2}{k-1}}
=\fracc{ (\beta_1-1)! (\beta_2-1)!(q-1)}{(\beta_2+\beta_1-1)!}
.}{sum_idenq}

This is proven in Appendix~\ref{app:idenq}. Plugging this result  into~\eqref{sum_nq1}, we arrive at
\all{
n(q)=\fracc{2(\beta_1+\beta_2)(\beta_1+\beta_2+1)}{q(q+1)(q+2)} u(q-\beta_1-\beta_2)
,}{nqfin_1}
where $u(\cdot) $ is the Heaviside step function. This is identical to the expression for a single-layer network growing under growth parameter ${\beta_1+\beta_2}$. In other words, the aggregated network is tantamount to one which grows under the preferential attachment mechanism where each incoming node establishes $\beta_1+\beta_2$ links to existing nodes, which is intuitively expected.

\section{Model 2: Uniform Attachment in  Two  Layers}

In this model, we assume that each incoming node establishes its link in both layers by selecting existing nodes uniformly at random. 
The rate equation~\eqref{rate_2} in the case of uniform attachment transforms into
\all{
& N_{t+1}(k, \ell )
= N_t({k, \ell})
+
\beta_1 \fracc{ N_t(k-1,\ell)- N_t(k,\ell) }{N(0)+ t}
\nonumber \\
&
+
\beta_2 \fracc{ N_t(k,\ell-1)- N_t(k,\ell) }{N(0)+ t}
+
\delta_{k \beta_1} \delta_{\ell \beta_2} 
.}{rate_2_u}

Using the substitution $n_{k , \ell}(t) = \frac{N_t(k,\ell)}{N(0)+t}$, this becomes
\all{
& \big[ N(0)+t \big] \big[ n_{t+1}(k,\ell) - n_t(k,\ell) \big] 
+ n _{t+1}(k,\ell) 
= 
\nonumber \\
&
+
\beta_1 \fracc{ N_t(k-1,\ell)- N_t(k,\ell) }{N(0)+ t}
\nonumber \\
&
+
\beta_2 \fracc{ N_t(k,\ell-1)- N_t(k,\ell) }{N(0)+ t}
+
\delta_{k \beta_1} \delta_{\ell \beta_2} 
,}{rate_3_u}

which simplifies to the following difference equation in the limit as ${t \rightarrow \infty}$, that is, the steady state:
\all{
n(k,\ell)=&\beta_1 \fracc{n(k-1,\ell)-n(k,\ell)}{1}+\beta_2\fracc{n (k,\ell-1)-n(k,\ell)}{1}
\nonumber \\ &
+ \delta_{k,\beta_1} \delta_{\ell,\beta_2}.
}{difference_2_temp}
This can be simplified and equivalently expressed as follows
\all{
\resizebox{\linewidth}{!}{$
n(k,\ell)= \fracc{\beta_1}{1+\beta_1+\beta_2} n(k-1,\ell) + \fracc{\beta_2}{1+\beta_1+\beta_2} n(k,\ell-1)
+ \fracc{ \delta_{k,\beta_1} \delta_{\ell,\beta_2}}{1+\beta_1+\beta_2}.
$}
}{difference_2}
This difference equation is solved in Appendix~\ref{app:sol_2}. The solution is
\all{
n(k, \ell) = \fracc{ \beta_1^{k-\beta_1} \beta_2^{\ell-\beta_2} \CC{k-\beta_1+\ell-\beta_2}{k-\beta_1}}{(1+\beta_1+\beta_2)^{k-\beta_1+\ell-\beta_2+1}}
.}{nkl_FIN_2}
This result agrees with the findings in~\cite{nameni}, and those in~\cite{Bian1,Bian2}---in the special case of $\beta_1=\beta_2$.

\begin{figure}[h]
\centering
\includegraphics[width=.95 \columnwidth]{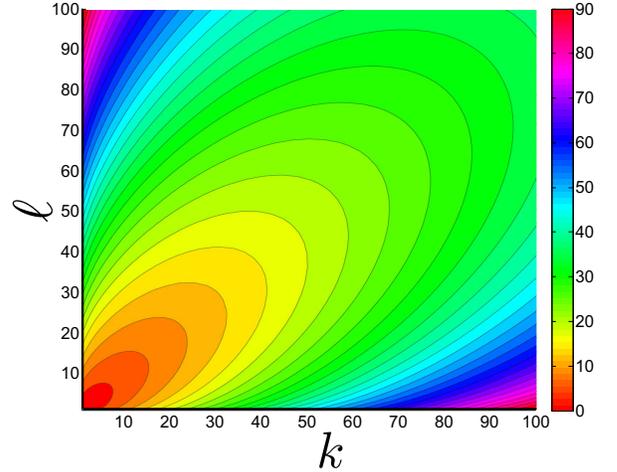}
\caption[Figure ]%
{(Color online) The   joint degree distribution for uniform growth with $\beta_1=3$ and $\beta_2=3$. Theoretical result is presented in~\eqref{nkl_FIN_2}. Since the values decay fast in $k$ and $\ell$, we have depicted the logarithm of the inverse of this function, for a smoother output and better visibility. The joint distribution attains its maximum at $k=\beta_1$ and $\ell=\beta_2$. The contours are symmetric with respect to the bisector because $\beta_1$ and $\beta_2$ are equal. Note that each axis begins at its corresponding value of $\beta$, so the origin is $\beta_1,\beta_2$.
}
\label{U_3_3}
\end{figure}

\begin{figure}[h]
\centering
\includegraphics[width=.95 \columnwidth]{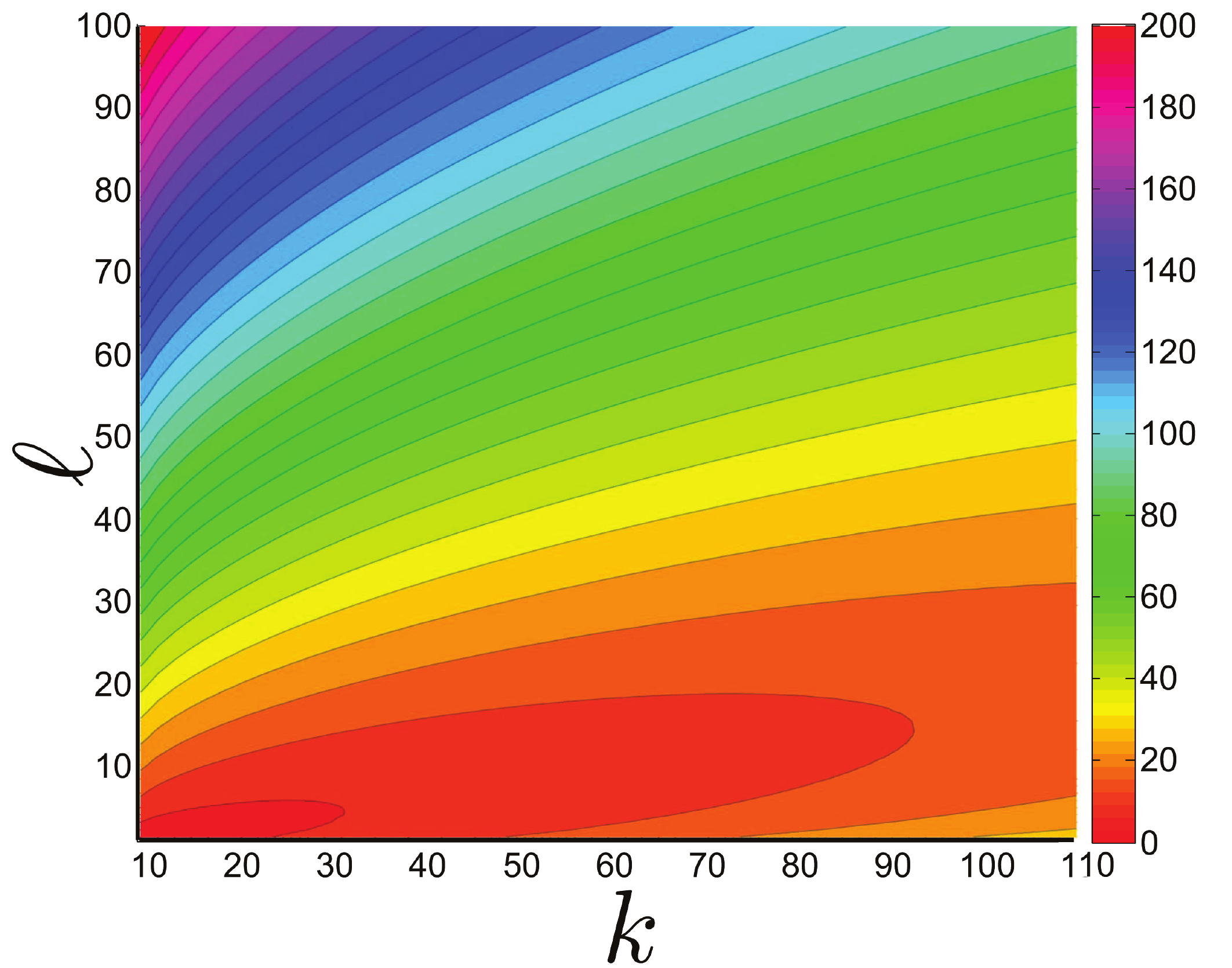}
\caption[Figure ]%
{(Color online) logarithm of the inverse of the  joint degree distribution for uniform growth with $\beta_1=10$ and $\beta_2=2$. Theoretical solution is presented in~\eqref{nkl_FIN_1}. The joint distribution attains its maximum at $k=\beta_1$ and $\ell=\beta_2$. The contours are skewed towards the $k$ axis because $\beta_1>\beta_2$.  Note that each axis begins at its corresponding value of $\beta$, so the origin is $\beta_1,\beta_2$.
}
\label{U_10_2}
\end{figure}

Figure~\ref{U_3_3} represents the joint degree distribution for the case of $\beta_1=\beta_2=3$. We plotted the logarithm of the inverse of this function for smoothness and visibility purposes. Figure~\ref{U_10_2} depicts the joint distribution for the case of $\beta_1=10$ and $\beta_2=2$. Note that in both cases, since the joint degree distribution is nonzero only for $k>\beta_1$ and $\ell>\beta_2$, the origin is situated at $\beta_1,\beta_2$.

\begin{figure}[h]
\centering
\includegraphics[width=.95 \columnwidth]{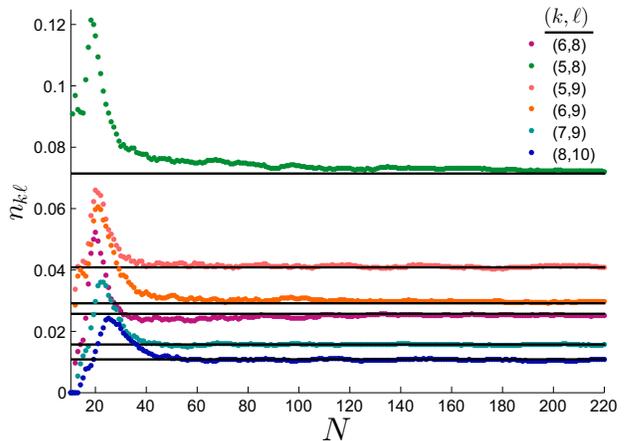}
\caption[Figure ]%
{(Color online) Comparison between the steady-state joint degree distribution and the theoretical prediction of~\eqref{nkl_FIN_2} for growth under uniform attachment with growth parameters $\beta_1=5$ and $\beta_2=8$.  The horizontal lines represent the steady-state solution given by~\eqref{nkl_FIN_2}, and markers represent simulation results. The initial seed networks are rings with 10 nodes in both layers.   The simulation results are averaged over 100 Monte Carlo trials. 
}
\label{ss_uniform_b1_5_b2_8_FIN}
\end{figure}

\begin{figure}[h]
\centering
\includegraphics[width=.95 \columnwidth]{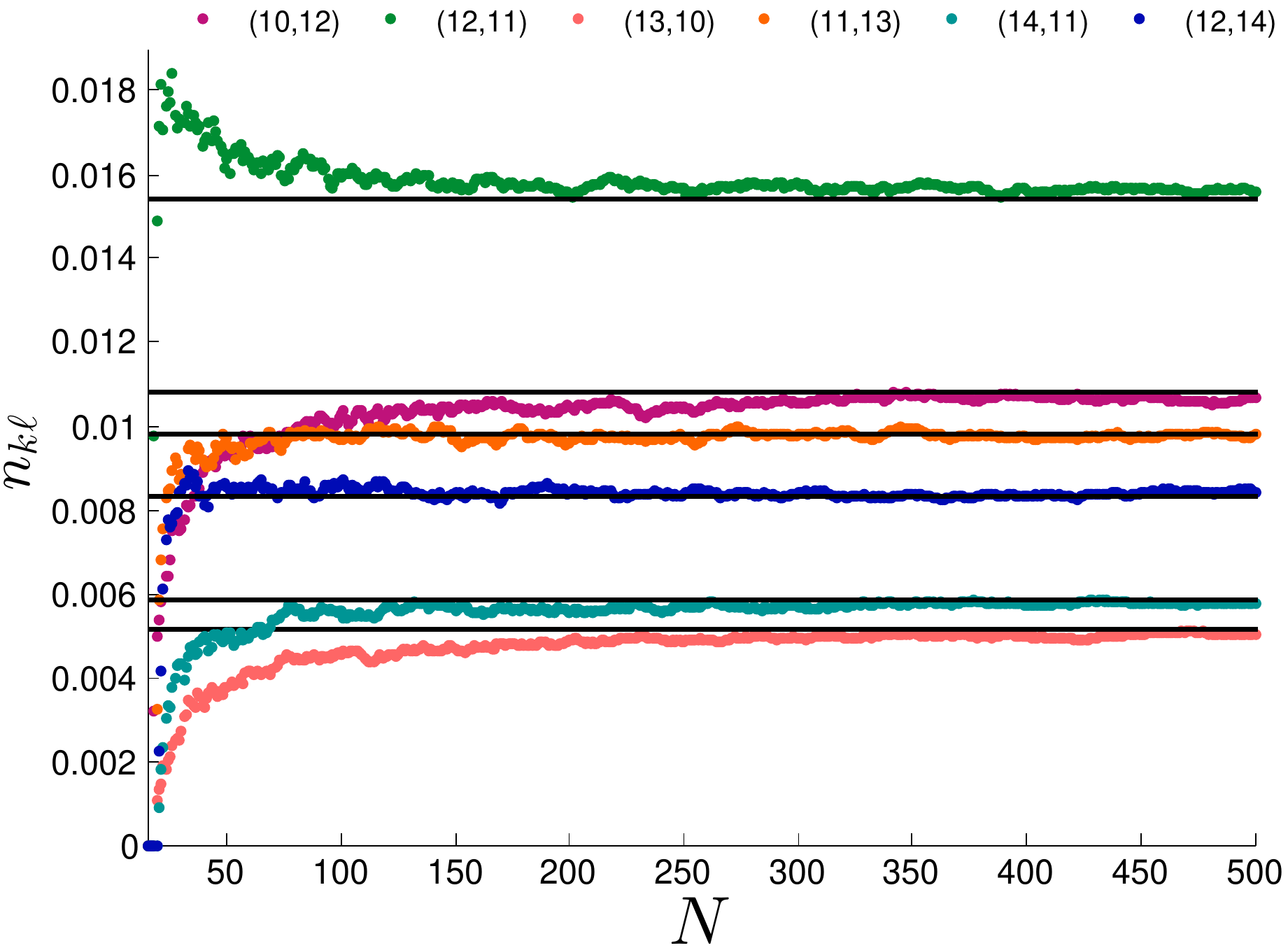}
\caption[Figure ]%
{(Color online) Comparison between the steady-state joint degree distribution and the theoretical prediction of~\eqref{nkl_FIN_2} for growth under uniform attachment with growth parameters $\beta_1=\beta_2=10$.  The horizontal lines represent the steady-state asymptotes which accord with~\eqref{nkl_FIN_2}, and markers represent simulation results. The initial seed networks are  complete graphs with 15 nodes in both layers.   Several example value of $(k,\ell)$ are chosen for illustrative purposes.  The results are averaged over 100 Monte Carlo trials. 
}
\label{ss_uniform_complete_N015_b1_10_b2_10}
\end{figure}

Let us verify~\eqref{nkl_FIN_2} via simulations. For the first setup, we consider $\beta_1=5$ and $\beta_2=8$. We plot $n(k,\ell)$ for multiple example instances  of $k$ ad $\ell$.The initial seed networks in both layers are rings  with 10 nodes.  The results are illustrated in Figure~\ref{ss_uniform_b1_5_b2_8_FIN}.  The simulation results visibly converge to the theoretical predictions. Moreover, equilibrium emerges relatively early, that is, even when the size of the network is 200, which is only 20 times the size of the initial seed network, the effects of the initial nodes are already negligible. For the second example, we consider the symmetric case $\beta_1=\beta_2=10$. For the initial seed  network, we consider  complete graphs  with $15$ nodes in both layers. The results are depicted in Figure~\ref{ss_uniform_complete_N015_b1_10_b2_10}. The results again verify the steady-state prediction, and confirm that equilibrium emerges relatively early on in the process, namely, when the size of the network is 20 times the size of the seed network.

To find the conditional average degree, we first need the degree distribution of single layers. This is found previously for example in~\cite{ME_PRE,Bian1}. The degree distribution in the first layer is
\all{
n_k= \frac{1}{\beta_1 } \left( \fracc{\beta_1}{\beta_1+1} \right)^{k-\beta_1+1}
}{nk_1_uu}

We need to compute
\all{
\bar{\ell}(k) &= \DD \sum_{\ell} \ell n(\ell|k)= \DD \sum_{\ell} \ell \fracc{n(k,\ell)}{n_{k}}
\nonumber \\
&=
\DD \sum_{\ell} \ell \fracc{ \fracc{ \beta_1^{k-\beta_1} \beta_2^{\ell-\beta_2} \CC{k-\beta_1+\ell-\beta_2}{k-\beta_1}}{(1+\beta_1+\beta_2)^{k-\beta_1+\ell-\beta_2+1}}}{\frac{1}{\beta_1 } \left( \frac{\beta_1}{\beta_1+1} \right)^{k-\beta_1+1}}
\nonumber \\ &
= 
\fracc{ (\beta_1+1)^{k-\beta_1+1}}{(\beta_1+\beta_2+1)^{k-\beta_1+1}}
\DD \sum_{\ell} \ell
\frac{ \beta_2^{\ell-\beta_2} \CC{k-\beta_1+\ell-\beta_2}{k-\beta_1 } }{(1+\beta_1+\beta_2)^{ \ell-\beta_2}}
}{lbar_2}

We have performed this summation in Appendix~\ref{app:nk_2}. 
The result is
\all{
\bar{\ell}(k) =
\frac{\beta_2}{\beta_1+1}(k+2)
.}{lbark_uu}

This is identical to~\eqref{lbar_1}. For the special case of $\beta_1=\beta_2$, this result (that the expected degree has the same expression under preferential and uniform attachment schemes) is also obtained in~\cite{Bian1}---see Equations (S.40) and (S.45) therein.

Finally, to obtain the distribution of total degree, we undertake the steps similar to those in the previous section. We have: 
\all{
n(q)&=\DD \sum_k n(k,q-k)
\nonumber \\ & =
\DD \sum_k 
  \fracc{ \beta_1^{k-\beta_1} \beta_2^{q-k-\beta_2} \CC{q-\beta_1-\beta_2}{k-\beta_1}}{(1+\beta_1+\beta_2)^{q-\beta_1 -\beta_2+1}}
  \nonumber \\ &
  \resizebox{0.85\linewidth}{!}{$
   =
 \fracc{  \beta_2^{q -\beta_2-\beta_1} }{(1+\beta_1+\beta_2)^{q-\beta_1 -\beta_2+1}}
\DD \sum_k 
  \beta_1^{k-\beta_1} \beta_2^{-k+\beta_1} \CC{q-\beta_1-\beta_2}{k-\beta_1}
  $}
  \nonumber \\ &
   =
 \fracc{  \beta_2^{q -\beta_2-\beta_1} }{(1+\beta_1+\beta_2)^{q-\beta_1 -\beta_2+1}}
\DD\left( 1+\frac{\beta_1}{\beta_2} \right)^{q-\beta_1-\beta_2}
  \nonumber \\ &
   =
\fracc{1}{\beta_1+\beta_2}
\DD\left( \fracc{\beta_1+\beta_2}{1+\beta_1+\beta_2}\right)^{q-\beta_1-\beta_2}
.}{nq2}

This is similar to~\eqref{nk_1_uu}. This means that the degree distribution of the aggregated network is identical to that of a uniformly growing network in which each newcomer forms $\beta_1+\beta_2$ links to existing nodes.

\section{Generalization of Model 1 to $M$ Layers} \label{prefM}

Now let us consider the case of $M$ layers, where  the network possesses one set of nodes and $M$ distinct  sets of links. Each incoming node establishes $\beta_m$ links in layer $m$ to existing nodes in that layer, where ${m \in \{1,2,\ldots,M \} }$. The initial number of links at layer $m$ is denoted by $L_m(0)$. 
In this model, the degree of each node can be conveniently represented by a vector of length $M$. The degree vector of node  $x$, denoted by $\mathbf{k}_x$, stores the degree of node $x$ in layer $m$ as its $m$-th component. Let $N(\mathbf{k})$ denote the number of nodes whose vector of degrees is $\mathbf{k}$. Let $n(\mathbf{k})$ denote the fraction of those nodes. 
If node $x$ receives a link in layer $m$, then its degree will change to 
${\mathbf{k}+1 \mathbf{e}_m}$, where $\mathbf{e}_m$ is the unit vector in $m$-th dimension, that is, it is a vector whose elements are all zero except its $m$-th element, which is unity. Let us denote the $m$th component of vector $\mathbf{k}$ by $k_m$. The rate equation for $N(\mathbf{k})$ reads
\all{
N_{t+1}(\mathbf{k}) 
= &N_t(\mathbf{k}) + \DD \prod_{m=1}^M \delta_{k_m,\beta_m}
\nonumber \\ &
+ \DD \sum_{m=1}^M \beta_m \fracc{(k_m-1) N_t(\mathbf{k}-\mathbf{e}_m)-(k_m)N_t(\mathbf{k})}{L_m(0)+2\beta_m t}
}{rate_gen_1}

Now we use $N_t(\mathbf{k}) = \frac{n_t(\mathbf{k})}{N(0)+t}$, and the following limit:
\all{
\lim_{t \rightarrow \infty} \beta_m \fracc{N(0)+t}{L_m(0)+2\beta_m t}= \frac{1}{2}
.}{lim_gen1}

With these substitutions, we can rewrite~\eqref{rate_gen_1} in the steady state as follows

\all{
n(\mathbf{k}) 
= &
\frac{1}{2}
\DD \sum_{m=1}^M \bigg[ (k_m-1) n(\mathbf{k}-\mathbf{e}_m)-(k_m) n(\mathbf{k}) \bigg]
\nonumber \\ &
+ \DD \prod_{m=1}^M \delta_{k_m,\beta_m}
}{rate_gen_2}

This can be rearranged and expressed equivalently as follows

\all{
n(\mathbf{k}) 
= &
\DD \sum_{m=1}^M \bigg[ \fracc{ k_m-1}{2+\sum_{m=1}^M k_m} n(\mathbf{k}-\mathbf{e}_m) \bigg]
\nonumber \\ &
+\fracc{2 }{2+\sum_{m=1}^M\beta_m}\DD \prod_{m=1}^M \delta_{k_m,\beta_m}
}{rate_gen_2_app}

%In Appendix~\ref{app:sol_gen_1}, we have solved this difference equation. The result is

%
%\all{
%n(\mathbf{k}) = & 
%\fracc{\prod_{m=1}^M (k_m-1)!}{\prod_{m=1}^M (\beta_m-1)!} 
%\fracc{(2+\sum_{m=1}^M \beta_m)!}{(2+\sum_{m=1}^M k_m)!}
%\nonumber \\ & \times
%\fracc{2 }{2+\sum_{m=1}^M\beta_m}
%\DD \prod_{n=2}^M \CC{\sum_{m=1}^n (k_m - \beta_m) }{k_n - \beta_n}
%}{n_gen1_FIN}

We first define the following auxiliary function:
\all{
\phi(\mathbf{k}) \deff \fracc{(2+\sum_{m=1}^M k_m)!}{\prod_{m=1}^M (k_m-1)!} n(\mathbf{k})
}{phi_gen_def}

Inserting this into~\eqref{rate_gen_2_app} yields the following simplified equation
\all{
\phi(\mathbf{k}) &= \sum_{m=0}^M \phi(\mathbf{k}-\mathbf{e}_m)
+2 
\fracc{(2+\sum_{m=1}^M k_m)!}{\prod_{m=1}^M (k_m-1)!}
\DD \prod_{m=1}^M \delta_{k_m,\beta_m}
\nonumber \\&
= 
\sum_{m=0}^M \phi(\mathbf{k}-\mathbf{e}_m)
+2 
\fracc{(2+\sum_{m=1}^M \beta_m)!}{\prod_{m=1}^M (\beta_m-1)!}
\DD \prod_{m=1}^M \delta_{k_m,\beta_m} 
}{m_diff_gen_1}

We define the $M$-dimensional Z-transform as follows:
\all{
\Phi(\mathbf{z}) \deff \sum_{k_1=0}^{\infty} \sum_{k_2=0}^{\infty} \ldots \sum_{k_M=0}^{\infty} 
\phi(\mathbf{z}) z^{-k1-k2-\ldots-k_M}
.}{psi_gen_def}

Plugging this into~\eqref{m_diff_gen_1}, we obtain
\all{
\Phi(\mathbf{z}) 
= 2 
\fracc{(2+\sum_{m=1}^M \beta_m)!}{\prod_{m=1}^M (\beta_m-1)!}
\fracc{
\DD \prod_{m=1}^M z_m^{-\beta_m} }{1-\sum_{m=0}^M z_m^{-1} }
.}{psi_gen_1}

The inverse of this generating function is given by

\all{
\resizebox{\linewidth}{!}{$
\phi(\mathbf{k}) = 
2
\fracc{(2+\sum_{m=1}^M \beta_m)!}{\prod_{m=1}^M (\beta_m-1)!}
\oint 
\DD \prod_{m=1}^M
\fracc{
z_m^{k_m-\beta_m-1} }{1-\sum_{m=0}^M z_m^{-1} } 
\frac{ d \mathbf{z_m}} {(2 \pi i)^M}
$}
.}{phi_inv_gen_1}

\begin{widetext}
\all{
&\fracc{1}{ {(2 \pi i)^M}} 
\DD \oint 
\DD \prod_{m=1}^M
\fracc{
z_m^{k_m-\beta_m-1} }{1-\sum_{m=0}^M z_m^{-1} } d z_m
= 
\fracc{1}{ {(2 \pi i)^M}} 
\DD \oint 
\DD \prod_{m=1}^M
z_m^{k_m-\beta_m-1} d z_m
\sum_{n=0}^{\infty} 
\left[ \sum_{m=0}^M z_m^{-1} \right]^n
\nonumber \\ &
=
\fracc{1}{ {(2 \pi i)^M}} 
\DD \oint \DD \prod_{m=1}^M
z_m^{k_m-\beta_m-1} d z_m
\sum_{n=0}^{\infty} 
\sum_{r_1+\ldots +r_M=n} \CC{n}{r_1,\ldots,r_M} 
\prod_{s=1}^M 
  z_1^{-r_s}
\nonumber \\ &
=
\DD
\sum_{n=0}^{\infty} 
 \sum_{r_1+\ldots +r_M=n} \CC{n}{r_1,\ldots,r_M} \prod_{s=1}^M
 \delta\big[r_s-(k_m-\beta_m)\big]
 = \CC{\sum_{m=1}^M (k_m-\beta_m)}{k_1-\beta_1,\ldots,k_m-\beta_M}
.}{M_inv}

\end{widetext}

Inserting this  result into~\eqref{phi_inv_gen_1}, we arrive at
\all{
\phi(\mathbf{k}) = 
2
\fracc{(2+\sum_{m=1}^M \beta_m)!}{\prod_{m=1}^M (\beta_m-1)!} 
\DD  \CC{\sum_{m=1}^M (k_m-\beta_m)}{k_1-\beta_1,\ldots,k_m-\beta_M}
.}{phi_gen_app}

Finally, from~\eqref{phi_gen_def} we obtain
\all{
n(\mathbf{k}) = & 
\fracc{\prod_{m=1}^M (k_m-1)!}{\prod_{m=1}^M (\beta_m-1)!} 
\fracc{(2+\sum_{m=1}^M \beta_m)!}{(2+\sum_{m=1}^M k_m)!}
\nonumber \\ & \times \fracc{2 }{2+\sum_{m=1}^M\beta_m}
\DD \CC{\sum_{m=1}^M (k_m-\beta_m)}{k_1-\beta_1,\ldots,k_m-\beta_M}
.}{n_gen1_FIN}

Note that   this expression holds for ${k_m\geq \beta_m}$, for all $M$ layers. If any of the $k_m$ values are smaller than the corresponding $\beta_m$ value, then $n(\mathbf{k})$ is zero. This is true because every node has an initial degree of $\beta_m$ in layer $m$ upon birth, and throughout the growth process, the degrees cannot decrease. 

Now let us consider some special cases in order to gain insight into the pattern~\eqref{n_gen1_FIN} follows. For $M=1$, we have:
\all{
n(k)=\fracc{(k-1)!}{(\beta-1)!} \fracc{(2+\beta)!}{(2+k)!} \times \fracc{2}{2+\beta} \CC{k-\beta}{k-\beta}
.}{special_1}

It can be easily simplified and rewritten as follows: 
\all{
n(k)=\fracc{(k-1)!}{(\beta-1)!}\fracc{(\beta+2)!}{(k+2)!}\fracc{2}{\beta+2}=\fracc{2\beta(\beta+1)}{k(k+1)(k+2)}
.}{1D}
This agrees with the degree distribution of the preferential attachment model for a single layer, obtained for example in~\cite{dorog_nk,redner2}.

Now let us consider the special case of $M=2$  and verify that~\eqref{n_gen1_FIN} indeed reduces to~\eqref{nkl_FIN_1_simple}. For $M=2$ we have: 
\all{
n(k_1,k_2) &= 
\fracc{  (k_1-1)!(k_2-1)!}{(\beta_1-1)!(\beta_2-1)!} 
\fracc{(2+\beta_1+\beta2 )!}{(2+k_1+k_2)!}
\nonumber \\ & \times \fracc{2 }{2+ \beta_1+\beta_2}
\fracc{(k_1-\beta_1+k_2-\beta_2)!}{(k_1-\beta_1)!(k_2-\beta_2)!}.
}{special_2}
Note that the factor ${(2+\beta_1+\beta_2)!}$ that exists in the numerator of the second fraction on the right hand side   simplifies into ${(1+\beta+1+\beta_2)!}$, due to the existence of the factor ${2+\beta_1+\beta_2}$ in the denominator of the third fraction. Applying this change, the result becomes identical  to the expression obtained in~\eqref{nkl_FIN_1_simple}.
% 
%Also note that the last product on the right hand equals unity for the case of $M=1$. 
% 
%In this case, we have

Finally, let us inspect the form of~\eqref{n_gen1_FIN} for $M=3$. For three layers, we have: 
\all{
&n(k_1,k_2,k_3) \nonumber \\ &
= 
\fracc{  (k_1-1)!(k_2-1)! (k_3-1)!}{(\beta_1-1)!(\beta_2-1)!(\beta_3-1)!} 
\fracc{(2+\beta_1+\beta2+\beta_3 )!}{(2+k_1+k_2+k_3)!}
\nonumber \\ & 
\times \fracc{2 }{2+ \beta_1+\beta_2+\beta_3}
\fracc{(k_1-\beta_1+k_2-\beta_2+k_3-\beta_3)!}{(k_1-\beta_1)!(k_2-\beta_2)!(k_3-\beta_3)!}.
}{special_3}
The pattern for general $M$ is clear from these three cases. 

\begin{figure}[h]
\centering
\includegraphics[width=.95 \columnwidth]{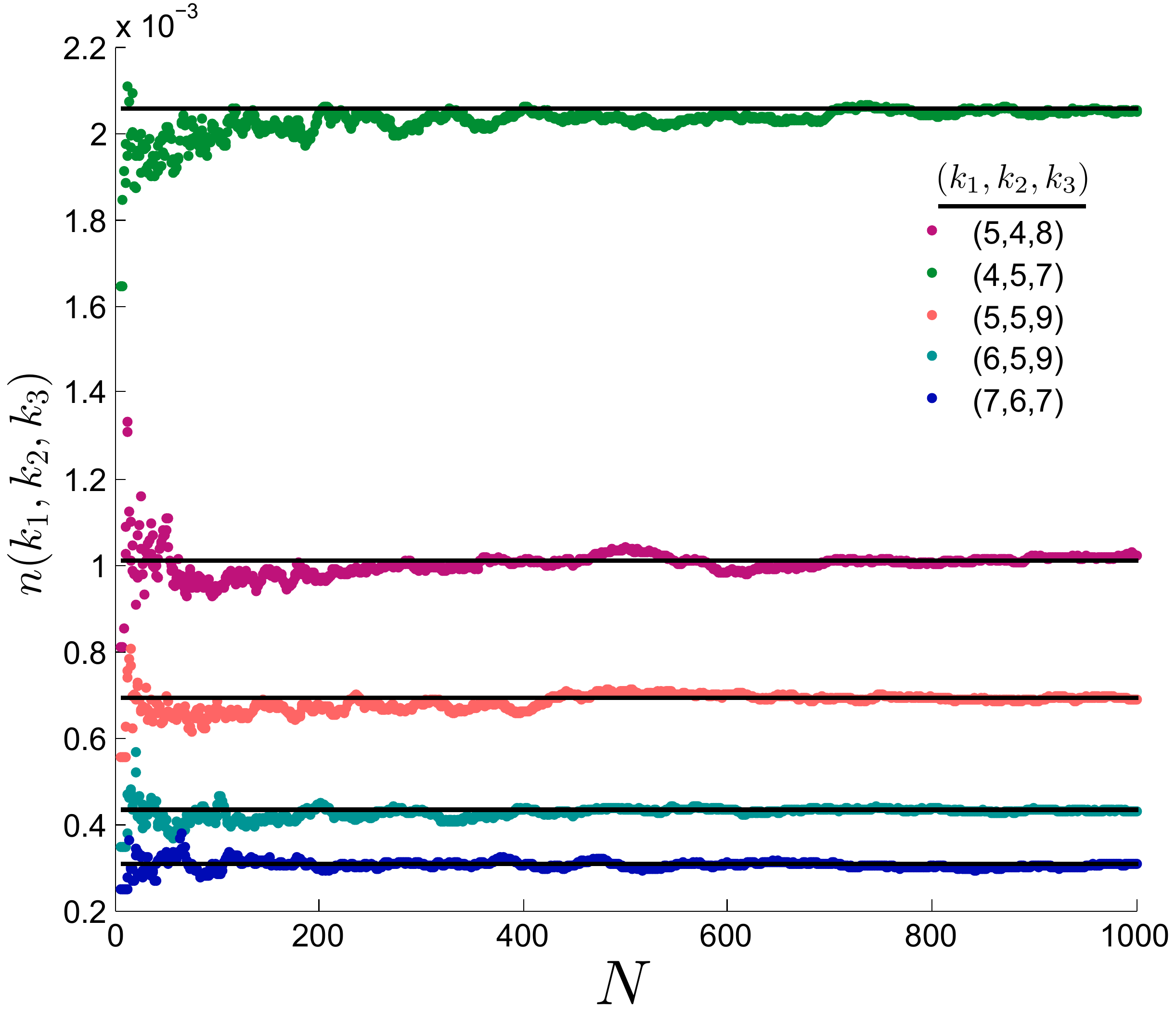}
\caption[Figure ]%
{(Color online) Comparison between the steady-state three-layer joint degree distribution and the theoretical  result of~\eqref{n_gen1_FIN} for the case of preferential attachment.  The horizontal lines represent the steady-state asymptotes  predicted by~\eqref{n_gen1_FIN}, and markers represent simulation results, which are averaged over 100 Monte Carlo trials. The initial seed networks have 5 nodes, and the topologies in the three layers are ring, star, and complete, respectively. The growth parameters are $\beta_1=2$, $\beta_2=3$, and $\beta_3=4$. Several example values of $(k_1,k_2,k_3)$ are chosen for illustrative purposes.  
}
\label{ss_threeLayer_pref_ring_star_complete_N0_5_b1_2_b2_3_b3_4}
\end{figure}

Let us verify the theoretical prediction in~\eqref{n_gen1_FIN} via simulations. We consider three layers, and an initial seed network with 5 nodes. In layer 1, the nodes  are situated on a ring. In layer 2, the topology is a star. In layer 3, we have a complete graph. The growth parameters are $\beta_1=2$, $\beta_2=3$, and $\beta_3=4$. Figure~\ref{ss_threeLayer_pref_ring_star_complete_N0_5_b1_2_b2_3_b3_4} depicts the results for multiple example     ${(k_1,k_2,k_3)}$ triplets. In all cases, the simulation results visibly converge to the horizontal asymptotes predicted by~\eqref{n_gen1_FIN}.

\section{Generalization of Model 2 to M Layers}
Now we assume that there are $M$ layers, all of them growing under uniform attachment. The rate equation reads
\all{
N_{t+1}(\mathbf{k}) 
= &N_t(\mathbf{k}) + \DD \prod_{m=1}^M \delta_{k_m,\beta_m}
\nonumber \\ &
+ \DD \sum_{m=1}^M \beta_m \fracc{ N_t(\mathbf{k}-\mathbf{e}_m)- N_t(\mathbf{k})}{N(0)+ t}
}{rate_gen_2}
For $n_t(\mathbf{k})$, this transforms into the following recurrence relation in the steady state
\all{
n(\mathbf{k}) 
= \DD \prod_{m=1}^M \delta_{k_m,\beta_m}
+ \DD \sum_{m=1}^M \beta_m \bigg[ n_t(\mathbf{k}-\mathbf{e}_m)- n_t(\mathbf{k}) \bigg]
}{rate_gen_2}

This can be rearranged and recast as

\all{
n(\mathbf{k}) 
= \fracc{ \DD \prod_{m=1}^M \delta_{k_m,\beta_m}}{1+\sum_{m=1}^M \beta_m }
+ \DD \sum_{m=1}^M \beta_m \fracc{\bigg[ n_t(\mathbf{k}-\mathbf{e}_m) \bigg]}{1+\sum_{m=1}^M \beta_m}
}{difference_4}

%We solve this difference equation in Appendix~\ref{app:sol_4}. The solution is
%
%\all{
%n(\mathbf{k}) = \frac{\beta_1^{k_1-\beta_1} \DD \prod_{n=2}^M \beta_n^{k_n-\beta_n} \CC{\sum_{m=1}^n (k_m - \beta_m)}{k_n-\beta_n}}{\left(1+\sum_{m=1}^M \beta_m \right)^{1+\sum_{m=1}^M (k_m-\beta_m) }} 
%.}{nkl_FIN_gen2}

Let us define
\eqq{
\begin{cases}
B \deff 1+\sum_{m=1}^M \beta_m \\
q_m \deff \fracc{\beta_m} { B}
\end{cases}
}{qdef_gen}
Taking the generating function of two sides of~\eqref{difference_4}, we get

\all{
\psi(\mathbf{z}) 
= \fracc{1}{ B } \DD \prod_{m=1}^M z_m^{-\beta_m}
+
\psi(\mathbf{z}) 
\DD \sum_{m=1}^M q_m z_m^{-1}.
}{psi3_1}

This can be equivalently expressed as follows

\all{
\psi(\mathbf{z}) 
= 
\fracc{1}{ B } \fracc{ \DD \prod_{m=1}^M z_m^{-\beta_m}}{1 - \DD \sum_{m=1}^M q_m z_m^{-1}}
.
}{psi3_gen_2_0}
The inverse of this generating function yields the desired degree distribution:
\all{
n(\mathbf{k})=
\fracc{1}{ B } \fracc{1}{(2\pi i)^M} \oint \fracc{ \DD \prod_{m=1}^M z_m^{-\beta_m} z_m^{k_m-1}}{1 - \DD \sum_{m=1}^M q_m z_m^{-1} d z_m}
.}{psi3_gen_2}

We can perform the integration by undertaking the following successive steps:
\begin{widetext}
\all{
&\fracc{1}{(2\pi i)^M} 
\oint \fracc{ \DD \prod_{m=1}^M z_m^{-\beta_m} z_m^{k_m-1} d z_m}{1 - \DD \sum_{m=1}^M q_m z_m^{-1}} 
=
\fracc{1}{(2\pi i)^M} 
\oint \prod_{m=1}^M z_m^{k_m-\beta_m-1} d z_m
\sum_{n=0}^{\infty} \bigg[ \sum_{m=1}^M q_m z_m^{-1} \bigg]^n 
\nonumber \\ &
=
\fracc{1}{(2\pi i)^M} 
\oint \prod_{m=1}^M z_m^{k_m-\beta_m-1} d z_m
\sum_{n=0}^{\infty} 
\sum_{r_1+\ldots+r_M=n} 
\CC{n}{r_1,\ldots,r_M}
\prod_{s=1}^M 
  q_m^{r_s} z_m^{-r_s}  
  \nonumber \\ &
=
\sum_{n=0}^{\infty} 
\sum_{r_1+\ldots+r_M=n} 
\CC{n}{r_1,\ldots,r_M}
\prod_{s=1}^M 
  q_m^{r_s} \delta_{r_s,(k_m-\beta_m)}
    \nonumber \\ &
    =
\CC{\sum_{m=1}^M (k_m-\beta_m)}{k_1-\beta_1,\ldots,k_M-\beta_M}  \prod_{m=1}^M q_m^{k_m-\beta_m}
.}{int_model_2}
\end{widetext}

Plugging this result into~\eqref{psi3_gen_2}, we get
\all{
n(\mathbf{k}) = \frac{1}{ B} 
\CC{\sum_{m=1}^M (k_m-\beta_m)}{k_1-\beta_1,\ldots,k_M-\beta_M}  \prod_{m=1}^M q_m^{k_m-\beta_m}
.}{n3_gen_app}

Using the definition of $B$ as given in~\eqref{qdef_gen}, this can be expressed equivalently as follows: 

\all{
n(\mathbf{k}) = \frac{ \CC{\sum_{m=1}^M (k_m-\beta_m)}{k_1-\beta_1,\ldots,k_M-\beta_M}    \DD \prod_{n=1}^M \beta_n^{k_n-\beta_n}  }{\left(1+\sum_{m=1}^M \beta_m \right)^{1+\sum_{m=1}^M (k_m-\beta_m) }} 
.}{nkl_FIN_gen2}
Note that, similar to the case of preferential growth,  this expression holds for ${k_m\geq \beta_m}$, for all $M$ layers. If any of the $k_m$ values are smaller than the corresponding $\beta_m$ value, then $n(\mathbf{k})$ is zero.

Now we consider the special cases of $M=1,2,3$ and simplify this expression to gain more insight into the pattern it follows for general $M$, for convenience of interpretation. 

 For $M=1$, we have
 \all{
n(k) = \frac{ \CC{   k-\beta }{k  - \beta }    \DD   \beta^{k-\beta}  }{\left(1+ \beta  \right)^{1+  (k -\beta ) }} 
.}{uspecial1}
 The binomial coefficient equals unity, and the result become identical to Equation~\eqref{nk_1_uu}, which is also obtained for example   in~\cite{ME_PRE,Bian1}.
  
Now we focus on the special case of $M=2$, and confirm that it agrees with~\eqref{nkl_FIN_2}. Note that the multinomial coefficient in~\eqref{nkl_FIN_gen2} becomes the ordinary binomial coefficient in this case. So for $M=2$ we have:

\all{
n(k_1,k_2) = \frac{ \CC{  k_1-\beta_1+k_2-\beta_2}{k_1-\beta_1 }    \DD   \beta_1^{k_1-\beta_1} \beta_2^{k_2\beta_2} }{\left(1+\beta_1+\beta_2 \right)^{1+k_1-\beta_1+k_2-\beta_2 }} 
,}{uspecial2}
which  is identical to Equation~\eqref{nk_1_uu}. 

Finally, let us consider the case of $M=3$. We expand the multinomial coefficient into factorials to render the pattern for general $M$ more apparent. We have:
\all{
\resizebox{\linewidth}{!}{$
n(k_1,k_2,k_3) = \frac{ \fracc{(k_1-\beta_1+k_2-\beta_2+k_3-\beta_3)!}{(k_1-\beta_1)!(k_2-\beta_2)!(k_3-\beta_3)!}    \DD  \beta_1^{k_1-\beta_1} \beta_2^{k_2-\beta_2} \beta_3^{k_3-\beta_3}  }{\DD \left(1+ \beta_1+\beta_2+\beta_3 \right)^{\DD 1+k_1-\beta_1+k_2-\beta_2+k_3-\beta_3 }} 
.$}
}{uspecial3}

\begin{figure}[h]
\centering
\includegraphics[width=.95 \columnwidth]{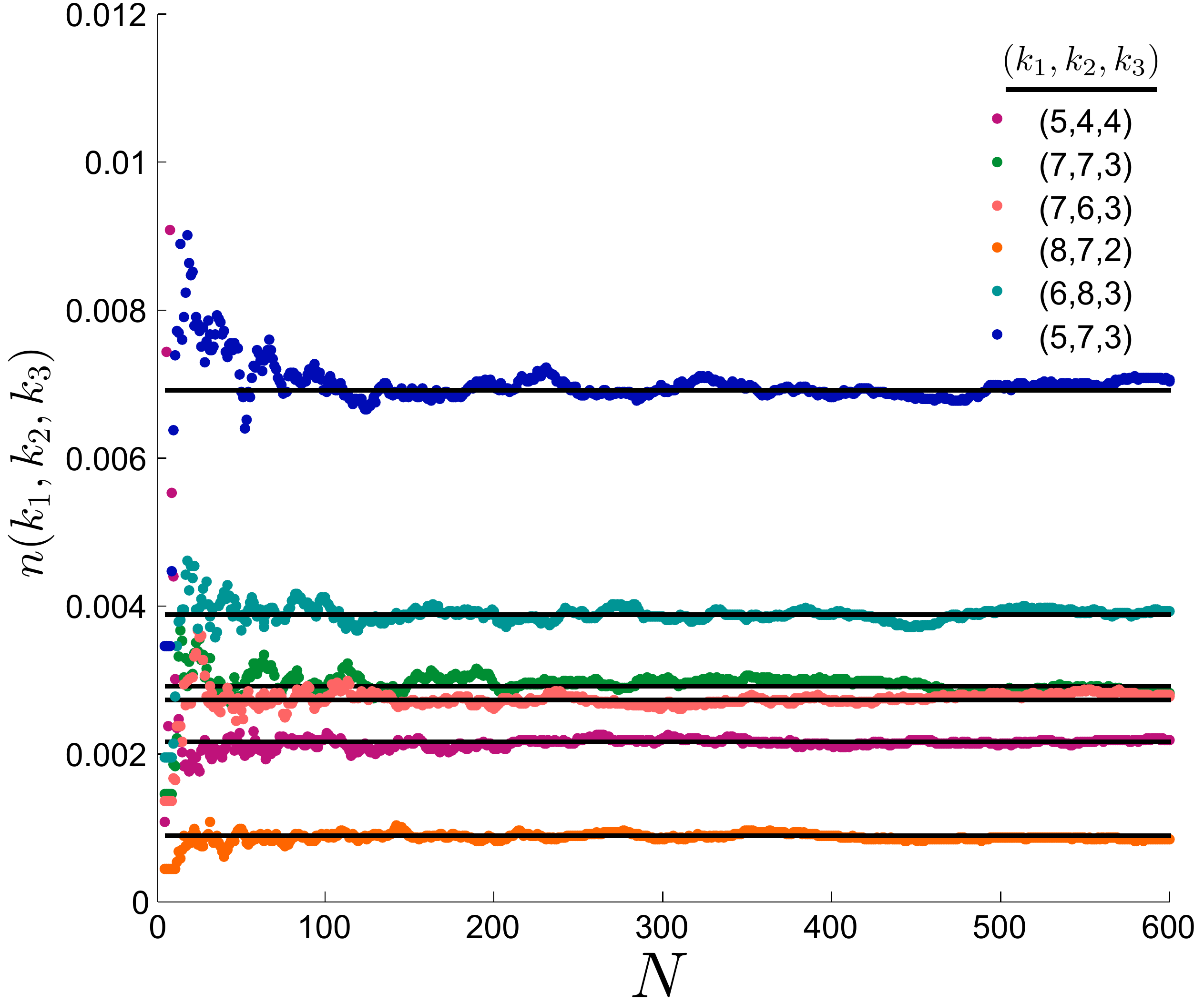}
\caption[Figure ]%
{(Color online) Comparison between the steady-state three-layer joint degree distribution and the theoretical  result of~\eqref{nkl_FIN_gen2} for the case of uniform attachment.  The horizontal lines represent the steady-state asymptotes  predicted by~\eqref{nkl_FIN_gen2}, and markers represent simulation results, which are averaged over 100 Monte Carlo trials. The initial seed networks have 8 nodes, and the topologies in the three layers are complete, star, and complete, respectively. The growth parameters are $\beta_1=3$, $\beta_2=4$, and $\beta_3=2$. Several example values of $(k_1,k_2,k_3)$ are chosen for illustrative purposes.  
}
\label{ss_threeLayer_uniform_complete_N0_5_b1_3_b2_4_b3_2}
\end{figure}

Let us verify the theoretical  result of~\eqref{nkl_FIN_gen2} via simulations. For the initial seed graph, consider 8 nodes which form complete graphs in layers one and three, and let them form a star graph in layer 2. Let the growth parameters be $\beta_1=3$, $\beta_2=4$, and $\beta_3=2$. 
 Figure~\ref{ss_threeLayer_uniform_complete_N0_5_b1_3_b2_4_b3_2} depicts the results for multiple example     ${(k_1,k_2,k_3)}$ triplets. In all cases, it is visible that the simulation results converge to the horizontal asymptotes predicted by~\eqref{nkl_FIN_gen2}.

\section{Temporal Evolution of the Joint Degree Distribution} \label{temporal}

We now move beyond the steady-state analysis and focus  on the temporal evolution of the structure of multiplex networks. Consider a given multiplex  network with $N(0)$ nodes, whose joint degree distribtuion is given by $n_0(\mathbf{k})$.  How would $n(\mathbf{k})$ evolve if new nodes are added to the network? In this section, we seek  a time-dependent solution for $n_t(\mathbf{k})$. We only consider the case where attachments are uniformly at random in each layer. Let us focus on the evolution of $N_t(\mathbf{k})$. At time $t$, upon the introduction of a single new node, we can rewrite the time-continuous analog of Equation~\eqref{rate_gen_1} as follows:

\all{
\rond{N_{t}(\mathbf{k}) }{t}
= &  \DD \prod_{m=1}^M \delta_{k_m,\beta_m}
\nonumber \\ &
+ \DD \sum_{m=1}^M \beta_m \fracc{  N_t(\mathbf{k}-\mathbf{e}_m)- N_t(\mathbf{k})}{N(0)+  t}
}{rate_gen_7}

Let  us define the following generating function:
\all{
\Psi_t(z_1,\ldots,z_M) \stackrel{\text{def}}{=}
\sum_{k_1} \ldots \sum_{k_M} N_t(\mathbf{k}) z_1^{-k_1} \ldots z_M^{-k_M}
.}{psi_t_def}

We now multiply both sides of~\eqref{rate_gen_7} by ${z_1^{-k_1} \ldots z_M^{-k_M}}$, and sum over all $k_i$. The result  is 
\all{
&\rond{\Psi_t(z_1,\ldots,z_M)}{t}= \prod_{i=1}^M z_i^{-\beta_i} 
\nonumber \\ &
+ 
\sum_{i=1}^M  \fracc{\beta_i}{N(0)+t} (z_{i}^{-1}-1)  \Psi_t(z_1,\ldots,z_M)
.}{PSI_dot0}

We can rearrange the terms and express this equation equivalently as follows:
\all{
\DD 
\resizebox{\linewidth}{!}{$
 \DD  \rond{\Psi_t(z_1,\ldots,z_M)}{t}  
- \DD
  \sum_{i=1}^M   \fracc{\beta_i(z_{i}^{-1}-1) \Psi_t(z_1,\ldots,z_M) }{N(0)+t}
= \prod_{i=1}^M z_i^{-\beta_i} 
.
$}
}{PSI_dot}

%We employ the method of characteristic curves to solve this linear partial differential equation. We need to solve the following system of equations: 
%\all{
%\fracc{dt}{1}= \fracc{-d z_i}{\frac{\beta_i(z_i-1)}{N(0)+t}}=\fracc{d \Psi_t(z_1,\ldots,z_M)}{z_i^{-\beta_i}},~~i=1\ldots M
%.}{system_2}
%
%For the $i$th equation of the type on the left,  it is straightforward separation of variables and integration to acquire characterstic curves: 
%\all{
%\fracc{dt}{1}= \fracc{-d z_i}{\frac{\beta_i(z_i-1)}{N(0)+t}}
%\Longrightarrow 
%(z_i-1)^{\beta_i} (N(0)+t)  = C_i,
%}{integrate_1}
%where $C_i$s are constants. 
This is a linear first order equation~\cite{boyce1992elementary} which can be solved  by the multiplication of both sides by the following integration factor: 

\all{
&
\mu_t(z_1,\ldots,z_M) = \DD \exp \int - \DD
  \sum_{i=1}^M   \fracc{\beta_i(z_{i}^{-1}-1)   }{N(0)+t} dt
  \nonumber \\ &
    = \exp \Bigg[- \ln (N(0)+t)     \sum_{i=1}^M  \beta_i (z_{i}^{-1}-1)  \Bigg] 
    \nonumber \\ &
    = (N(0)+t)^{- \sum_{i=1}^M  \beta_i (z_{i}^{-1}-1) }
  }{mu_def}

Using the integration factor, the solution is given by
\all{
& \Psi_t(z_1,\ldots,z_M) = \mu^{-1} \DD \Bigg[\int^t   \prod_{i=1}^M z_i^{-\beta_i}  \mu dt + C  \Bigg] 
\nonumber \\ &
=  \mu^{-1} \left[ \prod_{i=1}^M z_i^{-\beta_i}  \fracc{ (N(0)+t)^{1- \sum_{i=1}^M  \beta_i (z_{i}^{-1}-1) }}{1 - \sum_{i=1}^M  \beta_i (z_{i}^{-1}-1) } + C \right]
\nonumber \\ &
= (N(0)+t) \fracc{ \prod_{i=1}^M z_i^{-\beta_i} }{1- \sum_{i=1}^M  \beta_i (z_{i}^{-1}-1) } \nonumber \\ &
+ C \DD (N(0)+t)^{  \sum_{i=1}^M  \beta_i (z_{i}^{-1}-1) }
.}{psi_40}
In this equation, $C$ is a constant in time, but can depend on ${z_1,\ldots,z_M}$. It must be determined  in order to satisfy the initial conditions.  Note that the generating function of the joint degree distribution  of the initial network  is known. So we can use   $\Psi_0(z_1,\ldots,z_M)$ in order to identify $C$. Setting $t=0$ in~\eqref{psi_40}, and rearranging the terms, we arrive at
\all{
C= &\Psi_0(z_1,\ldots,z_M) \times
 N(0) ^{ - \sum_{i=1}^M  \beta_i (z_{i}^{-1}-1) }
 \nonumber \\ &
- 
\fracc{  \prod_{i=1}^M z_i^{-\beta_i} }{1- \sum_{i=1}^M  \beta_i (z_{i}^{-1}-1) }
N(0) ^{1 - \sum_{i=1}^M  \beta_i (z_{i}^{-1}-1) }
.}{C_find}
Finally, plugging this into~\eqref{psi_40} and rearranging the terms, we obtain the generating function of the joint degree distribution at arbitrary times: 
\all{
& 
\resizebox{.9\linewidth}{!}{$
\Psi_t(z_1,\ldots,z_M) = 
(N(0)+t) \fracc{ \prod_{i=1}^M z_i^{-\beta_i} }{1- \sum_{i=1}^M  \beta_i (z_{i}^{-1}-1) } 
$}
\nonumber \\ &
+ \Psi_0(z_1,\ldots,z_M) \times \left(1+\fracc{ t}{N(0)}\right)^{\sum_{i=1}^M  \beta_i (z_{i}^{-1}-1) }
\nonumber \\ & 
-
\resizebox{.85\linewidth}{!}{$
 N(0) 
\left(1+\fracc{ t}{N(0)}\right)^{\sum_{i=1}^M  \beta_i (z_{i}^{-1}-1) }
\fracc{  \prod_{i=1}^M z_i^{-\beta_i} }{1- \sum_{i=1}^M  \beta_i (z_{i}^{-1}-1) }
$}
.}{PSI_FIN}

Now we need to take the inverse of this transform. 
First let us take the inverse transform of ${G_1 \stackrel{\text{def}}{=} \frac{1}{1- \sum_{i=1}^M  \beta_i (z_{i}^{-1}-1) }}$. We again denote  $1+\sum_{m=1}^M \beta_m$ by $B$, and we denote $\frac{\beta_m}{B}$ by $q_m$,  for brevity of notation.  Using straightforward properties of geometric series and multinational expansions, we have:
\all{
&
G_1 =\frac{1}{B} \fracc{1}{1-\sum_i q_i z_i^{-1}}
=\frac{1}{B} \sum_{m=1}^{\infty} (\sum_i q_i z_i^{-1})^{m}
\nonumber \\ &
=\frac{1}{B} \sum_{m} \sum_{r_1+\ldots +r_M=m} \CC{m}{r_1,\ldots,r_M} \prod_{s=1}^M q_s^{r_s} z_s^{-r_s} 
.}{INV_1}

The inverse transform of the expressions $z_s^{-r_s}$ is  given by the delta function $\DD \delta_{\DD k_s,r_s}$. Using this result, we get
\all{
G_1 \xrightarrow {\mathcal{Z}^{-1}}
\frac{1}{B}
\sum_{m} \sum_{r_1+\ldots +r_M=m} \CC{m}{r_1,\ldots,r_M} \prod_{s=1}^M \delta_{k_s,r_s} q_s^{k_s}
.}{G1_INV0}

Applying the delta functions, all the sums reduce to a single nonvanishing term: 

\all{
G_1 \xrightarrow {\mathcal{Z}^{-1}}
\frac{1}{B}  \CC{\sum_{m=1}^M k_m}{k_1,\ldots,k_M}   
  \prod_{r=1}^M \fracc{\beta_r^{k_r}}{B^{k_r}}
.}{G1_INV1}

The multiplication by $\prod_{i=1}^M z_i^{-\beta_i} $ merely results in shifts, that is, every $k_r$ changes to ${k_r-\beta_r}$. So we obtain
\all{
\resizebox{.95 \linewidth}{!}{$
\fracc{ \prod_{i=1}^M z_i^{-\beta_i} }{1- \sum_{i=1}^M  \beta_i (z_{i}^{-1}-1) } 
\xrightarrow {\mathcal{Z}^{-1}}
\frac{1}{B} \CC{\sum_{m=1}^M k_m-\beta_m}{k_1-\beta_1,\ldots,k_M-\beta_M}   
  \prod_{r=1}^M \fracc{\beta_r^{k_r-\beta_r}}{B^{k_r-\beta_r}}
 .
 $}}{G1_INV}

Now let us take the inverse transform of 
 ${ G_2 \stackrel{\text{def}}{=} 
 \left(1+\frac{ t}{N(0)}\right)^{\sum_{i=1}^M  \beta_i  z_{i}^{-1}  } }$. Using elementary properties of the exponential function, we have
 \all{
& G_2= \prod_{m=1}^M 
 \left(1+\frac{ t}{N(0)}\right)^{\beta_m z_m^{-1}}
 \nonumber \\&
 = \prod_{m=1}^M \exp \Bigg[\beta_m z_m^{-1} \ln   \left(1+\frac{ t}{N(0)}\right)\Bigg] 
 \nonumber \\ &
 = \prod_{m=1}^M \sum_{r=1}^{\infty} 
\fracc{ \beta_m^r 
 \left[ \ln   \left(1+\frac{ t}{N(0)}\right)\right] ^r}{r!}
 z_m^{-r } 
.}{G2_INV0}
The inverse transform of the $z_m^{-r}$ terms yield  delta functions, which helps us eliminate the sums. The result is
\all{
&G_2 \xrightarrow {\mathcal{Z}^{-1}}
\prod_{m=1}^M \beta_m^{k_m} 
\fracc{ 
 \left[ \ln   \left(1+\frac{ t}{N(0)}\right)\right] ^{k_m}}{k_m!}
.}{G2_INV}

Plugging the expressions for the inverse transforms found in~\eqref{G1_INV} and~\eqref{G2_INV} into~\eqref{PSI_FIN}, and noting  that  the multiplication in the $z$ domain is equivalent to convolution in the $k$ domain, we arrive at the following expression for the inverse transform of ${\Psi_t(z_1,\ldots,z_M)}$, which  is equal to $N_t(z_1,\ldots,z_M)$ by definition. Dividing the result by the number of nodes at time $t$, which is ${N(0)+t}$, we obtain the joint degree distribution as:

\begin{widetext}
\all{
&  n_t(k_1,\ldots,k_M) =  
%\resizebox{.9\linewidth}{!}{$
\fracc{1}{B} \CC{\sum_{m=1}^M k_m-\beta_m}{k_1-\beta_1,\ldots,k_M-\beta_M}   
  \prod_{r=1}^M \fracc{\beta_r^{k_r-\beta_r}}{B^{k_r-\beta_r}}
%$}
\nonumber \\ &
+  \left(1+\fracc{t}{N(0)}\right)^{-B } 
\DD \sum_{\xi_1,\ldots,\xi_M}
n_0( \xi_1,\ldots, \xi_M)
 \prod_{m=1}^M \beta_m^{k_m-\xi_m} 
\fracc{ 
 \left[ \ln   \left(1+\frac{ t}{N(0)}\right)\right] ^{k_m-\xi_m}}{(k_m-\xi_m)!}
\nonumber \\ & 
%\resizebox{.9 \linewidth}{!}{$
-\fracc{ 1}{B} \left(1+\fracc{t}{N(0)}\right)^{-B }
 \prod_{m=1}^M  \beta_r^{k_r-\beta_r} B^{\beta_m}
 \DD 
 \sum_{\xi_1=\beta_1}^{k_1}
 \ldots
 \sum_{\xi_M=\beta_M}^{k_M}
 \CC{\sum_{m=1}^M \xi_m-\beta_m}{\xi_1-\beta_1,\ldots,\xi_M-\beta_M}   
 \prod_{m=1}^M  
%\beta_m^{k_m-\xi_m}
\fracc{ 
 \left[ \ln   \left(1+\frac{ t}{N(0)}\right)\right] ^{k_m-\xi_m}}{B^{\xi_m}(k_m-\xi_m)! }
% $}
.}{PT_FIN}
%We can also express the convolutions in extensive form: 
%\all{
%  N_t(z_1,\ldots,z_M) &=  
%%\resizebox{.9\linewidth}{!}{$
%(N(0)+t)  \CC{\sum_{m=1}^M k_m-\beta_m}{k_1-\beta_1,\ldots,k_M-\beta_M}   
%  \prod_{r=1}^M \fracc{\beta_r^{k_r-\beta_r}}{B^{k_r-\beta_r}}
%%$}
%\nonumber \\ &
%+ \left(1+\fracc{t}{N(0)}\right)^{-B}
%\sum_{r_1,\ldots,r_M} 
%\left[
%\prod_{m=1}^M \beta_m^{k_m} 
%\fracc{ 
% \left[ \ln   \left(1+\frac{ t}{N(0)}\right)\right] ^{k_m}}{k_m!}
% \right]
%   N_0(k_1-r_1,\ldots,k_M-r_M)  
%\nonumber \\ & 
%- N(0) \left(1+\fracc{t}{N(0)}\right)^{-B}
%%\resizebox{.85\linewidth}{!}{$
%%\times
%\sum_{r_1,\ldots,r_M} 
% \CC{\sum_{m=1}^M k_m-\beta_m}{k_1-\beta_1,\ldots,k_M-\beta_M}   
% \left[ \prod_{r=1}^M \fracc{\beta_r^{k_r-\beta_r}}{B^{k_r-\beta_r}}
%\right]  
%\left[
% \prod_{m=1}^M \beta_m^{k_m-r_m} 
%\fracc{ 
% \left[ \ln   \left(1+\frac{ t}{N(0)}\right)\right] ^{k_m-r_m}}{(k_m-r_m)!}
% \right]
%%$}
%.}{PSI_FIN_2}
\end{widetext}

The first term on the right hand side of~\eqref{PT_FIN} is independent of time, and is the steady-state solution which was obtained in~\eqref{nkl_FIN_gen2}. The second term characterizes the effect of the initial network. Note that the factor $\left(1+\frac{t}{N(0)}\right)^{-B-1} $ makes this term vanish in the long-time limit. This is expected, because  in the limit as $t \rightarrow \infty$, almost every node is among those who were subsequently appended to the network, not  those who belonged to the initial network. The last term on the right hand side is also transient and vanishes as time goes to infinity. At $t=0$, note that the first and third terms on the right hand side of~\eqref{PT_FIN} cancel out, and only the second term remains. In the second term, all the individual terms in the sum  are zero (because the logarithm in the summand  become $\log[1+0]$, which is zero), except one term, which pertains to ${\xi_m=k_m \forall m}$ (and for the logarithm, $0^0=1$), which means that the result correctly reduces to $n_0(k_1,\ldots,k_M)$. 

For the special case of $M=2$, we get the following expression for the joint degree distribution of two layers: 
\begin{widetext} 
 \all{
 &  n_t(k_1,k_2)=  
%\resizebox{.9\linewidth}{!}{$
   \CC{k_1-\beta_1+k_2-\beta_2}{k_1-\beta_1 }   
 \fracc{\beta_1^{k_1-\beta_1}\beta_2^{k_2-\beta_2}}{(\beta_1+\beta_2+1)^{1+k_1-\beta_1+k_2-\beta_2}}
%$}
\nonumber \\ &
+ \left(1+\fracc{t}{N(0)}\right)^{-(\beta_1+\beta_2+1)}
\sum_{\xi_1,\xi_2} n_0( \xi_1, \xi_2)   \beta_1^{k_1-\xi_1} \beta_2^{k_2-\xi_2} 
\fracc{ 
 \left[ \ln   \left(1+\frac{ t}{N(0)}\right)\right] ^{k_1-\xi_1+ k_2-\xi_2}}{(k_1-\xi_1)! (k_2-\xi_2)!}
\nonumber \\ & 
  \resizebox{.95\linewidth}{!}{$
-   \left(1+\fracc{t}{N(0)}\right)^{-(\beta_1+\beta_2+1)}
\fracc{ \beta_1^{k_1-\beta_1}\beta_2^{k_2-\beta_2}}
{(\beta_1+\beta_2+1)^{1 -\beta_1 -\beta_2  }
}
\DD   \sum_{\xi_1=\beta_1}^{k_1}
\DD   \sum_{\xi_2=\beta_2}^{k_2}
   \fracc{ \DD \CC{\xi_1-\beta_1+\xi_2-\beta_2 }{\xi_1-\beta_1 } \left[ \ln   \left(1+\frac{ t}{N(0)}\right)\right] ^{ k_1-\xi_1+k_2-\xi_2}}{  (\beta_1+\beta_2+1)^{  \xi_1+  \xi_2 }(k_1-\xi_1) ! (k_2-\xi_2) !}  
  $}
.}{PKLT_FIN_2}

In the limit as $t \rightarrow \infty$, the second and third terms on the right hand side vanish, and the first term prevails, which coincides with the result obtained in~\eqref{nkl_FIN_2}. 

\end{widetext}

\begin{figure}[h]
\centering
\includegraphics[width=.95 \columnwidth]{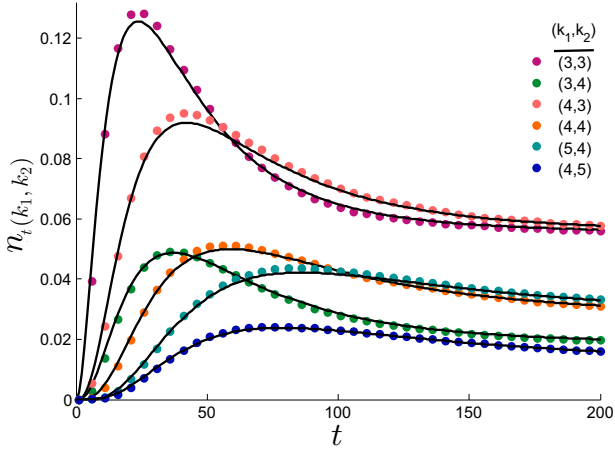}
\caption[Figure ]%
{(Color online) Temporal evolution of the joint degree distribution for two layers. The growth parameters are $\beta_1=3$ and $\beta_2=2$.  The solid curves represent theoretical prediction given by~\eqref{PKLT_FIN_2}. The  markers represent simulation results, averaged over 100 Monte Carlo trials. 
}
\label{ring_ring_N0_50_b1_3_b2_2}
\end{figure}

 Now let us verify the theoretical prediction for the time-dependent joint distribution via simulations.  The first setup we consider is as follows. Consider  50  nodes with identical connectivity in both layers at time $t=0$, forming a ring. For the growth mechanism, consider $\beta_1=3$ and $\beta_2=2$. Figure~\ref{ring_ring_N0_50_b1_3_b2_2} depicts $n_t(k_1,k_2)$ for some example test values for $k_1$ and $k_2$. It can be observed that simulation results are in good agreement with theoretical predictions.

\begin{figure}[h]
\centering
\includegraphics[width=.95 \columnwidth]{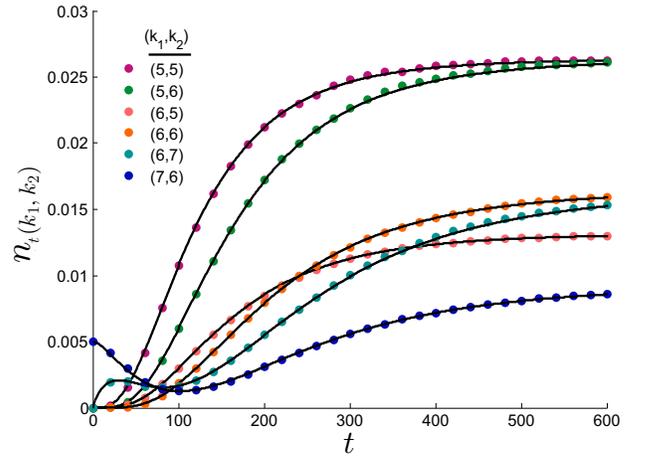}
\caption[Figure ]%
{(Color online) Temporal evolution of the joint degree distribution for two layers. The initial networks are ${SW(200,0.05,2)}$  and   ${SW(200,0.01,2)}$, respectively, as discussed in the text. The growth parameters are $\beta_1=3$ and $\beta_2=4$.
Solid lines represent theoretical predictions, given in~\eqref{PKLT_FIN_2}. Markers represent simulation results, averaged over 100 Monte Carlo trials. 
}
\label{SW_SW_p1_5sadom_p2_1sadom_d_ha_2_N0_200_b1_3_b2_4}
\end{figure}

 For the second example, we consider the initial networks to be small-world graphs~\cite{newman1999renormalization}. We construct the initial networks in the following way. Consider a network of $N(0)$ nodes.  Suppose that the nodes are situated on a circle and then each node is connected to $b$ immediate neighbors to the left and $b$ to the right. Then, each nonexistent link is established independently with probability $p$. We denote such a network by $SW(N(0),b,p)$, where $SW$ denotes small-world. For the second simulation setup, we consider a ${SW(200,0.05,2)}$ for the first layer and a ${SW(200,0.01,2)}$ for the second layer. We set the growth parameters to be $\beta_1=3$ and $\beta_2=4$. The results for several test cases of $k_1$ and $k_2$ are depicted in Figure~\ref{SW_SW_p1_5sadom_p2_1sadom_d_ha_2_N0_200_b1_3_b2_4}. Good agreement is observed between simulation results and theoretical predictions. 
% 
% For the illustration of the distribution at different instants of time, undertake the step similar to above; we fix the degree at one layer. Let us fix $k_2$ this time, at an example value of $k_2=8$. In Figure~\ref{}, we plot $n_t(k_1,8)$ for multiple instances of time. 
% 
% 

%
%
%Finally, let us consider one example with more than two layers.  Let the initial network have 200 nodes. In the first layer, we assume an Erd\H{o}s-R\'enyi (henceforth ER) network, where each link exists independently with probability 0.08. For the second layer, we consider a ring. For the third layer, we consider a Barab\'asi-Albert graph, in which the initial connectivity of each incoming node (that is, the number of links each node establishes upon arrival) is 5. Figure~\ref{} depicts the theoretical predictions and simulation results for several examples instances of $k_1$, $k_2$, and $k_3$. The high accuracy of the theoretical prediction is evident. 

\begin{figure}[h]
\centering
\includegraphics[width=.95 \columnwidth]{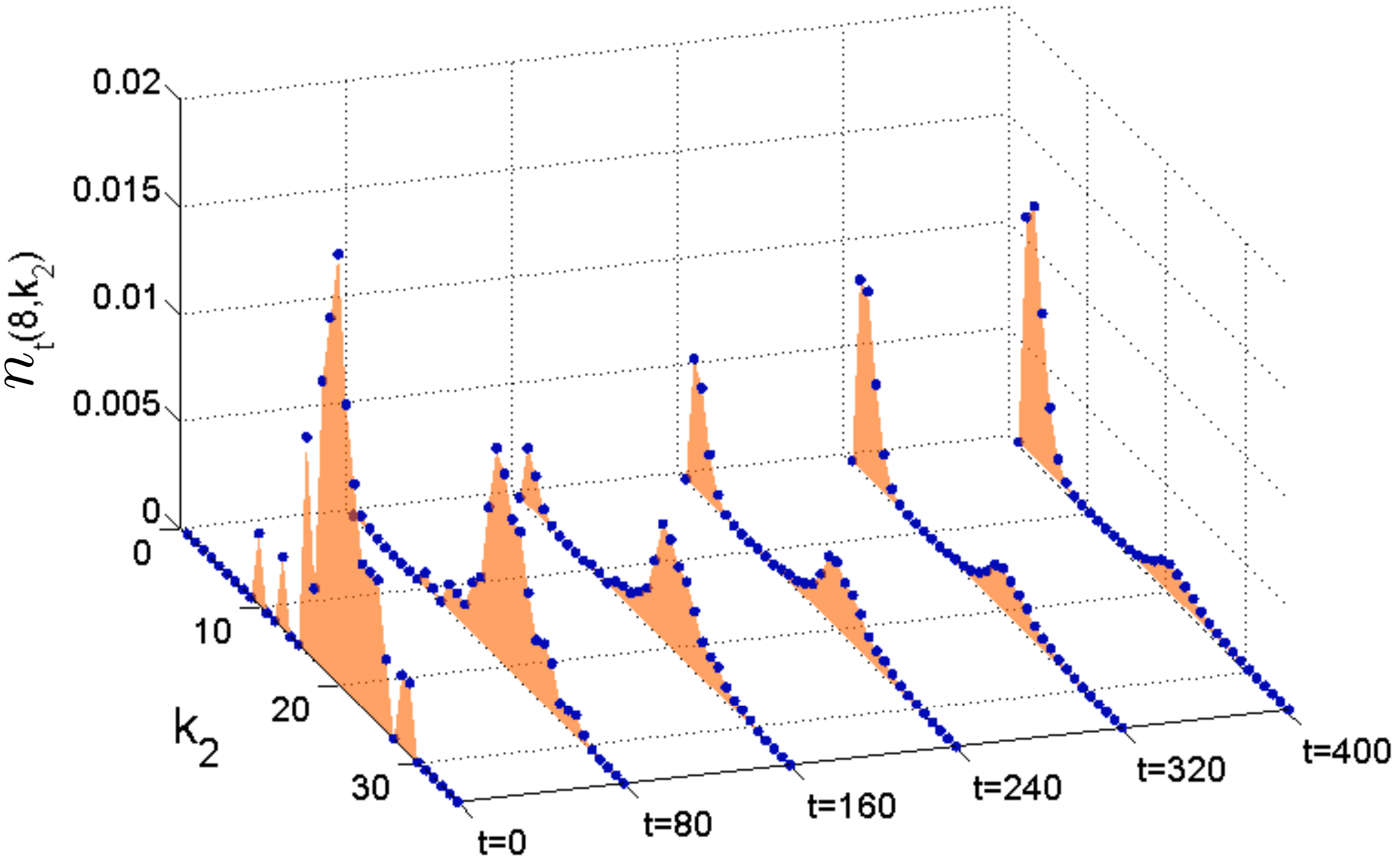}
\caption[Figure ]%
{(Color online) The  joint degree distribution $n_t(8,k_2)$ as a function of $k_2$ for different times. The  initial number of connections of incoming nodes  are $\beta_1=4$ and $\beta_2=2$.  The initial network in layer 1 is a $SW(300,2,0.02)$ and that of layer 2 is $SW(300,4,0.04)$. The shades represent theoretical predictions, given in~\eqref{PKLT_FIN_2}. Markers represent simulation results, averaged over 100 Monte Carlo trials. Shades are chosen instead of linear plot for better visibility.
}
\label{temporal_SW_2_sdahom_SW_4_4sadom_N0_300}
\end{figure}

 We can also illustrate the evolution of the joint distribution  by depicting it at different instants of time. To that end, since at most three dimensions can be plotted and $n_t(k_1,k_2)$ has four dimensions, we need to discard one dimension to enable us plot the distribution at different time points. To that end, we need to fix one dimension. Since the index of layers are arbitrary, without loss of generality, we fix $k_1$. For the simulation,  consider the  following setting. The initial network in layer 1 is a $SW(300,2,0.02)$ and that of layer 2 is $SW(300,4,0.04)$. The  initial number of connections of incoming nodes  are $\beta_1=4$ and $\beta_2=2$. 
 We  plot $n_t(k_1,k_2)$ at different time steps as a function of $k_2$, having fixed $k_1$ at the example value of $k_1=8$. In other words, we plot $n_t(8,k_2)$. The results are depicted in Figure~\ref{temporal_SW_2_sdahom_SW_4_4sadom_N0_300}.

\begin{figure}[h]
\centering
\includegraphics[width=.95 \columnwidth]{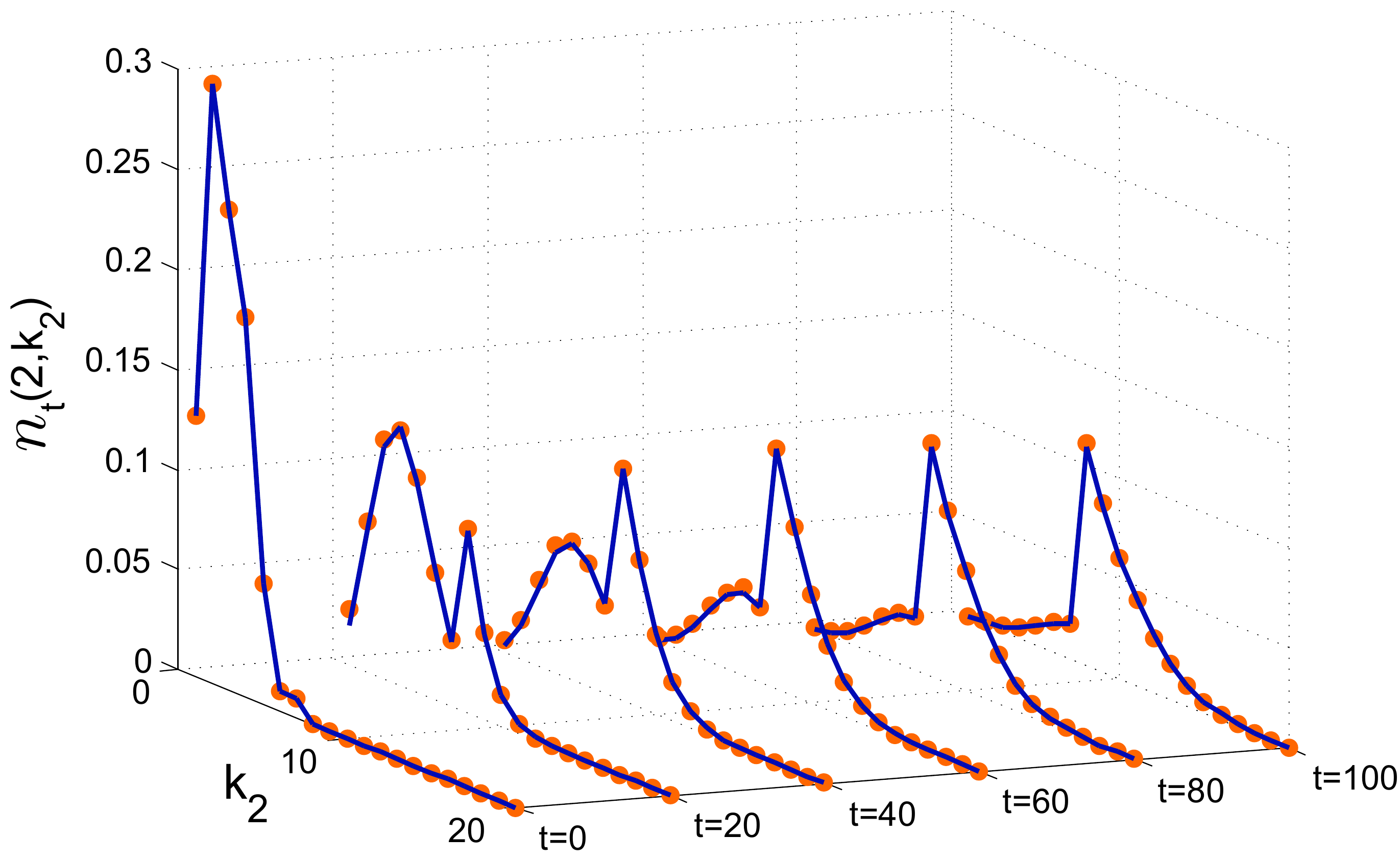}
\caption[Figure ]%
{(Color online) The  joint degree distribution $n_t(2,k_2)$ as a function of $k_2$  at different instants of time (i.e., we have fixed $k_1=2$ to enable graphical illustration). The growth parameters are $\beta_1=2$ and $\beta_2=8$. The initial network in the first layer is a ring with 100 nodes, and in the second layer the topology is an ER network with the link existence probability equal to 0.03. The  solid lines represent theoretical predictions, given in~\eqref{PKLT_FIN_2}. Markers represent simulation results, averaged over 100 Monte Carlo trials. Note that as time progresses, a peak at $k_2=\beta_2$ emerges, which is intuitively expected. 
}
\label{temporal_ring_ER_3sadom_b1_2_b2_8_N0_100_kfix_2}
\end{figure}

As another example, consider a ring with 100 nodes to be the initial network in the first layer, and let the nodes form an Erd\H{o}s-R\'enyi (henceforth ER) network~\cite{erdos1960evolution} in which the probability of existence of each link is 0.03. The results for the joint degree distribution with $k_1$ fixed at 2 is depicted in Figure~\ref{temporal_ring_ER_3sadom_b1_2_b2_8_N0_100_kfix_2}. In other words, Figure~\ref{temporal_ring_ER_3sadom_b1_2_b2_8_N0_100_kfix_2} depicts $n_t(2,k_2)$ as a function of $k_2$ at multiple instants of time. As time progresses, a peak at $k_2=\beta_2$ emerges. This is because as more new nodes enter the network, the number of newcomers who have degree $\beta_2$ increases and they constitute the majority of the network at long times. This causes the initial lump in Figure~\ref{temporal_ring_ER_3sadom_b1_2_b2_8_N0_100_kfix_2}---which pertained to the initial network---to lose its mass and become smaller. The mass moves towards $k_2=\beta_2$ and construct a peak there.

\begin{figure}[h]
\centering
\includegraphics[width=.95 \columnwidth]{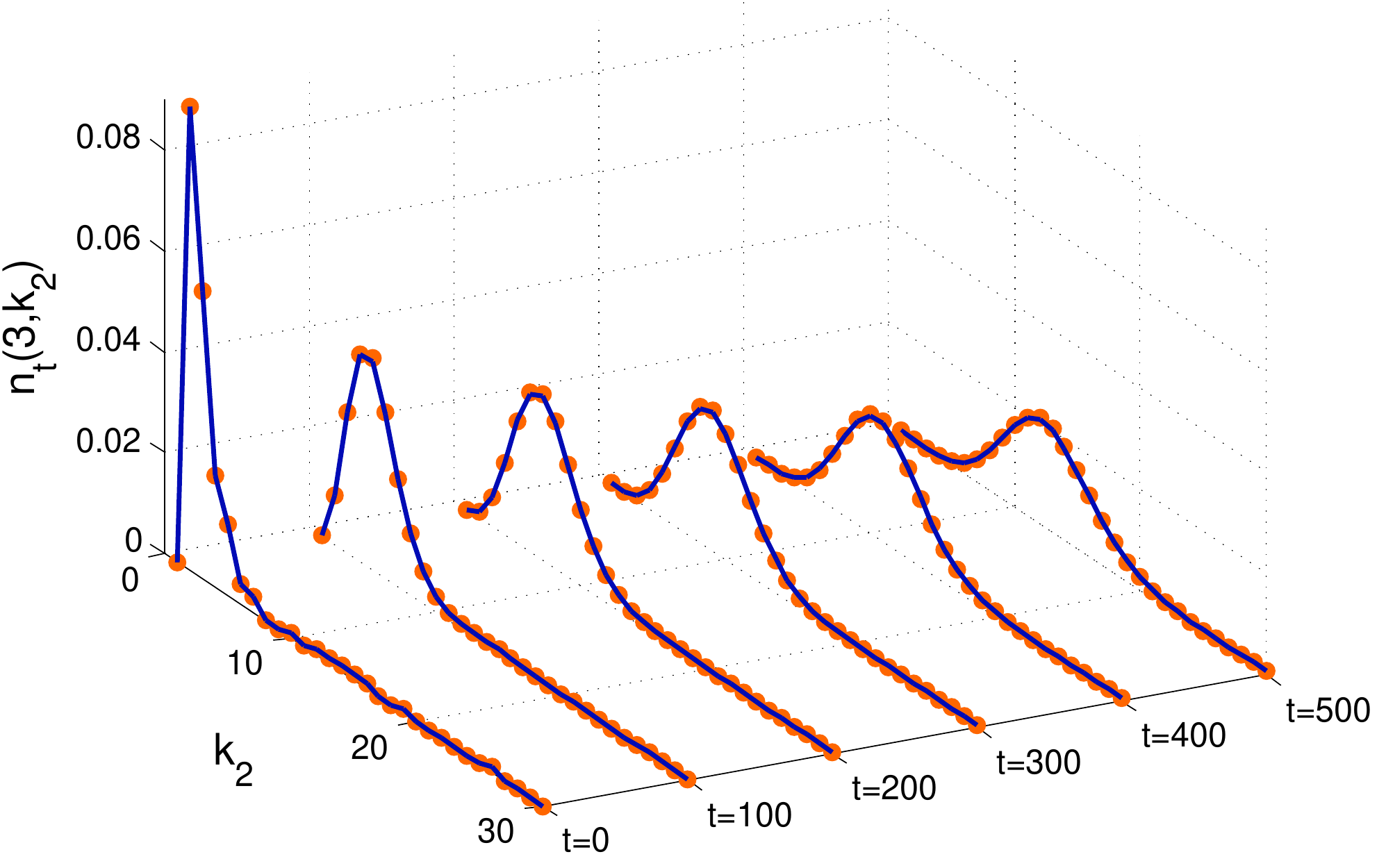}
\caption[Figure ]%
{(Color online) The  joint degree distribution $n_t(3,k_2)$ as a function of $k_2$  at different instants of time (i.e., we have fixed $k_1=3$ to enable graphical illustration). The growth parameters  are $\beta_1=1$ and $\beta_2=25$.  The  solid lines represent theoretical predictions, given in~\eqref{PKLT_FIN_2}. Markers represent simulation results, averaged over 100 Monte Carlo trials. The initial networks are BA networks in both layers, as described in the text. 
}
\label{temporal_BA_BA_b01_2_b02_2_beta1_1_beta2_25_N0_1000}
\end{figure}

For the next example, we consider two layers with Barab\'asi-Albert (henceforth BA) topology~\cite{barabasi1999emergence}. The initial networks have 300 nodes, and the initial connectivity of incoming nodes are 2 for both layers. So in the resultant network, the majority of nodes   have degree 2, and the degree distribution has a power-law tail with exponent 3, as is expected in BA graphs. After the BA network with 300 nodes has been constructed under the said mechanism, we start the growth process with growth parameters $\beta_1=1$ and $\beta_2=25$. This means that the second layer will be more densely connected as compared to layer 1. We fix the layer-1 degree at $k_1=3$
 and plot ${n_t(3,k_2)}$ as a function of $k_2$ at different instants of time. The results are presented in Figure~\ref{temporal_BA_BA_b01_2_b02_2_beta1_1_beta2_25_N0_1000}.  Note that the initial distribution has a peak at $k_2=2$, which is expected, because as mentioned above, the majority of nodes in the initial BA graph have degree 2 in both layers. As the growth process proceeds, the center of the peak moves towards higher values of $k_2$. This is because the newcomers have initial degree $\beta_2=25$ in the second layer, and as time progresses, the number of such  nodes increase.

\begin{figure}[h]
\centering
\includegraphics[width=.95 \columnwidth]{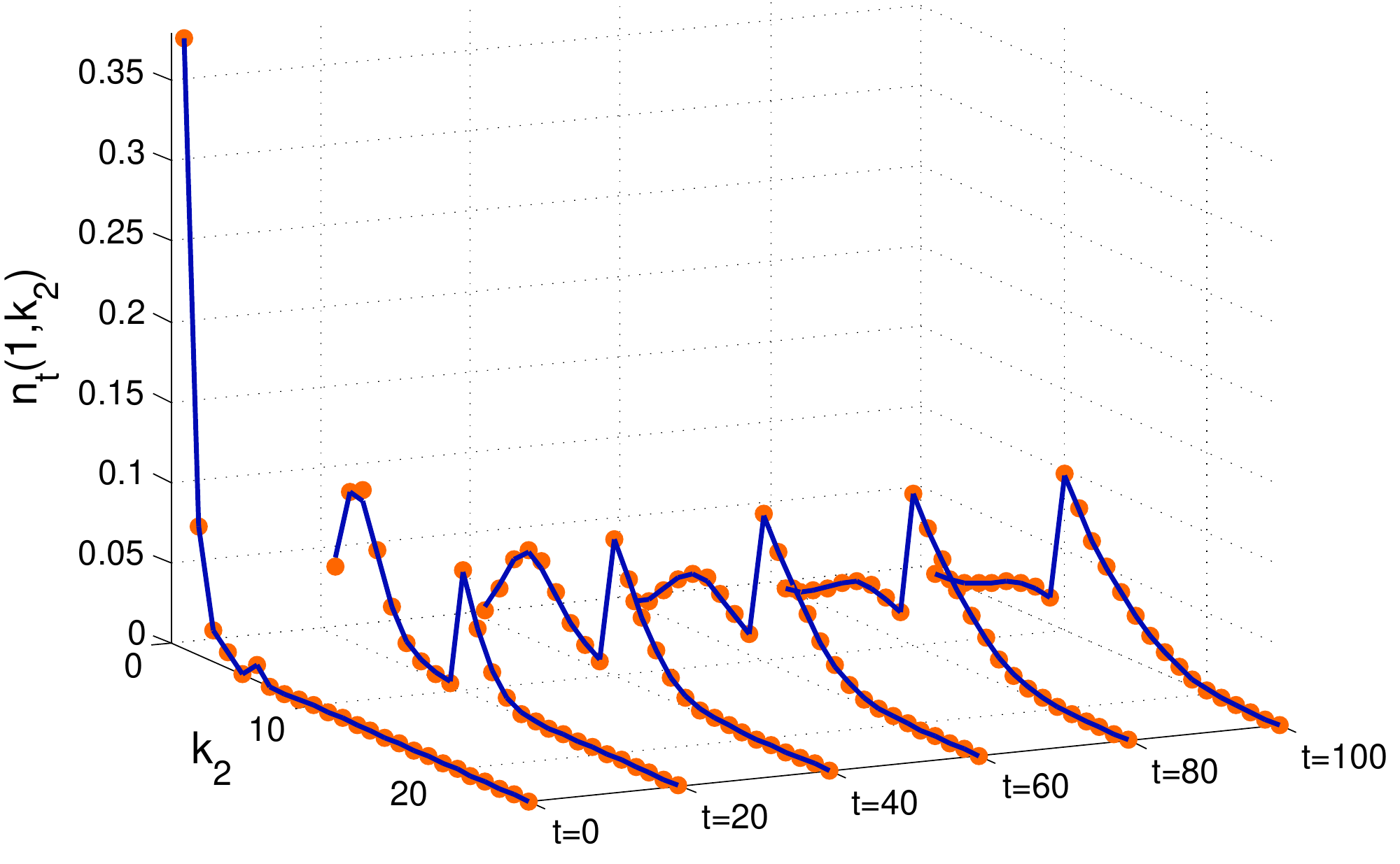}
\caption[Figure ]%
{(Color online) The  joint degree distribution $n_t(1,k_2)$ as a function of $k_2$  at different instants of time (i.e., we have fixed $k_1=1$ to enable graphical illustration). The growth parameters are $\beta_1=1$ and $\beta_2=10$. The  solid lines represent theoretical predictions, given in~\eqref{PKLT_FIN_2}. Markers represent simulation results, averaged over 100 Monte Carlo trials. The initial networks are RRT networks in both layers, as described in the text. 
}
\label{temporal_RRT_RRT_b01_1_b02_1_N0_100_b1_1_b2_10}
\end{figure}

Let us also investigate   the effect of sudden change in the growth parameters through one example. Consider a random recursive tree~\cite{dorogovtsev2008transition}, which is constructed as follows. We begin by a single node, and then new nodes enter the system sequentially, and each newcomer randomly chooses one existing node and connects to it. The result is a tree, called a random recursive tree (hereinafter RRT). We choose  the initial networks in both layers to be RRTs with 100 nodes. We then apply to this substrate a growth process with $\beta_1=1$ (which would be a continuation of the growth mechanism that gave rise to the RRT in the first layer) and $\beta_2=10$. This means that in the second layer, the  growth mechanism is undergoing a regime shift, as the number of connections per incoming node abruptly jumps from 1  to 10. The results for the temporal evolution of ${n_t(1,k_2)}$ is plotted in Figure~\ref{temporal_RRT_RRT_b01_1_b02_1_N0_100_b1_1_b2_10}. We observe that a new peak emerges at $k_2=10$, and the initial lump that peaked at $k_2=1$ at the outset gradually loses its mass to the new lump concentrated at $k_2=10$, which embodies the nodes that are introduced to the network in the   second phase of the growth process.

\begin{figure}[h]
\centering
\includegraphics[width=.95 \columnwidth]{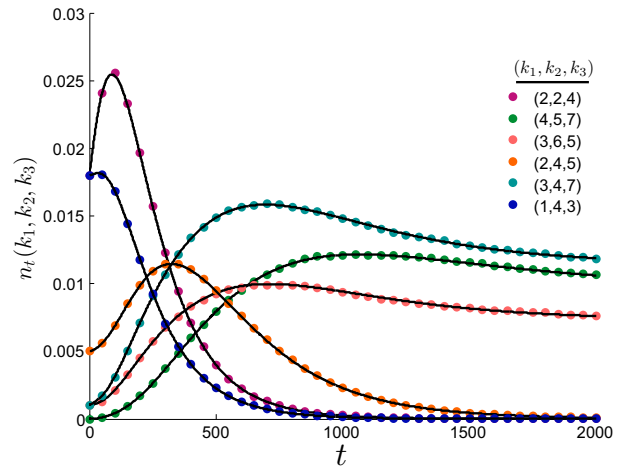}
\caption[Figure ]%
{(Color online) The  joint degree distribution $n_t(k_1,k_2,k_3)$ for several example triplets $(k_1,k_2,k_3)$. The growth parameters are $\beta_1=1$, $\beta_2=4$, and $\beta_3=5$. The solid lines  represent theoretical predictions, and the markers represent simulation results averaged over 1000 Monte Carlo trials. The initial networks are BA networks in all three layers, as discussed in the text. 
}
\label{rangirangi_3_layers_BA_BA_BA_b01_1_b02_2_b03_3_b1_3_b2_4_b3_5_N0_1000}
\end{figure}

Let us also consider one example to verify accuracy of the theoretical prediction of the joint degree distribution  for more than two layers. Let us consider the following setting for the initial network: a BA network of 100 nodes  with parameter 1 (that is, the initial number of links each incoming node establishes upon  birth is 1) in the first layer, a BA network with parameter 2 in the second layer, and a BA network with parameter 3 in the third layer. For the growth mechanism, we consider $\beta_1=1$, $\beta_2=4$, and $\beta_3=5$. The results for several example triplets of $(k_1,k_2,k_3)$ are depicted in Figure~\ref{rangirangi_3_layers_BA_BA_BA_b01_1_b02_2_b03_3_b1_3_b2_4_b3_5_N0_1000}
. 
 simulation results are visibly in good agreement with theoretical predictions.

\section{Summary and Open Problems}
%Previous theoretical treatment of growing multiplex networks have hitherto focus on homogeneous growth. 
In this paper we focused on heterogeneous growth of multiplex networks. We considered both the preferential and uniform growth mechanisms. We analyzed the problem for $M>2$ layers and obtained the joint degree distributions. For the case of uniform growth, we analyzed the problem for arbitrary times, and obtained the joint degree distribution for arbitrary number of layers at arbitrary times, for given initial conditions.  We verified our findings through Monte Carlo simulations, and observed that the theoretical predictions are remarkably accurate.

%We observed that in the two-layer case, depending on $\beta_1,\beta_2$, the degree of a node in one layer can be smaller or greater than its expected degree in the other layer. If $\beta_1>\beta_2$, then we have $k>\overline{\ell}(k)$ regardless of the value of $k$. Similarly, $\beta_1<\beta_2$, we have $k<\overline{\ell}$, regardless of $k$. If $\beta_1=\beta_2$, then there is a cutoff degree $k_c$, and we have $k<\overline{\ell}$ if $k$ is smaller than $k_c$, $k>\overline{\ell}$ if $k$ is greater than $k_c$, and $k=\overline{\ell}$ if $k=k_c$. This is true for both uniform and preferential attachment mechanisms. 

An immediate generalization of this work is to consider nonzero couplings in the preferential growth mechanism, so that the link reception probability of a node in each layer depends on its degrees in all layers, and then find the joint degree distribution via the rate equation approach (nonzero couplings are envisaged in~\cite{Bian1,Bian2,coevolution}, but the rate equation approach remains unsolved and no closed-form solution for the joint degree distribution exists in the literature). For example, in the two-layer case, the connection kernel would depend on the degrees of existing nodes in both layers. Consider an existing node $x$, who has degree $k_x$ in layer 1 and degree $\ell_x$ in layer 2 at time $t$. Then its probability of receiving a link in layer 1 from the incoming node would be proportional to $g_{11}  k_{x} + g_{12} \ell_x$ , and its probability of receiving a link in layer 2 would be proportional to $g_{21} k_{x} + g_{22} \ell_{x}$. Our solution presented in this paper pertains to the special case of $g_{11}=g_{22}=1$ and $g_{12}=g_{21}=0$. The rate equation corresponding to the general case (counterpart of Equation~\eqref{rate_1}, with all four $g_{ij}$s nonzero, would be as follows:

\all{
& N_{k, \ell} (t+1)
= N_{k, \ell} (t)+
\delta_{k \beta_1} \delta_{\ell \beta_2} 
\nonumber \\
&
+
\beta_1 \fracc{\bigg[g_{11} (k-1 )+g_{12}\ell \bigg] N_t(k-1,\ell) }{g_{11} L_1(0)+g_{12} L_2(0)+ 2(g_{11} \beta_1+g_{12}\beta_2) t}
\nonumber \\
&
+
\beta_2 \fracc{\bigg[ g_{21} k + g_{22} (\ell-1) \bigg] N_t(k,\ell-1) }{g_{21} L_1(0)+g_{22} L_2(0)+ 2(g_{21} \beta_1+g_{22}\beta_2) t}
\nonumber \\
&
-
\beta_1 \fracc{\bigg[g_{11} k+g_{12}\ell \bigg] N_{k, \ell}(t) }{g_{11} L_1(0)+g_{12} L_2(0)+ 2(g_{11} \beta_1+g_{12}\beta_2) t}
\nonumber \\
&
-
\beta_2 \fracc{\bigg[g_{21} k+g_{22}\ell \bigg] N_{k, \ell}(t) }{g_{21} L_1(0)+g_{22} L_2(0)+ 2(g_{21} \beta_1+g_{22}\beta_2) t}
.}{rate_couple}

It is straightforward to simplify this equation in the steady state, but solving the resulting difference equation is more intricate than the one studied in this paper.

Furthermore, a related quantity which is ubiquitous in the studying of epidemics and various diffusion processes over networks is the nearest-neighbor degree distribution~\cite{Eguiluz_PRL,Barthelemy_PRL,Castellano_PRL,Castellano_PRL_2}. Let us denote this quantity by $p_1(k,\ell|q,r)$. For a node with degree $k$ in layer 1 and degree $\ell$ in layer 2, if we select one of its neighbors in layer 1 randomly, then $p_1(k,\ell|q,r)$ would be the probability that this neighbor has degree $q$ in layer 1 and degree $r$ in layer 2. Similarly, $p_2(k,\ell|q,r)$ would be the probability that a randomly selected layer-2 neighbor of a node with degrees $k,\ell$ will have degrees $q,r$. This quantity is essential for studying dynamics on networks, and to obtain it for multiplex networks, one also needs exact expressions for the degree distributions---which the present paper focused on.

%
%\section{Acknowledgment}
%This work was funded in part by the Natural Sciences and Engineering Research Council of Canada. 

\appendix

\section{Solving Difference Equation~\eqref{difference_1}}\label{app:sol_1}

We need to solve
\all{
n(k,\ell) = &\fracc{k-1}{k+\ell+2} n(k-1,\ell) 
\fracc{\ell-1}{k+\ell+2} n(k, \ell-1) 
\nonumber \\ &
+ 
\fracc{2\delta_{k \beta_1} \delta_{\ell \beta_2} }{2+\beta_1+\beta_2}
.}{difference_1_app}

We define the new sequence
\all{
m_{k \ell} \deff \fracc{(k+\ell+2)!}{(k-1)!(\ell-1)!} n(k,\ell)
.}{m_1_def}

The following holds
\eq{
\resizebox{\linewidth}{!}{$
\begin{cases}
\fracc{k-1}{k+\ell+2} n(k-1,\ell) = \fracc{(k-1)!(\ell-1)!} n(k,\ell){(k+\ell+2)!} m_{k-1,\ell} \\ \\
\fracc{\ell-1}{k+\ell+2} n(k, \ell-1) = \fracc{(k-1)!(\ell-1)!} n(k,\ell){(k+\ell+2)!} m_{k,\ell-1}.
\end{cases}
$}
}
Plugging these into~\eqref{difference_1_app}, we can recast it as
\all{
m_{k \ell} = m_{k-1,\ell}+m_{k,\ell-1}+2 \fracc{(\beta_1+\beta_2+1)!}{(\beta_1-1)!(\beta_2-1)!}\delta_{k,\beta_1}\delta_{\ell,\beta_2}.
}{m_diff_1}

Now define the Z-transform of sequence $m_{k,\ell}$ as follows: 
\eqq{
\begin{cases}
\psi(z,y) \deff \DD \sum_k \sum_{\ell} m_{k,\ell} z^{-k} y^{-\ell} \\ \\
m_{k,\ell} = \fracc{1}{(2 \pi i)^2} \DD \oint \oint \psi(z,y) z^{k-1} y^{\ell-1} dz dy
.
\end{cases}
}{psi_def}

Taking the Z transform of every term in~\eqref{m_diff_1}, we arrive at
\all{
\psi(z,y)=&
z^{-1} \psi(z,y)+y^{-1}\psi(z,y)
\nonumber \\
&+2 \fracc{(\beta_1+\beta_2+1)!}{(\beta_1-1)!(\beta_2-1)!} z^{-\beta_1} y^{-\beta_2}
.}{psi_eq_1}
This can be rearranged and rewritten as follows
\all{
\psi(z,y) = \fracc{2}{1-z^{-1}-y^{-1}} 
\fracc{(\beta_1+\beta_2+1)!}{(\beta_1-1)!(\beta_2-1)!} z^{-\beta_1} y^{-\beta_2}}{psi_eq_2}

The inverse transform is given by
\all{
&m_{k,\ell} = 
\fracc{2(\beta_1+\beta_2+1)!}{(\beta_1-1)!(\beta_2-1)! } \oint \oint \fracc{z^{k-\beta_1-1}y^{\ell-\beta_2-1} dz dy}{(-4\pi^2)(1-z^{-1}-y^{-1})} 
\nonumber \\
& 
= 
\fracc{2(\beta_1+\beta_2+1)!}{(\beta_1-1)!(\beta_2-1)!} \oint \oint \fracc{z^{k-\beta_1}y^{\ell-\beta_2} dz dy}{(-4\pi^2)(zy-z-y)} 
\nonumber \\
& 
= 
\fracc{2(\beta_1+\beta_2+1)!}{(\beta_1-1)!(\beta_2-1)!} \oint \oint \fracc{z^{k-\beta_1}y^{\ell-\beta_2} dz dy}{(-4\pi^2)(y-1) \big[ z-\frac{y}{y-1} \big] } 
.}{m_1_1}

First we integrate over $z$. We get

\all{
m_{k,\ell} &= 
\fracc{2(\beta_1+\beta_2+1)!}{(\beta_1-1)!(\beta_2-1)!} \oint \fracc{\left(\frac{y}{y-1}\right)^{k-\beta_1}y^{\ell-\beta_2} dy}{(2\pi i)(y-1) } 
\nonumber \\ &
= \fracc{2(\beta_1+\beta_2+1)!}{(\beta_1-1)!(\beta_2-1)!} \oint \fracc{ y^{k-\beta_1+\ell-\beta_2} dy}{(2\pi i)(y-1)^{k-\beta_1+1} } 
.}{m_1_2}

Now note that the residue of $\fracc{f(y)}{(y-1)^n}$ for positive integer equals $\fracc{f^{(n-1)}(1)}{(n-1)!}$, where the numerator denotes the ${(n-1)}$th derivative of the function $f(y)$, evaluated at ${y=1}$. Also, note that the $m$-th derivative of the function $y^n$, for integer $n$ and $m$, equals $\fracc{m!}{(n-m)!}y^{n-m}$. Combining these two facts, we obtain

\all{
m_{k,\ell} = 
\fracc{2(\beta_1+\beta_2+1)!}{(\beta_1-1)!(\beta_2-1)!} \CC{k-\beta_1+\ell-\beta_2}{k-\beta_1}
.}{m_1_3}

Using~\eqref{m_1_def}, we arrive at
\all{
n(k,\ell) = 
\fracc{2(\beta_1+\beta_2+1)!}{(\beta_1-1)!(\beta_2-1)!}
\fracc{1}{k(k+1)\ell(\ell+1)}
\fracc{ \CC{k-\beta_1+\ell-\beta_2}{k-\beta_1}}{\CC{k+\ell+2}{k+1}}
.}{m_1_app_FIN}

This can be equivalently expressed as follows: 
\all{
\resizebox{\linewidth}{!}
{$ n(k,\ell) = 
\fracc{2\beta_1(\beta_1+1) \beta_2(\beta_2+1)}{(\beta_1+\beta_2+2) k(k+1)\ell(\ell+1)}
\fracc{\CC{\beta_1+\beta_2+2}{\beta_1+1} }{\CC{k+\ell+2}{k+1}}
\CC{k-\beta_1+\ell-\beta_2}{k-\beta_1}
$}
}{nkl_FIN_1_app}

\section{Performing the Summation in~\eqref{lbar_k_1}}\label{app:nk_1}

We need to calculate

\all{
\bar{\ell}(k) = 
\DD \sum_{\ell} 
\frac{ \beta_2(\beta_2+1) }{(2+\beta_1+\beta_2) }
\frac{\CC{\beta_1+\beta_2+2}{\beta_1+1}\CC{k-\beta_1+\ell-\beta_2}{k-\beta_1 } }{\CC{k+\ell+2}{\ell}}
}{lbar_k_1_app}

We use the following identity:
\all{
\fracc{1}{\CC{n}{m}}= (n+1) \DD \int_0^1 t^m (1-t)^{n-m} dt,
}{iden_gamma}
to rewrite the binomial reciprocal of the coefficient as follows
\all{
\fracc{1}{\CC{k+\ell+2}{\ell}} = (k+\ell+3) \DD \int_0^1 t^{\ell} (1-t)^{k+2}
dt
.}{recip_binom}

Also, from Taylor expansion, it is elementary to show that
\eqq{
S_1(x,n) \deff \DD \sum_m x^m \CC{m}{n} = \fracc{x^{n}}{(1-x)^{n+1}}.
}{taylor_0}
This identity will be used in the steps below. 

Plugging~\eqref{recip_binom} into~\eqref{lbar_k_1_app}, we have
\begin{widetext}
\all{
\bar{\ell}(k) & = 
\DD \sum_{\ell} 
\frac{ \beta_2(\beta_2+1) }{(2+\beta_1+\beta_2) }
\frac{\CC{\beta_1+\beta_2+2}{\beta_1+1}\CC{k-\beta_1+\ell-\beta_2}{k-\beta_1 } }{\CC{k+\ell+2}{\ell}}
\nonumber \\ &
= 
\frac{ \beta_2(\beta_2+1) }{(2+\beta_1+\beta_2) } \CC{\beta_1+\beta_2+2}{\beta_1+1}
\DD \sum_{\ell} (k+\ell+3) \CC{k-\beta_1+\ell-\beta_2}{k-\beta_1 } \DD \int_0^1 t^{\ell} (1-t)^{k+2} dt
\nonumber \\&
= 
\frac{ \beta_2(\beta_2+1) }{(2+\beta_1+\beta_2) } \CC{\beta_1+\beta_2+2}{\beta_1+1}
\DD \int_0^1 (1-t)^{k+2} t^{-k-2} \DD \sum_{\ell} (k+\ell+3) t^{k+\ell+2} \CC{k-\beta_1+\ell-\beta_2}{k-\beta_1 } dt
\nonumber \\&
= 
\frac{ \beta_2(\beta_2+1) }{(2+\beta_1+\beta_2) } \CC{\beta_1+\beta_2+2}{\beta_1+1}
\DD \int_0^1 (1-t)^{k+2} t^{-k-2} 
\fracc{d}{dt} \left[ \DD \sum_{\ell} t^{k+\ell+3} \CC{k-\beta_1+\ell-\beta_2}{k-\beta_1 } \right] dt
\nonumber \\&
= 
\frac{ \beta_2(\beta_2+1) }{(2+\beta_1+\beta_2) } \CC{\beta_1+\beta_2+2}{\beta_1+1}
\DD \int_0^1 (1-t)^{k+2} t^{-k-2} 
\fracc{d}{dt} \left[t^{3+\beta_1+\beta_2} \DD \sum_{\ell} t^{k-\beta_1+\ell-\beta_2} \CC{k-\beta_1+\ell-\beta_2}{k-\beta_1 } \right] dt
\nonumber \\&
\stackrel{\textnormal{\eqref{taylor_0}}}{=}
\frac{ \beta_2(\beta_2+1) }{(2+\beta_1+\beta_2) } \CC{\beta_1+\beta_2+2}{\beta_1+1}
\DD \int_0^1 (1-t)^{k+2} t^{-k-2} 
\fracc{d}{dt} \left[t^{3+\beta_1+\beta_2} \fracc{t^{k-\beta_1}}{(1-t)^{k-\beta_1+1}} \right] dt
\nonumber \\&
=
\frac{ \beta_2(\beta_2+1) }{(2+\beta_1+\beta_2) } \CC{\beta_1+\beta_2+2}{\beta_1+1}
\DD \int_0^1 (1-t)^{k+2} t^{-k-2} 
\fracc{d}{dt} \left[ \fracc{t^{k+\beta_2+3}}{(1-t)^{k-\beta_1+1}} \right] dt
\nonumber \\&
=
\frac{ \beta_2(\beta_2+1) }{(2+\beta_1+\beta_2) } \CC{\beta_1+\beta_2+2}{\beta_1+1}
\DD \int_0^1 (1-t)^{\beta_1} t^{\beta_2} 
\left[ k+\beta_2+3-(1+\beta_1+\beta_2) t \right] dt
\nonumber \\&
=
\frac{ \beta_2(\beta_2+1) }{(2+\beta_1+\beta_2) } \CC{\beta_1+\beta_2+2}{\beta_1+1}
\left[ (k+\beta_2+3) \DD \int_0^1 (1-t)^{\beta_1} t^{\beta_2} dt 
-(1+\beta_1+\beta_2) \DD \int_0^1 (1-t)^{\beta_1} t^{\beta_2+1} dt 
\right] 
\nonumber \\&
\stackrel{\textnormal{\eqref{recip_binom}}}{=}
\frac{ \beta_2(\beta_2+1) }{(2+\beta_1+\beta_2) } \CC{\beta_1+\beta_2+2}{\beta_1+1}
\left[ (k+\beta_2+3) \fracc{\beta_1! \beta_2! }{(\beta_1+\beta_2+1)!}
-(1+\beta_1+\beta_2) \fracc{\beta_1!( \beta_2+1)! }{(\beta_1+\beta_2+2)!}
\right] 
\nonumber \\&
=
\frac{ \beta_2(\beta_2+1) \beta_1 ! \beta_2 !}{(2+\beta_1+\beta_2) (1+\beta_1+\beta_2)!} \CC{\beta_1+\beta_2+2}{\beta_1+1}
\left[ (k+\beta_2+3) 
-(\beta_2+1)
\right] 
\nonumber \\&
=
\fracc{\beta_2}{\beta_1+1} (k+2)
}{lbar_k_1_app}
\end{widetext}

\section{Proving the Identity Given in~\eqref{sum_idenq}}\label{app:idenq}

We use the property of the Gamma integral given in~\eqref{iden_gamma} in order to deal with the binomial coefficient that is in the denominator of the summand. We have:

\all{
& \DD \sum_k
\fracc{  \CC{q-\beta_1 -\beta_2}{k-\beta_1}}{\CC{q-2}{k-1}}
\nonumber \\ &
=
 \DD \sum_k
  \CC{q-\beta_1 -\beta_2}{k-\beta_1} (q-1) 
  \int_0^1 (1-t)^{k-1} t^{q-k-1}  dt
  \nonumber \\ &
  \resizebox{\linewidth}{!}{$
= (q-1)
\DD \int_0^1 t^{q-1-\beta_1 } (1-t)^{\beta_1-1}  \DD \sum_k
  \CC{q-\beta_1 -\beta_2}{k-\beta_1} 
 \left(\fracc{ 1-t}{t}\right)^{k-\beta_1}  dt$}
   \nonumber \\ &
  \resizebox{\linewidth}{!}{$
= (q-1)
\DD \int_0^1 t^{q-1-\beta_1 } (1-t)^{\beta_1-1}  \DD 
\left( 1+\fracc{1-t}{t} \right)^{q-\beta_1-\beta_2}  dt$}
\nonumber \\ &
= (q-1) \DD \int_0^1 t^{\beta_2-1 } (1-t)^{\beta_1-1}  
\nonumber \\ &
\stackrel{\textnormal{\eqref{iden_gamma}}}{=}
\fracc{(q-1)}{(\beta_2+\beta_1-1) \CC{\beta_2+\beta_1-2}{\beta_2-1 }}
\nonumber \\ &
=\fracc{ (\beta_1-1)! (\beta_2-1)!(q-1)}{(\beta_2+\beta_1-1)!}
.}{sum_iden}

\section{Solving Difference Equation~\eqref{difference_2}}\label{app:sol_2}

Let us repeat the equation we need to solve for easy reference
\all{
\resizebox{\linewidth}{!}{$
n(k,\ell)= \fracc{\beta_1}{1+\beta_1+\beta_2} n(k-1,\ell) + \fracc{\beta_2}{1+\beta_1+\beta_2} n_{k,\ell-1}
+ \fracc{ \delta_{k,\beta_1} \delta_{\ell,\beta_2} }{1+\beta_1+\beta_2}
$}
.}{difference_2_app}

Let us define the following quantities from brevity:
\eqq{
\begin{cases}
q_1 &\deff \fracc{\beta_1}{1+\beta_1+\beta_2} \\ 
q_2 &\deff \fracc{\beta_2}{1+\beta_1+\beta_2}
\end{cases}
}{qs_def}

Taking the Z transform from both sides of~\eqref{difference_2_app}, we get
\all{
\psi(z,y)= q_1 z^{-1} \psi(z,y) + q_2 y^{-1} \psi(z,y) + \fracc{ z^{-\beta_1} y^{-\beta_2}}{1+\beta_1+\beta_2}
.}{psi2_temp_1}
This can be rearranged and recast as
\all{
\psi(z,y) = \fracc{1}{1-q_1 z^{-1} - q_2 y^{-1}} \fracc{z^{-\beta_1} y^{-\beta_2}}{1+\beta_1+\beta_2}
.}{psi2_temp_2}

This can be inverted through the following steps
\all{
n(k,\ell) &=
\fracc{1}{(1+\beta_1+\beta_2)(2 \pi i)^2} \DD \oint \psi(z,y) z^{k-1} y^{\ell-1} dz dy
\nonumber \\ &
= \fracc{1}{(1+\beta_1+\beta_2)(2 \pi i)^2} 
\DD \oint \oint \fracc{ z^{k-\beta_1-1} y^{\ell-\beta_2-1}}{1-q_1 z^{-1} - q_2 y^{-1}} dz dy
\nonumber \\ &
= \fracc{1}{(1+\beta_1+\beta_2)(2 \pi i)^2} 
\DD \oint \oint \fracc{ z^{k-\beta_1} y^{\ell-\beta_2}}{z y - y q_1 - z q_2} dz dy
\nonumber \\ &
= \fracc{1}{(1+\beta_1+\beta_2)(2 \pi i)^2} 
\DD \oint \oint \fracc{ z^{k-\beta_1} y^{\ell-\beta_2}}{z - \frac{y q_1}{y-q_2} } \fracc{1}{y-q_2} dz dy
\nonumber \\ &
= \fracc{1}{(1+\beta_1+\beta_2)(2 \pi i) } 
\DD \oint \fracc{ y^{\ell-\beta_2}}{y-q_2} \left( \frac{y q_1}{y-q_2} \right)^{k-\beta_1} dz dy
\nonumber \\ &
= \fracc{ q_1^{k-\beta_1}}{(1+\beta_1+\beta_2)(2 \pi i) } 
\DD \oint \fracc{ y^{k-\beta_1+\ell-\beta_2}}{(y-q_2)^{k-\beta_1+1}} dz dy
\nonumber \\ &
= \fracc{ q_1^{k-\beta_1} }{(1+\beta_1+\beta_2)}
\DD \fracc{1}{(k-\beta_1)!} \fracc{(k-\beta_1+\ell-\beta_2)!}{(\ell-\beta_2)!} q_2^{\ell-\beta_2}
\nonumber \\ &
=\fracc{ q_1^{k-\beta_1} q_2^{\ell-\beta_2}}{(1+\beta_1+\beta_2)} \CC{k-\beta_1+\ell-\beta_2}{k-\beta_1}
.
}{nkl2_app}
After inserting the expressions for $q_1,q_2$ from~\eqref{qs_def}, this becomes
\all{
n(k,\ell)= \fracc{ \beta_1^{k-\beta_1} \beta_2^{\ell-\beta_2} \CC{k-\beta_1+\ell-\beta_2}{k-\beta_1}}{(1+\beta_1+\beta_2)^{k-\beta_1+\ell-\beta_2+1}}.
}{nkl2_app_FIN}

\section{Performing the Summation in~\eqref{lbar_2}}\label{app:nk_2}
We need to perform the following summation:
\all{
\bar{\ell}(k) &= 
\fracc{ (\beta_1+1)^{k-\beta_1+1}}{(\beta_1+\beta_2+1)^{k-\beta_1+1}}
\DD \sum_{\ell} \ell
\frac{ \beta_2^{\ell-\beta_2} \CC{k-\beta_1+\ell-\beta_2}{k-\beta_1 } }{(1+\beta_1+\beta_2)^{ \ell-\beta_2}}
}{lbar_2_app}

Let us denote $k-\beta_1$ by $k'$ and $\ell-\beta_2$ by $\ell'$. Also let us denote $\frac{\beta_2}{1+\beta_1+\beta_2}$ by $x$. De need to evaluate the following sum: ${\sum_{\ell'} (\ell'+\beta_2) x^{\ell'} \CC{k'+\ell'}{k'}}$. Let us use~\eqref{taylor_0} and define 
\eqq{
S_1(x,n) \deff \DD \sum_m x^m \CC{m}{n} = \fracc{x^{n}}{(1-x)^{n+1}}
}{taylor_1}
We have: 
\all{
&\sum_{\ell'}(\beta_2 + \ell' ) x^{\ell'} \CC{k'+\ell'}{k'}
\nonumber \\ &
= \beta_2 x^{-k'} S_1(x,k') + x \sum_{\ell'}\ell' x^{\ell'-1} \CC{k'+\ell'}{k'} 
\nonumber \\ &
= \beta_2 x^{-k'} S_1(x,k') + x \fracc{d}{dx} \Big( x^{-k'} S_1(x,k') \Big) 
\nonumber \\ &
= \beta_2 x^{-k'} \fracc{x^{k'}}{(1-x)^{k'+1}} + x \fracc{d}{dx} \Big( \fracc{x^{k'}}{(1-x)^{k'+1}}\Big) 
\nonumber \\ &
= \fracc{1}{(1-x)^{k'+2}} \big[\beta_2+ x(k'+1-\beta_2) \big] 
.}{sum11}

Replacing $x$ with $\frac{\beta_2}{1+\beta_1+\beta_2}$ and $k'$ by $k-\beta_1$ and inserting this result into~\eqref{lbar_2_app}, we get
\all{
&\resizebox{\linewidth}{!}{$
\fracc{1}{[1-(\frac{\beta_2}{1+\beta_1+\beta_2})]^{k-\beta_1+2}}
\big[\beta_2+\frac{\beta_2}{1+\beta_1+\beta_2} (k-\beta_1+1-\beta_2) \big] 
$}
\nonumber \\ 
&\resizebox{\linewidth}{!}{$
= \fracc{ (1+\beta_1+\beta_2)^{k-\beta_1+2} }{(1+\beta_1)^{k-\beta_1+2}}
\big[\beta_2+\frac{\beta_2}{1+\beta_1+\beta_2} (k-\beta_1+1-\beta_2) \big] 
$}
\nonumber \\ 
&
= \fracc{ (1+\beta_1+\beta_2)^{k-\beta_1+2} }{(1+\beta_1)^{k-\beta_1+2}}
\big[\frac{\beta_2(k+2)}{1+\beta_1+\beta_2} \big] 
}{sum_11_2}
Plugging this into~\eqref{lbar_2_app}, we get

\all{
\bar{\ell}(k) &= 
\frac{\beta_2(k+2)}{1+\beta_1}
}{lbar2_app_FIN}


\begin{thebibliography}{46}%
\makeatletter
\providecommand \@ifxundefined [1]{%
 \@ifx{#1\undefined}
}%
\providecommand \@ifnum [1]{%
 \ifnum #1\expandafter \@firstoftwo
 \else \expandafter \@secondoftwo
 \fi
}%
\providecommand \@ifx [1]{%
 \ifx #1\expandafter \@firstoftwo
 \else \expandafter \@secondoftwo
 \fi
}%
\providecommand \natexlab [1]{#1}%
\providecommand \enquote  [1]{``#1''}%
\providecommand \bibnamefont  [1]{#1}%
\providecommand \bibfnamefont [1]{#1}%
\providecommand \citenamefont [1]{#1}%
\providecommand \href@noop [0]{\@secondoftwo}%
\providecommand \href [0]{\begingroup \@sanitize@url \@href}%
\providecommand \@href[1]{\@@startlink{#1}\@@href}%
\providecommand \@@href[1]{\endgroup#1\@@endlink}%
\providecommand \@sanitize@url [0]{\catcode `\\12\catcode `\$12\catcode
  `\&12\catcode `\#12\catcode `\^12\catcode `\_12\catcode `\%12\relax}%
\providecommand \@@startlink[1]{}%
\providecommand \@@endlink[0]{}%
\providecommand \url  [0]{\begingroup\@sanitize@url \@url }%
\providecommand \@url [1]{\endgroup\@href {#1}{\urlprefix }}%
\providecommand \urlprefix  [0]{URL }%
\providecommand \Eprint [0]{\href }%
\providecommand \doibase [0]{http://dx.doi.org/}%
\providecommand \selectlanguage [0]{\@gobble}%
\providecommand \bibinfo  [0]{\@secondoftwo}%
\providecommand \bibfield  [0]{\@secondoftwo}%
\providecommand \translation [1]{[#1]}%
\providecommand \BibitemOpen [0]{}%
\providecommand \bibitemStop [0]{}%
\providecommand \bibitemNoStop [0]{.\EOS\space}%
\providecommand \EOS [0]{\spacefactor3000\relax}%
\providecommand \BibitemShut  [1]{\csname bibitem#1\endcsname}%
\let\auto@bib@innerbib\@empty
%</preamble>
\bibitem [{\citenamefont {Menichetti}\ \emph {et~al.}(2014)\citenamefont
  {Menichetti}, \citenamefont {Remondini}, \citenamefont {Panzarasa},
  \citenamefont {Mondrag{\'o}n},\ and\ \citenamefont
  {Bianconi}}]{citation_aps}%
  \BibitemOpen
  \bibfield  {author} {\bibinfo {author} {\bibfnamefont {G.}~\bibnamefont
  {Menichetti}}, \bibinfo {author} {\bibfnamefont {D.}~\bibnamefont
  {Remondini}}, \bibinfo {author} {\bibfnamefont {P.}~\bibnamefont
  {Panzarasa}}, \bibinfo {author} {\bibfnamefont {R.~J.}\ \bibnamefont
  {Mondrag{\'o}n}}, \ and\ \bibinfo {author} {\bibfnamefont {G.}~\bibnamefont
  {Bianconi}},\ }\href@noop {} {\bibfield  {journal} {\bibinfo  {journal} {PloS
  one}\ }\textbf {\bibinfo {volume} {9}},\ \bibinfo {pages} {e97857} (\bibinfo
  {year} {2014})}\BibitemShut {NoStop}%
\bibitem [{\citenamefont {Nicosia}\ and\ \citenamefont
  {Latora}(2014)}]{aps_imdb}%
  \BibitemOpen
  \bibfield  {author} {\bibinfo {author} {\bibfnamefont {V.}~\bibnamefont
  {Nicosia}}\ and\ \bibinfo {author} {\bibfnamefont {V.}~\bibnamefont
  {Latora}},\ }\href@noop {} {\bibfield  {journal} {\bibinfo  {journal} {arXiv
  preprint arXiv:1403.1546}\ } (\bibinfo {year} {2014})}\BibitemShut {NoStop}%
\bibitem [{\citenamefont {Lewis}\ \emph {et~al.}(2008)\citenamefont {Lewis},
  \citenamefont {Kaufman}, \citenamefont {Gonzalez}, \citenamefont {Wimmer},\
  and\ \citenamefont {Christakis}}]{facebook}%
  \BibitemOpen
  \bibfield  {author} {\bibinfo {author} {\bibfnamefont {K.}~\bibnamefont
  {Lewis}}, \bibinfo {author} {\bibfnamefont {J.}~\bibnamefont {Kaufman}},
  \bibinfo {author} {\bibfnamefont {M.}~\bibnamefont {Gonzalez}}, \bibinfo
  {author} {\bibfnamefont {A.}~\bibnamefont {Wimmer}}, \ and\ \bibinfo {author}
  {\bibfnamefont {N.}~\bibnamefont {Christakis}},\ }\href@noop {} {\bibfield
  {journal} {\bibinfo  {journal} {Social networks}\ }\textbf {\bibinfo {volume}
  {30}},\ \bibinfo {pages} {330} (\bibinfo {year} {2008})}\BibitemShut
  {NoStop}%
\bibitem [{\citenamefont {Magnani}\ and\ \citenamefont {Rossi}(2011)}]{online}%
  \BibitemOpen
  \bibfield  {author} {\bibinfo {author} {\bibfnamefont {M.}~\bibnamefont
  {Magnani}}\ and\ \bibinfo {author} {\bibfnamefont {L.}~\bibnamefont
  {Rossi}},\ }in\ \href@noop {} {\emph {\bibinfo {booktitle} {Advances in
  Social Networks Analysis and Mining (ASONAM), 2011 International Conference
  on}}}\ (\bibinfo {organization} {IEEE},\ \bibinfo {year} {2011})\ pp.\
  \bibinfo {pages} {5--12}\BibitemShut {NoStop}%
\bibitem [{\citenamefont {Bargigli}\ \emph {et~al.}(2015)\citenamefont
  {Bargigli}, \citenamefont {Di~Iasio}, \citenamefont {Infante}, \citenamefont
  {Lillo},\ and\ \citenamefont {Pierobon}}]{interbank}%
  \BibitemOpen
  \bibfield  {author} {\bibinfo {author} {\bibfnamefont {L.}~\bibnamefont
  {Bargigli}}, \bibinfo {author} {\bibfnamefont {G.}~\bibnamefont {Di~Iasio}},
  \bibinfo {author} {\bibfnamefont {L.}~\bibnamefont {Infante}}, \bibinfo
  {author} {\bibfnamefont {F.}~\bibnamefont {Lillo}}, \ and\ \bibinfo {author}
  {\bibfnamefont {F.}~\bibnamefont {Pierobon}},\ }\href@noop {} {\bibfield
  {journal} {\bibinfo  {journal} {Quantitative Finance}\ }\textbf {\bibinfo
  {volume} {15}},\ \bibinfo {pages} {673} (\bibinfo {year} {2015})}\BibitemShut
  {NoStop}%
\bibitem [{\citenamefont {Cardillo}\ \emph {et~al.}(2013)\citenamefont
  {Cardillo}, \citenamefont {G{\'o}mez-Garde{\~n}es}, \citenamefont {Zanin},
  \citenamefont {Romance}, \citenamefont {Papo}, \citenamefont {del Pozo},\
  and\ \citenamefont {Boccaletti}}]{air}%
  \BibitemOpen
  \bibfield  {author} {\bibinfo {author} {\bibfnamefont {A.}~\bibnamefont
  {Cardillo}}, \bibinfo {author} {\bibfnamefont {J.}~\bibnamefont
  {G{\'o}mez-Garde{\~n}es}}, \bibinfo {author} {\bibfnamefont {M.}~\bibnamefont
  {Zanin}}, \bibinfo {author} {\bibfnamefont {M.}~\bibnamefont {Romance}},
  \bibinfo {author} {\bibfnamefont {D.}~\bibnamefont {Papo}}, \bibinfo {author}
  {\bibfnamefont {F.}~\bibnamefont {del Pozo}}, \ and\ \bibinfo {author}
  {\bibfnamefont {S.}~\bibnamefont {Boccaletti}},\ }\href@noop {} {\bibfield
  {journal} {\bibinfo  {journal} {Sci. Rep.}\ }\textbf {\bibinfo {volume} {3}}
  (\bibinfo {year} {2013})}\BibitemShut {NoStop}%
\bibitem [{\citenamefont {Battiston}\ \emph
  {et~al.}(2015{\natexlab{a}})\citenamefont {Battiston}, \citenamefont
  {Iacovacci}, \citenamefont {Nicosia}, \citenamefont {Bianconi},\ and\
  \citenamefont {Latora}}]{battiston}%
  \BibitemOpen
  \bibfield  {author} {\bibinfo {author} {\bibfnamefont {F.}~\bibnamefont
  {Battiston}}, \bibinfo {author} {\bibfnamefont {J.}~\bibnamefont
  {Iacovacci}}, \bibinfo {author} {\bibfnamefont {V.}~\bibnamefont {Nicosia}},
  \bibinfo {author} {\bibfnamefont {G.}~\bibnamefont {Bianconi}}, \ and\
  \bibinfo {author} {\bibfnamefont {V.}~\bibnamefont {Latora}},\ }\href@noop {}
  {\bibfield  {journal} {\bibinfo  {journal} {arXiv preprint arXiv:1506.01280}\
  } (\bibinfo {year} {2015}{\natexlab{a}})}\BibitemShut {NoStop}%
\bibitem [{\citenamefont {Szell}\ \emph {et~al.}(2010)\citenamefont {Szell},
  \citenamefont {Lambiotte},\ and\ \citenamefont {Thurner}}]{game}%
  \BibitemOpen
  \bibfield  {author} {\bibinfo {author} {\bibfnamefont {M.}~\bibnamefont
  {Szell}}, \bibinfo {author} {\bibfnamefont {R.}~\bibnamefont {Lambiotte}}, \
  and\ \bibinfo {author} {\bibfnamefont {S.}~\bibnamefont {Thurner}},\
  }\href@noop {} {\bibfield  {journal} {\bibinfo  {journal} {Proc. Nat. Acad.
  Sci.}\ }\textbf {\bibinfo {volume} {107}},\ \bibinfo {pages} {13636}
  (\bibinfo {year} {2010})}\BibitemShut {NoStop}%
\bibitem [{\citenamefont {De~Domenico}\ \emph
  {et~al.}(2013{\natexlab{a}})\citenamefont {De~Domenico}, \citenamefont
  {Sol{\'e}-Ribalta}, \citenamefont {Cozzo}, \citenamefont {Kivel{\"a}},
  \citenamefont {Moreno}, \citenamefont {Porter}, \citenamefont {G{\'o}mez},\
  and\ \citenamefont {Arenas}}]{pion1}%
  \BibitemOpen
  \bibfield  {author} {\bibinfo {author} {\bibfnamefont {M.}~\bibnamefont
  {De~Domenico}}, \bibinfo {author} {\bibfnamefont {A.}~\bibnamefont
  {Sol{\'e}-Ribalta}}, \bibinfo {author} {\bibfnamefont {E.}~\bibnamefont
  {Cozzo}}, \bibinfo {author} {\bibfnamefont {M.}~\bibnamefont {Kivel{\"a}}},
  \bibinfo {author} {\bibfnamefont {Y.}~\bibnamefont {Moreno}}, \bibinfo
  {author} {\bibfnamefont {M.~A.}\ \bibnamefont {Porter}}, \bibinfo {author}
  {\bibfnamefont {S.}~\bibnamefont {G{\'o}mez}}, \ and\ \bibinfo {author}
  {\bibfnamefont {A.}~\bibnamefont {Arenas}},\ }\href@noop {} {\bibfield
  {journal} {\bibinfo  {journal} {Phys. Rev. X}\ }\textbf {\bibinfo {volume}
  {3}},\ \bibinfo {pages} {041022} (\bibinfo {year}
  {2013}{\natexlab{a}})}\BibitemShut {NoStop}%
\bibitem [{\citenamefont {Kivel\''a}\ \emph {et~al.}(2014)\citenamefont
  {Kivel\''a}, \citenamefont {Arenas}, \citenamefont {Barthelemy},
  \citenamefont {Gleeson}, \citenamefont {Moreno},\ and\ \citenamefont
  {Porter}}]{pion2}%
  \BibitemOpen
  \bibfield  {author} {\bibinfo {author} {\bibfnamefont {M.}~\bibnamefont
  {Kivel\''a}}, \bibinfo {author} {\bibfnamefont {A.}~\bibnamefont {Arenas}},
  \bibinfo {author} {\bibfnamefont {M.}~\bibnamefont {Barthelemy}}, \bibinfo
  {author} {\bibfnamefont {J.~P.}\ \bibnamefont {Gleeson}}, \bibinfo {author}
  {\bibfnamefont {Y.}~\bibnamefont {Moreno}}, \ and\ \bibinfo {author}
  {\bibfnamefont {M.~A.}\ \bibnamefont {Porter}},\ }\href@noop {} {\bibfield
  {journal} {\bibinfo  {journal} {J Complex Net.}\ }\textbf {\bibinfo {volume}
  {2}},\ \bibinfo {pages} {203} (\bibinfo {year} {2014})}\BibitemShut {NoStop}%
\bibitem [{\citenamefont {Sole-Ribalta}\ \emph {et~al.}(2013)\citenamefont
  {Sole-Ribalta}, \citenamefont {De~Domenico}, \citenamefont {Kouvaris},
  \citenamefont {Diaz-Guilera}, \citenamefont {Gomez},\ and\ \citenamefont
  {Arenas}}]{pion3}%
  \BibitemOpen
  \bibfield  {author} {\bibinfo {author} {\bibfnamefont {A.}~\bibnamefont
  {Sole-Ribalta}}, \bibinfo {author} {\bibfnamefont {M.}~\bibnamefont
  {De~Domenico}}, \bibinfo {author} {\bibfnamefont {N.~E.}\ \bibnamefont
  {Kouvaris}}, \bibinfo {author} {\bibfnamefont {A.}~\bibnamefont
  {Diaz-Guilera}}, \bibinfo {author} {\bibfnamefont {S.}~\bibnamefont {Gomez}},
  \ and\ \bibinfo {author} {\bibfnamefont {A.}~\bibnamefont {Arenas}},\
  }\href@noop {} {\bibfield  {journal} {\bibinfo  {journal} {Phy. Rev. E}\
  }\textbf {\bibinfo {volume} {88}},\ \bibinfo {pages} {032807} (\bibinfo
  {year} {2013})}\BibitemShut {NoStop}%
\bibitem [{\citenamefont {Son}\ \emph {et~al.}(2012)\citenamefont {Son},
  \citenamefont {Bizhani}, \citenamefont {Christensen}, \citenamefont
  {Grassberger},\ and\ \citenamefont {Paczuski}}]{epid1}%
  \BibitemOpen
  \bibfield  {author} {\bibinfo {author} {\bibfnamefont {S.-W.}\ \bibnamefont
  {Son}}, \bibinfo {author} {\bibfnamefont {G.}~\bibnamefont {Bizhani}},
  \bibinfo {author} {\bibfnamefont {C.}~\bibnamefont {Christensen}}, \bibinfo
  {author} {\bibfnamefont {P.}~\bibnamefont {Grassberger}}, \ and\ \bibinfo
  {author} {\bibfnamefont {M.}~\bibnamefont {Paczuski}},\ }\href@noop {}
  {\bibfield  {journal} {\bibinfo  {journal} {Eur. Phy. Lett.}\ }\textbf
  {\bibinfo {volume} {97}},\ \bibinfo {pages} {16006} (\bibinfo {year}
  {2012})}\BibitemShut {NoStop}%
\bibitem [{\citenamefont {Saumell-Mendiola}\ \emph {et~al.}(2012)\citenamefont
  {Saumell-Mendiola}, \citenamefont {Serrano},\ and\ \citenamefont
  {Bogu{\~n}{\'a}}}]{epid2}%
  \BibitemOpen
  \bibfield  {author} {\bibinfo {author} {\bibfnamefont {A.}~\bibnamefont
  {Saumell-Mendiola}}, \bibinfo {author} {\bibfnamefont {M.~{\'A}.}\
  \bibnamefont {Serrano}}, \ and\ \bibinfo {author} {\bibfnamefont
  {M.}~\bibnamefont {Bogu{\~n}{\'a}}},\ }\href@noop {} {\bibfield  {journal}
  {\bibinfo  {journal} {Phys. Rev. E}\ }\textbf {\bibinfo {volume} {86}},\
  \bibinfo {pages} {026106} (\bibinfo {year} {2012})}\BibitemShut {NoStop}%
\bibitem [{\citenamefont {Granell}\ \emph {et~al.}(2013)\citenamefont
  {Granell}, \citenamefont {G{\'o}mez},\ and\ \citenamefont
  {Arenas}}]{interplay}%
  \BibitemOpen
  \bibfield  {author} {\bibinfo {author} {\bibfnamefont {C.}~\bibnamefont
  {Granell}}, \bibinfo {author} {\bibfnamefont {S.}~\bibnamefont {G{\'o}mez}},
  \ and\ \bibinfo {author} {\bibfnamefont {A.}~\bibnamefont {Arenas}},\
  }\href@noop {} {\bibfield  {journal} {\bibinfo  {journal} {Phys. Rev. Lett.}\
  }\textbf {\bibinfo {volume} {111}},\ \bibinfo {pages} {128701} (\bibinfo
  {year} {2013})}\BibitemShut {NoStop}%
\bibitem [{\citenamefont {Granell}\ \emph {et~al.}(2014)\citenamefont
  {Granell}, \citenamefont {G{\'o}mez},\ and\ \citenamefont
  {Arenas}}]{interplay2}%
  \BibitemOpen
  \bibfield  {author} {\bibinfo {author} {\bibfnamefont {C.}~\bibnamefont
  {Granell}}, \bibinfo {author} {\bibfnamefont {S.}~\bibnamefont {G{\'o}mez}},
  \ and\ \bibinfo {author} {\bibfnamefont {A.}~\bibnamefont {Arenas}},\
  }\href@noop {} {\bibfield  {journal} {\bibinfo  {journal} {Physical Review
  E}\ }\textbf {\bibinfo {volume} {90}},\ \bibinfo {pages} {012808} (\bibinfo
  {year} {2014})}\BibitemShut {NoStop}%
\bibitem [{\citenamefont {Cellai}\ \emph {et~al.}(2013)\citenamefont {Cellai},
  \citenamefont {L{\'o}pez}, \citenamefont {Zhou}, \citenamefont {Gleeson},\
  and\ \citenamefont {Bianconi}}]{percolation1}%
  \BibitemOpen
  \bibfield  {author} {\bibinfo {author} {\bibfnamefont {D.}~\bibnamefont
  {Cellai}}, \bibinfo {author} {\bibfnamefont {E.}~\bibnamefont {L{\'o}pez}},
  \bibinfo {author} {\bibfnamefont {J.}~\bibnamefont {Zhou}}, \bibinfo {author}
  {\bibfnamefont {J.~P.}\ \bibnamefont {Gleeson}}, \ and\ \bibinfo {author}
  {\bibfnamefont {G.}~\bibnamefont {Bianconi}},\ }\href@noop {} {\bibfield
  {journal} {\bibinfo  {journal} {Phys. Rev. E}\ }\textbf {\bibinfo {volume}
  {88}},\ \bibinfo {pages} {052811} (\bibinfo {year} {2013})}\BibitemShut
  {NoStop}%
\bibitem [{\citenamefont {Baxter}\ \emph {et~al.}(2014)\citenamefont {Baxter},
  \citenamefont {Dorogovtsev}, \citenamefont {Mendes},\ and\ \citenamefont
  {Cellai}}]{percolation2}%
  \BibitemOpen
  \bibfield  {author} {\bibinfo {author} {\bibfnamefont {G.~J.}\ \bibnamefont
  {Baxter}}, \bibinfo {author} {\bibfnamefont {S.~N.}\ \bibnamefont
  {Dorogovtsev}}, \bibinfo {author} {\bibfnamefont {J.~F.}\ \bibnamefont
  {Mendes}}, \ and\ \bibinfo {author} {\bibfnamefont {D.}~\bibnamefont
  {Cellai}},\ }\href@noop {} {\bibfield  {journal} {\bibinfo  {journal} {Phys.
  Rev. E}\ }\textbf {\bibinfo {volume} {89}},\ \bibinfo {pages} {042801}
  (\bibinfo {year} {2014})}\BibitemShut {NoStop}%
\bibitem [{\citenamefont {De~Domenico}\ \emph
  {et~al.}(2013{\natexlab{b}})\citenamefont {De~Domenico}, \citenamefont
  {Sol{\'e}}, \citenamefont {G{\'o}mez},\ and\ \citenamefont {Arenas}}]{RW}%
  \BibitemOpen
  \bibfield  {author} {\bibinfo {author} {\bibfnamefont {M.}~\bibnamefont
  {De~Domenico}}, \bibinfo {author} {\bibfnamefont {A.}~\bibnamefont
  {Sol{\'e}}}, \bibinfo {author} {\bibfnamefont {S.}~\bibnamefont {G{\'o}mez}},
  \ and\ \bibinfo {author} {\bibfnamefont {A.}~\bibnamefont {Arenas}},\
  }\href@noop {} {\bibfield  {journal} {\bibinfo  {journal} {arXiv preprint
  arXiv:1306.0519}\ } (\bibinfo {year} {2013}{\natexlab{b}})}\BibitemShut
  {NoStop}%
\bibitem [{\citenamefont {Battiston}\ \emph
  {et~al.}(2015{\natexlab{b}})\citenamefont {Battiston}, \citenamefont
  {Nicosia},\ and\ \citenamefont {Latora}}]{RW2}%
  \BibitemOpen
  \bibfield  {author} {\bibinfo {author} {\bibfnamefont {F.}~\bibnamefont
  {Battiston}}, \bibinfo {author} {\bibfnamefont {V.}~\bibnamefont {Nicosia}},
  \ and\ \bibinfo {author} {\bibfnamefont {V.}~\bibnamefont {Latora}},\
  }\href@noop {} {\bibfield  {journal} {\bibinfo  {journal} {arXiv preprint
  arXiv:1505.01378}\ } (\bibinfo {year} {2015}{\natexlab{b}})}\BibitemShut
  {NoStop}%
\bibitem [{\citenamefont {Matamalas}\ \emph {et~al.}(2015)\citenamefont
  {Matamalas}, \citenamefont {Poncela-Casasnovas}, \citenamefont {G{\'o}mez},\
  and\ \citenamefont {Arenas}}]{cooperation3}%
  \BibitemOpen
  \bibfield  {author} {\bibinfo {author} {\bibfnamefont {J.~T.}\ \bibnamefont
  {Matamalas}}, \bibinfo {author} {\bibfnamefont {J.}~\bibnamefont
  {Poncela-Casasnovas}}, \bibinfo {author} {\bibfnamefont {S.}~\bibnamefont
  {G{\'o}mez}}, \ and\ \bibinfo {author} {\bibfnamefont {A.}~\bibnamefont
  {Arenas}},\ }\href@noop {} {\bibfield  {journal} {\bibinfo  {journal}
  {Scientific reports}\ }\textbf {\bibinfo {volume} {5}} (\bibinfo {year}
  {2015})}\BibitemShut {NoStop}%
\bibitem [{\citenamefont {G{\'o}mez-Garde{\~n}es}\ \emph
  {et~al.}(2012)\citenamefont {G{\'o}mez-Garde{\~n}es}, \citenamefont
  {Reinares}, \citenamefont {Arenas},\ and\ \citenamefont
  {Flor{\'\i}a}}]{cooperation1}%
  \BibitemOpen
  \bibfield  {author} {\bibinfo {author} {\bibfnamefont {J.}~\bibnamefont
  {G{\'o}mez-Garde{\~n}es}}, \bibinfo {author} {\bibfnamefont {I.}~\bibnamefont
  {Reinares}}, \bibinfo {author} {\bibfnamefont {A.}~\bibnamefont {Arenas}}, \
  and\ \bibinfo {author} {\bibfnamefont {L.~M.}\ \bibnamefont {Flor{\'\i}a}},\
  }\href@noop {} {\bibfield  {journal} {\bibinfo  {journal} {Scientific
  reports}\ }\textbf {\bibinfo {volume} {2}} (\bibinfo {year}
  {2012})}\BibitemShut {NoStop}%
\bibitem [{\citenamefont {Wang}\ \emph {et~al.}(2013)\citenamefont {Wang},
  \citenamefont {Szolnoki},\ and\ \citenamefont {Perc}}]{cooperation2}%
  \BibitemOpen
  \bibfield  {author} {\bibinfo {author} {\bibfnamefont {Z.}~\bibnamefont
  {Wang}}, \bibinfo {author} {\bibfnamefont {A.}~\bibnamefont {Szolnoki}}, \
  and\ \bibinfo {author} {\bibfnamefont {M.}~\bibnamefont {Perc}},\ }\href@noop
  {} {\bibfield  {journal} {\bibinfo  {journal} {Sci. Rep.}\ }\textbf {\bibinfo
  {volume} {3}} (\bibinfo {year} {2013})}\BibitemShut {NoStop}%
\bibitem [{\citenamefont {Wang}\ \emph {et~al.}(2015)\citenamefont {Wang},
  \citenamefont {Wang}, \citenamefont {Szolnoki},\ and\ \citenamefont
  {Perc}}]{matjaz}%
  \BibitemOpen
  \bibfield  {author} {\bibinfo {author} {\bibfnamefont {Z.}~\bibnamefont
  {Wang}}, \bibinfo {author} {\bibfnamefont {L.}~\bibnamefont {Wang}}, \bibinfo
  {author} {\bibfnamefont {A.}~\bibnamefont {Szolnoki}}, \ and\ \bibinfo
  {author} {\bibfnamefont {M.}~\bibnamefont {Perc}},\ }\href@noop {} {\bibfield
   {journal} {\bibinfo  {journal} {The European Physical Journal B}\ }\textbf
  {\bibinfo {volume} {88}},\ \bibinfo {pages} {1} (\bibinfo {year}
  {2015})}\BibitemShut {NoStop}%
\bibitem [{\citenamefont {Gomez}\ \emph {et~al.}(2013)\citenamefont {Gomez},
  \citenamefont {Diaz-Guilera}, \citenamefont {Gomez-Garde{\~n}es},
  \citenamefont {Perez-Vicente}, \citenamefont {Moreno},\ and\ \citenamefont
  {Arenas}}]{diffusion}%
  \BibitemOpen
  \bibfield  {author} {\bibinfo {author} {\bibfnamefont {S.}~\bibnamefont
  {Gomez}}, \bibinfo {author} {\bibfnamefont {A.}~\bibnamefont {Diaz-Guilera}},
  \bibinfo {author} {\bibfnamefont {J.}~\bibnamefont {Gomez-Garde{\~n}es}},
  \bibinfo {author} {\bibfnamefont {C.~J.}\ \bibnamefont {Perez-Vicente}},
  \bibinfo {author} {\bibfnamefont {Y.}~\bibnamefont {Moreno}}, \ and\ \bibinfo
  {author} {\bibfnamefont {A.}~\bibnamefont {Arenas}},\ }\href@noop {}
  {\bibfield  {journal} {\bibinfo  {journal} {Phys. Rev. Lett.}\ }\textbf
  {\bibinfo {volume} {110}},\ \bibinfo {pages} {028701} (\bibinfo {year}
  {2013})}\BibitemShut {NoStop}%
\bibitem [{\citenamefont {Cozzo}\ \emph {et~al.}(2013)\citenamefont {Cozzo},
  \citenamefont {Banos}, \citenamefont {Meloni},\ and\ \citenamefont
  {Moreno}}]{information}%
  \BibitemOpen
  \bibfield  {author} {\bibinfo {author} {\bibfnamefont {E.}~\bibnamefont
  {Cozzo}}, \bibinfo {author} {\bibfnamefont {R.~A.}\ \bibnamefont {Banos}},
  \bibinfo {author} {\bibfnamefont {S.}~\bibnamefont {Meloni}}, \ and\ \bibinfo
  {author} {\bibfnamefont {Y.}~\bibnamefont {Moreno}},\ }\href@noop {}
  {\bibfield  {journal} {\bibinfo  {journal} {Phys. Rev. E}\ }\textbf {\bibinfo
  {volume} {88}},\ \bibinfo {pages} {050801} (\bibinfo {year}
  {2013})}\BibitemShut {NoStop}%
\bibitem [{\citenamefont {Boccaletti}\ \emph {et~al.}(2014)\citenamefont
  {Boccaletti}, \citenamefont {Bianconi}, \citenamefont {Criado}, \citenamefont
  {Del~Genio}, \citenamefont {G{\'o}mez-Garde{\~n}es}, \citenamefont {Romance},
  \citenamefont {Sendi{\~n}a-Nadal}, \citenamefont {Wang},\ and\ \citenamefont
  {Zanin}}]{survey}%
  \BibitemOpen
  \bibfield  {author} {\bibinfo {author} {\bibfnamefont {S.}~\bibnamefont
  {Boccaletti}}, \bibinfo {author} {\bibfnamefont {G.}~\bibnamefont
  {Bianconi}}, \bibinfo {author} {\bibfnamefont {R.}~\bibnamefont {Criado}},
  \bibinfo {author} {\bibfnamefont {C.}~\bibnamefont {Del~Genio}}, \bibinfo
  {author} {\bibfnamefont {J.}~\bibnamefont {G{\'o}mez-Garde{\~n}es}}, \bibinfo
  {author} {\bibfnamefont {M.}~\bibnamefont {Romance}}, \bibinfo {author}
  {\bibfnamefont {I.}~\bibnamefont {Sendi{\~n}a-Nadal}}, \bibinfo {author}
  {\bibfnamefont {Z.}~\bibnamefont {Wang}}, \ and\ \bibinfo {author}
  {\bibfnamefont {M.}~\bibnamefont {Zanin}},\ }\href@noop {} {\bibfield
  {journal} {\bibinfo  {journal} {Phys. Rep.}\ }\textbf {\bibinfo {volume}
  {544}},\ \bibinfo {pages} {1} (\bibinfo {year} {2014})}\BibitemShut {NoStop}%
\bibitem [{\citenamefont {Kenett}\ \emph {et~al.}(2015)\citenamefont {Kenett},
  \citenamefont {Perc},\ and\ \citenamefont {Boccaletti}}]{matjaz2}%
  \BibitemOpen
  \bibfield  {author} {\bibinfo {author} {\bibfnamefont {D.~Y.}\ \bibnamefont
  {Kenett}}, \bibinfo {author} {\bibfnamefont {M.}~\bibnamefont {Perc}}, \ and\
  \bibinfo {author} {\bibfnamefont {S.}~\bibnamefont {Boccaletti}},\
  }\href@noop {} {\bibfield  {journal} {\bibinfo  {journal} {Chaos, Solitons \&
  Fractals}\ }\textbf {\bibinfo {volume} {80}},\ \bibinfo {pages} {1} (\bibinfo
  {year} {2015})}\BibitemShut {NoStop}%
\bibitem [{\citenamefont {Nicosia}\ \emph {et~al.}(2014)\citenamefont
  {Nicosia}, \citenamefont {Bianconi}, \citenamefont {Latora},\ and\
  \citenamefont {Barthelemy}}]{Bian2}%
  \BibitemOpen
  \bibfield  {author} {\bibinfo {author} {\bibfnamefont {V.}~\bibnamefont
  {Nicosia}}, \bibinfo {author} {\bibfnamefont {G.}~\bibnamefont {Bianconi}},
  \bibinfo {author} {\bibfnamefont {V.}~\bibnamefont {Latora}}, \ and\ \bibinfo
  {author} {\bibfnamefont {M.}~\bibnamefont {Barthelemy}},\ }\href@noop {}
  {\bibfield  {journal} {\bibinfo  {journal} {Phys. Rev. E}\ }\textbf {\bibinfo
  {volume} {90}},\ \bibinfo {pages} {042807} (\bibinfo {year}
  {2014})}\BibitemShut {NoStop}%
\bibitem [{\citenamefont {Nicosia}\ \emph {et~al.}(2013)\citenamefont
  {Nicosia}, \citenamefont {Bianconi}, \citenamefont {Latora},\ and\
  \citenamefont {Barthelemy}}]{Bian1}%
  \BibitemOpen
  \bibfield  {author} {\bibinfo {author} {\bibfnamefont {V.}~\bibnamefont
  {Nicosia}}, \bibinfo {author} {\bibfnamefont {G.}~\bibnamefont {Bianconi}},
  \bibinfo {author} {\bibfnamefont {V.}~\bibnamefont {Latora}}, \ and\ \bibinfo
  {author} {\bibfnamefont {M.}~\bibnamefont {Barthelemy}},\ }\href {\doibase
  10.1103/PhysRevLett.111.058701} {\bibfield  {journal} {\bibinfo  {journal}
  {Phys. Rev. Lett.}\ }\textbf {\bibinfo {volume} {111}},\ \bibinfo {pages}
  {058701} (\bibinfo {year} {2013})}\BibitemShut {NoStop}%
\bibitem [{\citenamefont {Kim}\ and\ \citenamefont {Goh}(2013)}]{coevolution}%
  \BibitemOpen
  \bibfield  {author} {\bibinfo {author} {\bibfnamefont {J.~Y.}\ \bibnamefont
  {Kim}}\ and\ \bibinfo {author} {\bibfnamefont {K.-I.}\ \bibnamefont {Goh}},\
  }\href {\doibase 10.1103/PhysRevLett.111.058702} {\bibfield  {journal}
  {\bibinfo  {journal} {Phys. Rev. Lett.}\ }\textbf {\bibinfo {volume} {111}},\
  \bibinfo {pages} {058702} (\bibinfo {year} {2013})}\BibitemShut {NoStop}%
\bibitem [{\citenamefont {Bianconi}(2015)}]{SUSY}%
  \BibitemOpen
  \bibfield  {author} {\bibinfo {author} {\bibfnamefont {G.}~\bibnamefont
  {Bianconi}},\ }\href {\doibase 10.1103/PhysRevE.91.012810} {\bibfield
  {journal} {\bibinfo  {journal} {Phys. Rev. E}\ }\textbf {\bibinfo {volume}
  {91}},\ \bibinfo {pages} {012810} (\bibinfo {year} {2015})}\BibitemShut
  {NoStop}%
\bibitem [{\citenamefont {Fotouhi}\ and\ \citenamefont
  {Momeni}(2015)}]{nameni}%
  \BibitemOpen
  \bibfield  {author} {\bibinfo {author} {\bibfnamefont {B.}~\bibnamefont
  {Fotouhi}}\ and\ \bibinfo {author} {\bibfnamefont {N.}~\bibnamefont
  {Momeni}},\ }in\ \href@noop {} {\emph {\bibinfo {booktitle} {Complex Networks
  VI}}}\ (\bibinfo  {publisher} {Springer},\ \bibinfo {year} {2015})\ pp.\
  \bibinfo {pages} {159--170}\BibitemShut {NoStop}%
\bibitem [{\citenamefont {Halu}\ \emph {et~al.}(2014)\citenamefont {Halu},
  \citenamefont {Mukherjee},\ and\ \citenamefont {Bianconi}}]{india}%
  \BibitemOpen
  \bibfield  {author} {\bibinfo {author} {\bibfnamefont {A.}~\bibnamefont
  {Halu}}, \bibinfo {author} {\bibfnamefont {S.}~\bibnamefont {Mukherjee}}, \
  and\ \bibinfo {author} {\bibfnamefont {G.}~\bibnamefont {Bianconi}},\
  }\href@noop {} {\bibfield  {journal} {\bibinfo  {journal} {Physical Review
  E}\ }\textbf {\bibinfo {volume} {89}},\ \bibinfo {pages} {012806} (\bibinfo
  {year} {2014})}\BibitemShut {NoStop}%
\bibitem [{\citenamefont {Barigozzi}\ \emph {et~al.}(2010)\citenamefont
  {Barigozzi}, \citenamefont {Fagiolo},\ and\ \citenamefont
  {Garlaschelli}}]{trade}%
  \BibitemOpen
  \bibfield  {author} {\bibinfo {author} {\bibfnamefont {M.}~\bibnamefont
  {Barigozzi}}, \bibinfo {author} {\bibfnamefont {G.}~\bibnamefont {Fagiolo}},
  \ and\ \bibinfo {author} {\bibfnamefont {D.}~\bibnamefont {Garlaschelli}},\
  }\href@noop {} {\bibfield  {journal} {\bibinfo  {journal} {Physical Review
  E}\ }\textbf {\bibinfo {volume} {81}},\ \bibinfo {pages} {046104} (\bibinfo
  {year} {2010})}\BibitemShut {NoStop}%
\bibitem [{\citenamefont {Dorogovtsev}\ \emph {et~al.}(2000)\citenamefont
  {Dorogovtsev}, \citenamefont {Mendes},\ and\ \citenamefont
  {Samukhin}}]{dorog_nk}%
  \BibitemOpen
  \bibfield  {author} {\bibinfo {author} {\bibfnamefont {S.~N.}\ \bibnamefont
  {Dorogovtsev}}, \bibinfo {author} {\bibfnamefont {J.~F.~F.}\ \bibnamefont
  {Mendes}}, \ and\ \bibinfo {author} {\bibfnamefont {A.~N.}\ \bibnamefont
  {Samukhin}},\ }\href@noop {} {\bibfield  {journal} {\bibinfo  {journal}
  {Phys. Rev. Lett.}\ }\textbf {\bibinfo {volume} {85}},\ \bibinfo {pages}
  {4633} (\bibinfo {year} {2000})}\BibitemShut {NoStop}%
\bibitem [{\citenamefont {Krapivsky}\ \emph {et~al.}(2000)\citenamefont
  {Krapivsky}, \citenamefont {Redner},\ and\ \citenamefont
  {Leyvraz}}]{redner2}%
  \BibitemOpen
  \bibfield  {author} {\bibinfo {author} {\bibfnamefont {P.}~\bibnamefont
  {Krapivsky}}, \bibinfo {author} {\bibfnamefont {S.}~\bibnamefont {Redner}}, \
  and\ \bibinfo {author} {\bibfnamefont {F.}~\bibnamefont {Leyvraz}},\
  }\href@noop {} {\bibfield  {journal} {\bibinfo  {journal} {Phys. Rev. Lett.}\
  }\textbf {\bibinfo {volume} {85}},\ \bibinfo {pages} {4629} (\bibinfo {year}
  {2000})}\BibitemShut {NoStop}%
\bibitem [{\citenamefont {Fotouhi}\ and\ \citenamefont
  {Rabbat}(2013)}]{ME_PRE}%
  \BibitemOpen
  \bibfield  {author} {\bibinfo {author} {\bibfnamefont {B.}~\bibnamefont
  {Fotouhi}}\ and\ \bibinfo {author} {\bibfnamefont {M.~G.}\ \bibnamefont
  {Rabbat}},\ }\href@noop {} {\bibfield  {journal} {\bibinfo  {journal}
  {Physical Review E}\ }\textbf {\bibinfo {volume} {88}},\ \bibinfo {pages}
  {062801} (\bibinfo {year} {2013})}\BibitemShut {NoStop}%
\bibitem [{\citenamefont {Boyce}\ \emph {et~al.}(1992)\citenamefont {Boyce},
  \citenamefont {DiPrima},\ and\ \citenamefont {Haines}}]{boyce1992elementary}%
  \BibitemOpen
  \bibfield  {author} {\bibinfo {author} {\bibfnamefont {W.~E.}\ \bibnamefont
  {Boyce}}, \bibinfo {author} {\bibfnamefont {R.~C.}\ \bibnamefont {DiPrima}},
  \ and\ \bibinfo {author} {\bibfnamefont {C.~W.}\ \bibnamefont {Haines}},\
  }\href@noop {} {\emph {\bibinfo {title} {Elementary differential equations
  and boundary value problems}}},\ Vol.~\bibinfo {volume} {9}\ (\bibinfo
  {publisher} {Wiley New York},\ \bibinfo {year} {1992})\BibitemShut {NoStop}%
\bibitem [{\citenamefont {Newman}\ and\ \citenamefont
  {Watts}(1999)}]{newman1999renormalization}%
  \BibitemOpen
  \bibfield  {author} {\bibinfo {author} {\bibfnamefont {M.~E.}\ \bibnamefont
  {Newman}}\ and\ \bibinfo {author} {\bibfnamefont {D.~J.}\ \bibnamefont
  {Watts}},\ }\href@noop {} {\bibfield  {journal} {\bibinfo  {journal} {Physics
  Letters A}\ }\textbf {\bibinfo {volume} {263}},\ \bibinfo {pages} {341}
  (\bibinfo {year} {1999})}\BibitemShut {NoStop}%
\bibitem [{\citenamefont {Erd{\"o}s}\ and\ \citenamefont
  {R{\'e}nyi}(1960)}]{erdos1960evolution}%
  \BibitemOpen
  \bibfield  {author} {\bibinfo {author} {\bibfnamefont {P.}~\bibnamefont
  {Erd{\"o}s}}\ and\ \bibinfo {author} {\bibfnamefont {A.}~\bibnamefont
  {R{\'e}nyi}},\ }\href@noop {} {\bibfield  {journal} {\bibinfo  {journal}
  {Publ. Math. Inst. Hung. Acad. Sci}\ }\textbf {\bibinfo {volume} {5}},\
  \bibinfo {pages} {17} (\bibinfo {year} {1960})}\BibitemShut {NoStop}%
\bibitem [{\citenamefont {Barab{\'a}si}\ and\ \citenamefont
  {Albert}(1999)}]{barabasi1999emergence}%
  \BibitemOpen
  \bibfield  {author} {\bibinfo {author} {\bibfnamefont {A.-L.}\ \bibnamefont
  {Barab{\'a}si}}\ and\ \bibinfo {author} {\bibfnamefont {R.}~\bibnamefont
  {Albert}},\ }\href@noop {} {\bibfield  {journal} {\bibinfo  {journal}
  {science}\ }\textbf {\bibinfo {volume} {286}},\ \bibinfo {pages} {509}
  (\bibinfo {year} {1999})}\BibitemShut {NoStop}%
\bibitem [{\citenamefont {Dorogovtsev}\ \emph {et~al.}(2008)\citenamefont
  {Dorogovtsev}, \citenamefont {Krapivsky},\ and\ \citenamefont
  {Mendes}}]{dorogovtsev2008transition}%
  \BibitemOpen
  \bibfield  {author} {\bibinfo {author} {\bibfnamefont {S.~N.}\ \bibnamefont
  {Dorogovtsev}}, \bibinfo {author} {\bibfnamefont {P.~L.}\ \bibnamefont
  {Krapivsky}}, \ and\ \bibinfo {author} {\bibfnamefont {J.}~\bibnamefont
  {Mendes}},\ }\href@noop {} {\bibfield  {journal} {\bibinfo  {journal} {EPL
  (Europhysics Letters)}\ }\textbf {\bibinfo {volume} {81}},\ \bibinfo {pages}
  {30004} (\bibinfo {year} {2008})}\BibitemShut {NoStop}%
\bibitem [{\citenamefont {Eguiluz}\ and\ \citenamefont
  {Klemm}(2002)}]{Eguiluz_PRL}%
  \BibitemOpen
  \bibfield  {author} {\bibinfo {author} {\bibfnamefont {V.~M.}\ \bibnamefont
  {Eguiluz}}\ and\ \bibinfo {author} {\bibfnamefont {K.}~\bibnamefont
  {Klemm}},\ }\href@noop {} {\bibfield  {journal} {\bibinfo  {journal} {Phys.
  Rev. Lett.}\ }\textbf {\bibinfo {volume} {89}},\ \bibinfo {pages} {108701}
  (\bibinfo {year} {2002})}\BibitemShut {NoStop}%
\bibitem [{\citenamefont {Barth{\'e}lemy}\ \emph {et~al.}(2004)\citenamefont
  {Barth{\'e}lemy}, \citenamefont {Barrat}, \citenamefont {Pastor-Satorras},\
  and\ \citenamefont {Vespignani}}]{Barthelemy_PRL}%
  \BibitemOpen
  \bibfield  {author} {\bibinfo {author} {\bibfnamefont {M.}~\bibnamefont
  {Barth{\'e}lemy}}, \bibinfo {author} {\bibfnamefont {A.}~\bibnamefont
  {Barrat}}, \bibinfo {author} {\bibfnamefont {R.}~\bibnamefont
  {Pastor-Satorras}}, \ and\ \bibinfo {author} {\bibfnamefont {A.}~\bibnamefont
  {Vespignani}},\ }\href@noop {} {\bibfield  {journal} {\bibinfo  {journal}
  {Phys. Rev. Lett.}\ }\textbf {\bibinfo {volume} {92}},\ \bibinfo {pages}
  {178701} (\bibinfo {year} {2004})}\BibitemShut {NoStop}%
\bibitem [{\citenamefont {Castellano}\ and\ \citenamefont
  {Pastor-Satorras}(2006)}]{Castellano_PRL}%
  \BibitemOpen
  \bibfield  {author} {\bibinfo {author} {\bibfnamefont {C.}~\bibnamefont
  {Castellano}}\ and\ \bibinfo {author} {\bibfnamefont {R.}~\bibnamefont
  {Pastor-Satorras}},\ }\href@noop {} {\bibfield  {journal} {\bibinfo
  {journal} {Phys. Rev. Lett.}\ }\textbf {\bibinfo {volume} {96}},\ \bibinfo
  {pages} {038701} (\bibinfo {year} {2006})}\BibitemShut {NoStop}%
\bibitem [{\citenamefont {Castellano}\ and\ \citenamefont
  {Pastor-Satorras}(2008)}]{Castellano_PRL_2}%
  \BibitemOpen
  \bibfield  {author} {\bibinfo {author} {\bibfnamefont {C.}~\bibnamefont
  {Castellano}}\ and\ \bibinfo {author} {\bibfnamefont {R.}~\bibnamefont
  {Pastor-Satorras}},\ }\href@noop {} {\bibfield  {journal} {\bibinfo
  {journal} {Phys. Rev. Lett.}\ }\textbf {\bibinfo {volume} {100}},\ \bibinfo
  {pages} {148701} (\bibinfo {year} {2008})}\BibitemShut {NoStop}%
\end{thebibliography}
\end{document}